\documentclass[12pt]{article}

\usepackage{graphicx}

\def\be{\begin{equation}}
\def\ee{\end{equation}}
\def\ba{\begin{eqnarray}}
\def\ea{\end{eqnarray}}
\def\ra{\rightarrow}

\def\cp{{\cal CP}}

\begin{document}

\title{Phenomenological aspects of CP violation}

\author{Jo\~{a}o P.\ Silva\\
\\
  {\small Centro de F\'{\i}sica Te\'{o}rica de Part\'{\i}culas}\\
  {\small Instituto Superior T\'{e}cnico}\\
  {\small P-1049-001 Lisboa, Portugal}\\
\\
{\small and}\\
\\
  {\small Instituto Superior de Engenharia de Lisboa}\\
  {\small Rua Conselheiro Em\'{\i}dio Navarro}\\
  {\small 1900 Lisboa, Portugal}}

\date{}

\maketitle

\begin{abstract}
We present a pedagogical review of the phenomenology of CP violation,
with emphasis on $B$ decays.
Main topics include the phenomenology of neutral meson systems,
CP violation in the Standard Model of electroweak interactions,
and $B$ decays.
We stress the importance of the reciprocal basis,
sign conventions,
rephasing invariance,
general definitions of the CP transformation and the spurious phases
they bring about,
CP violation as originating from the clash of two contributions,
the $\rho - \eta$ plane,
the four phases of a generalized CKM matrix,
and the impact of discrete ambiguities.
Specific $B$ decays are included in order to illustrate
some general techniques used in extracting information from
$B$ physics experiments.
We include a series of simple exercises.
The style is informal.
\end{abstract}

\vspace{1.2in}

\begin{centering}
Lectures presented at the\\
Central European School in Particle Physics\\
Faculty of Mathematics and Physics, Charles University, Prague\\
September 14-24, 2004\\
\end{centering}

\newpage

\tableofcontents 

\newpage

\section{\label{ch:overview}Overview}

This set of lectures is meant as a primer on CP violation, with
special emphasis on $B$ decays.
As a result,
we include some ``fine-details'' usually glanced over
in more extensive and/or advanced presentations.
Some are mentioned in the text; some are relegated to the
exercises (which are referred to in the text by \textbf{Ex}),
collected in appendix~\ref{appendixEx}.
The other appendices can be viewed as slightly longer exercises
which have been worked out explicitly.
It is hoped that,
after going through this text and the corresponding exercises,
the students will be able to read more advanced articles
and books on the subject.
Part of what is treated here is discussed
in detail in the book ``CP violation''
by Gustavo Castelo Branco, 
Lu\'{\i}s Lavoura,
and Jo\~{a}o P.\ Silva \cite{BLS},
where a large number of other topics can be found.
We will often refer to it.

Chapter~\ref{ch:intro} includes
a brief summary of the landmark experiments
and of typical difficulties faced by theoretical interpretations
of CP violation experiments.
Chapter~\ref{ch:ph_mixing} contains a complete
description of neutral meson mixing,
including the need for the reciprocal basis,
the need for invariance under rephasing of the state vectors 
(covered in more detail in appendix~\ref{appendix-rephasing}),
and CPT violation
(relegated to appendix~\ref{appendix-CPT}, whose simple formulation
allows the trivial discussion of propagation in matter
suggested in \textbf{(Ex-37)}).
The production and decay of a neutral meson system
is covered in chapter~\ref{ch:producao},
where we point out that a fourth type of CP violation
exists,
has not been measured,
and, before it is measured,
it must be taken into account as a source of
systematic uncertainty in the extraction
of the CKM phase $\gamma$ from $B \rightarrow D$ decays,
due to $D^0 - \overline{D^0}$ mixing.
Section~\ref{sec:checklist} compiles a list of expressions whose
sign convention should be checked when comparing
different articles.
We review the Standard Model (SM) of electroweak interactions
in chapter~\ref{ch:SM},
with emphasis on CP violating quantities which are
invariant under rephasing of the quark fields.
This is used to stress that
CP violation lies not in the charged $W$ interactions,
nor in the Yukawa couplings;
but rather on the ``clash'' between the two.
We stress that there are only two large phases in
the CKM matrix -- $\beta$ and $\gamma$
($\alpha = \pi - \beta - \gamma$ \textit{by definition\/}) --
and that the interactions of the usual quarks
with $W^\pm$ require only two further phases,
regardless of the model in question
-- this is later used in section~\ref{sec:simplified_q/p}
in order to parametrize a class of new physics models
with non-unitary CKM matrix and new phases in 
$B - \overline{B}$ mixing.
We point out that the ``unitarity triangle'' provides
a comparison between information involving mixing and
information obtained exclusively from decay,
but we stress that this is only one of many
tests on the CKM matrix.
On the contrary,
the strategy of placing all CKM constraints
on the $\rho - \eta$ plane,
looking for inconsistencies,
is a generic and effective method to search for new physics.
In chapter~\ref{ch:road},
we concentrate on generic properties of $B$ decays.
We describe weak phases, strong phases,
and also the impact of the spurious phases
brought about by CP transformations.
We describe in detail the invariance of the observable $\lambda_f$
under the rephasing of both hadronic kets and
quark field operators and,
complemented in subsection~\ref{subsec:rephasing_invariance},
show how the spurious phases drop out of this physical observable.
Chapter~\ref{ch:B_decays} contains a description of
some important $B_d$ decays.

Throughout,
the emphasis is not on the detailed numerical
analysis of the latest experimental announcements
(although some such information is included)
but,
rather,
on generic lessons and strategies that may be learned from
some classes of methods used in interpreting $B$ decays.

Finally,
the usual warnings:
given its size, only a few topics could be included
in this text and their choice was mostly driven by personal taste;
also, only those references used in preparing the lectures
have been mentioned.
A more complete list can be found,
for example,
in the following books \cite{Sac,W,Jar,Sto,BS}
and reviews \cite{Gri88,Fri88,Nir92,BabarPhysBook,Nir01,RunII,CKMfitter-04}.

\section{\label{ch:intro}Introduction}

These lectures concern the behavior of elementary particles
and interactions under the following discrete symmetries:
C--charge which transforms a particle into its
antiparticle;
P--parity which reverses the spatial axis;
and T--time-reversal which inverts the time axis.
As far as we know,
all interactions except the weak interaction are invariant under
these transformations\footnote{We will ignore the strong
CP problem in these lectures.}.
The fact that C and P are violated was included in 1958
into the V-A form of the weak Lagrangian \cite{V-A}.
The interest in CP violation grew out of a 
1964 experiment by Christenson, Cronin, Fitch, and Turlay \cite{Chr64}.
The basic idea behind this experiment is quite simple:
if you find that a given particle can decay into two
CP eigenstates which have opposite CP eigenvalues,
you will have established the existence of CP violation.

There are two neutral kaon states which are eigenvectors
of the strong Lagrangian:
$K^0$, made out of $\bar s\, d$ quarks,
and $\overline{K^0}$, composed of the $s\, \bar d$ quarks.
A generic state with one neutral kaon will necessarily be
a linear combination of these two states.
Clearly,
the charge transformation (C) exchanges
$K^0$ with $\overline{K^0}$,
while the parity transformation (P) inverts the 3-momentum.
Therefore,
the composed transformation CP acting on the state $K^0(\vec{p})$
yields the state $\overline{K^0}(-\vec{p})$.
Given that the physical states correspond to kets which
are defined up to a phase \cite{Dirac},
we may write
\begin{equation}
{\cal CP} | K^0(\vec{p}) \rangle =
e^{i \xi} | \overline{K^0}(-\vec{p}) \rangle.
\label{CPonkaon}
\end{equation}
We name $\xi$ the
``spurious phase brought about by the CP transformation''.
From now on we will consider the kaon's rest frame,
suppressing the reference to the kaon momentum.
The states
\be
| K_\pm \rangle =
\frac{1}{\sqrt{2}} \left(
| K^0 \rangle \pm e^{i \xi} | \overline{K^0} \rangle
\right)
\label{K_pm}
\ee
are eigenstates of CP, corresponding to the eigenvalues
$\pm 1$, 
respectively.
Let us start by assuming that CP is a good symmetry of the
total Hamiltonian.
Then,
the eigenstates of the Hamiltonian are simultaneously eigenstates
of CP, and they should only decay into final states with the same
CP eigenvalue.

On the other hand,
the states of two and three pions obtained from the
decay of a neutral kaon obey
\textbf{(Ex-1)}
\ba
{\cal CP} |\pi \pi \rangle &=& |\pi \pi \rangle,
\nonumber\\
{\cal CP} {|\pi \pi \pi \rangle}_0 &=&
-\, {|\pi \pi \pi \rangle}_0\ ,
\label{CP-of-pipi}
\ea
where ${|\pi \pi \pi \rangle}_0$ denotes
the ground state of the three pion system.

Therefore,
if we continue to assume CP conservation, we are forced
to conclude that
$|K_+\rangle$ can only decay into two pions
(or to some excited state of the three pion system).
In contrast,
$|K_-\rangle$ cannot decay into two pions,
but it can decay into the ground state of the
three pion system.\footnote{Of course, both states can decay
semileptonically.}
Since the phase-space for the decay into two pions is larger than
that for the decay into three pions
(whose mass almost adds up to the kaon mass),
we conclude that the lifetime of $|K_+\rangle$
should be smaller than that of $|K_-\rangle$.
As a result,
the hypothesis that CP is conserved by the total Hamiltonian,
leads to the correspondence
\ba
|K_+\rangle &\longrightarrow& |K_S\rangle
\nonumber\\
|K_-\rangle &\longrightarrow& |K_L\rangle
\ea
where $K_S$ ($K_L$) denotes the short-lived (long-lived) kaon.

Experimentally,
there are in fact two kaon states with widely different
lifetimes:
$\tau_S=(8.953\pm 0.006) \times 10^{-11}\, \mbox{s}$;
$\tau_L=(5.18\pm 0.04) \times 10^{-8}\, \mbox{s}$ \cite{PDG}.
This has the following interesting consequence:
given a kaon beam,
it is possible to extract the long-lived
component by waiting for the beam to ``time-evolve''
until times much larger than a few times $\tau_S$.
For those times,
the beam will contain only $K_L$,
which could decay into three pions.
What Christenson, Cronin, Fitch, and Turlay found was that
these $K_L$,
besides decaying into three pions, as expected,
also decayed occasionally into two pions.
This established CP violation.

Although this is a 1964 experiment,
we had to wait until 1999 for a different type of
CP violation to be agreed upon \cite{epsilon_prime};
and this still in the neutral kaon system.
Events soon accelerated with the announcement
in July 2000 by BABAR (at SLAC, USA) and Belle (at KEK, Japan)
of the first hints of CP violation in a completely different
neutral meson system \cite{Babarfirst,Bellefirst};
the $B_d$ meson system, which is a heavier ``cousin'' of the
kaon, involving the quarks $b$ and $d$.
The results obtained by July 2001 \cite{Babarnew,Bellenew}
already excluded CP conservation in the $B_d$ meson system
at the 99.99\% C.L.

Part of the current interest in this field stems from these
two facts:
we had to wait 37 years to detect CP violation outside
the kaon system;
and there are now a large number of results involving
CP violation in the $B$ system -- a number which is rapidly growing.
This allows us to probe deeper and deeper into 
the exact nature of CP violation.

\begin{figure}[htb]
\centerline{\includegraphics*[height=3in]{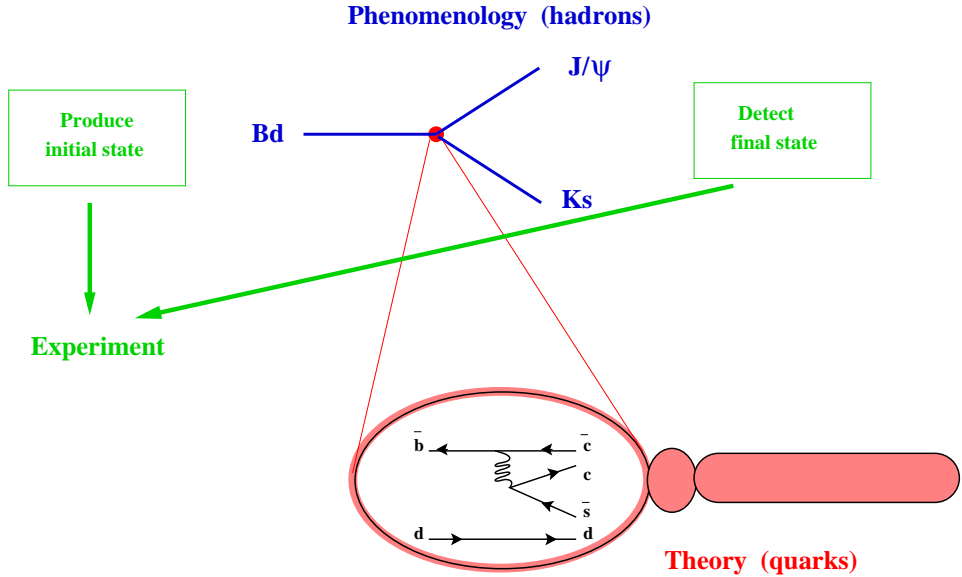}}
\caption{\label{generic_exp}Generic $B$ physics experiment.
}
\end{figure}

FIG.~\ref{generic_exp} shows a generic $B$ physics experiment.
One produces the initial state; it time-evolves; eventually it decays;
and the final state products are identified in the detector.
The experimental details involved in the production of $B$
mesons and in the detection of the final state products have taken
thousands of dedicated experimentalists years to perfect and cannot
possibly be discussed in these lectures.
However,
one should be aware that therein lie a number of aspects that
even theorists must cope with sooner or later:
Which final state particles are easy/difficult to detect?;
How do vertexing limitations affect our ability to follow the
time-dependent evolution of the initial neutral $B_d$ or $B_s$ meson?;
If we produce initially a $B_d - \overline{B_d}$ pair,
how are these two neutral mesons correlated?;
How can that correlation be used to tag the ``initial'' flavor of the
$B_d$ meson under study?;
and many, many others issues \dots \ 
Here, we will concentrate on the time-evolution and on the decay.

Unfortunately,
FIG.~\ref{generic_exp} already indicates
the most serious problem affecting our interpretation of these
experiments:
the theory is written in terms of the
fundamental quark fields;
while the experiment deals with hadrons.
Indeed,
we will need to calculate the matrix elements of
the effective quark-field operators when placed
in between initial and final states containing hadrons;
the so-called hadronic matrix elements:
\be
\langle\ \mbox{\small final state hadrons}\
|\ {\cal O}( \mbox{\small quark field operator})\ |\ 
\mbox{\small initial state hadrons}\ \rangle
\label{quarkVShadron}
\ee
Because knowing how the quarks combine into hadrons involves QCD
at low energies,
these quantities are plagued with uncertainties and are
sometimes known as the hadronic ``messy'' elements.

This is not to say that all is hopeless. Fortunately,
on the one hand,
there are a number of techniques which allow us to
have some control over these quantities in certain
special cases and, on the other hand,
there are certain decays which have particularly simple
interpretations in terms of the parameters in
the fundamental weak Lagrangian involving quarks.
But this crucial difficulty means that not all
experiments have clean theoretical interpretations
and forces us to seek information in as many different decays
as possible.

One final technical detail is worth mentioning.
As emphasized before,
the kets corresponding to physical states are defined
up to an overall phase \cite{Dirac};
we are free to rephase those kets at will.
Of course,
any phenomenological parameter describing CP violation must be
rephasing invariant and, thus,
must arise from the clash of two phases.
We will come back to this point in subsection~\ref{sec:usual_three}.
We are also free to rephase the quark field operators
which appear in our theoretical Lagrangian.
Again, this implies that all CP violating
quantities must arise from the clash of two phases
and that we must search for rephasing invariant
combinations of the parameters in the weak Lagrangian.
This is covered in section~\ref{sec:def_CPV_J}.
These two types of rephasing invariance have one further consequence.
Many authors write their expressions using specific
phase conventions: sometimes these choices are made explicit;
sometimes they are not.
As a result,
we must exercise great care when comparing expressions
in different articles and books.
Expressions where these questions become acutely critical
will be pointed out throughout these lectures.

\section{\label{ch:ph_mixing}Phenomenology of neutral meson mixing}

\subsection{Neutral meson mixing: the flavor basis}

We are interested in describing the mixing of a neutral meson $P^0$
with its antiparticle $\overline{P^0}$,
where $P$ stands for $K$, $D$, $B_d$ or $B_s$.
We will follow closely the presentation in \cite{reciprocal}.
In a given approximation \cite{LOY},
we may study the mixing in this particle--antiparticle
system separately from its subsequent decay.
The time evolution of the state $|\psi (t)\rangle$ 
describing the $P^0 - \overline{P^0}$ mixed state is given by
\be
i \frac{d}{dt} |\psi (t)\rangle =
\mbox{\boldmath $H$}\, |\psi (t)\rangle,
\label{time_evolution}
\ee
where $\mbox{\boldmath $H$}$ is a  $2 \times 2$ matrix
written in the $P^0 - \overline{P^0}$ rest frame,
and $t$ is the proper time.
It is common to break $\mbox{\boldmath $H$}$ into its
hermitian and anti-hermitian parts,
$\mbox{\boldmath $H$} = \mbox{\boldmath $M$} -i/2 \mbox{\boldmath $\Gamma$}$,
where
\begin{eqnarray}
\mbox{\boldmath $M$} &=& 
\left( \mbox{\boldmath $H$} + \mbox{\boldmath $H$}^\dagger \right)/2,
\nonumber\\
-i \mbox{\boldmath $\Gamma$}/2 &=& 
\left( \mbox{\boldmath $H$} - \mbox{\boldmath $H$}^\dagger \right)/2,
\end{eqnarray}
respectively.
Both $\mbox{\boldmath $M$}$ and $\mbox{\boldmath $\Gamma$}$ are hermitian.

The $\{| P^0 \rangle, | \overline{P^0} \rangle \}$ flavor
basis satisfies a number of common relations,
among which:
the orthonormality conditions
\begin{eqnarray}
\langle P^0 | \overline{P^0} \rangle =
\langle \overline{P^0} | P^0  \rangle 
&=& 0,
\nonumber\\
\langle P^0 | P^0 \rangle =
\langle \overline{P^0} | \overline{P^0} \rangle &=& 1;
\label{flavour:orthogonal}
\end{eqnarray}
the fact that $| P^0 \rangle \langle P^0|$ and
$|\overline{P^0} \rangle \langle \overline{P^0}|$
are projection operators;
the completeness relation
\begin{equation}
| P^0 \rangle \langle P^0|
+
|\overline{P^0} \rangle \langle \overline{P^0}|
=
1;
\label{flavour:completeness}
\end{equation}
and the decomposition of the effective Hamiltonian as
\begin{eqnarray}
{\cal H} 
&=&
| P^0 \rangle H_{11} \langle P^0|
+
| P^0 \rangle H_{12} \langle \overline{P^0}|
+
| \overline{P^0} \rangle H_{21} \langle P^0|
+
| \overline{P^0} \rangle H_{22} \langle \overline{P^0}|
\nonumber\\*[3mm]
&=&
\left(
\begin{array}{cc}
| P^0 \rangle, & | \overline{P^0} \rangle
\end{array}
\right)
\mbox{\boldmath $H$}
\left(
\begin{array}{c}
\langle P^0 | \\
\langle \overline{P^0}|
\end{array}
\right).
\label{decomposition}
\end{eqnarray}
All these relations involve the basis of flavor eigenkets
$\{| P^0 \rangle, | \overline{P^0} \rangle \}$
and the basis of the corresponding bras
$\{\langle P^0 |, \langle \overline{P^0}| \}$.

Under a CP transformation
\ba
H_{12} \equiv \langle P^0 |{\cal H}| \overline{P^0} \rangle
&\stackrel{{\cal CP}}{\longrightarrow}&
\langle P^0 |({\cal CP})^\dagger\ ({\cal CP})\,{\cal H}\,({\cal CP})^\dagger\
({\cal CP})| \overline{P^0} \rangle
\nonumber\\
& = &
\langle \overline{P^0} |e^{-i\xi}\, {\cal H}_{\rm cp}\,
e^{-i\xi}| P^0 \rangle
\nonumber\\
& = &
e^{-2i\xi}\ 
\langle \overline{P^0} | {\cal H}_{\rm cp}| P^0 \rangle
\\
H_{11} \equiv \langle P^0 |{\cal H}| P^0 \rangle
&\stackrel{{\cal CP}}{\longrightarrow}&
\langle P^0 |({\cal CP})^\dagger\ ({\cal CP})\,{\cal H}\,({\cal CP})^\dagger\
({\cal CP})| P^0 \rangle
\nonumber\\
& = &
\langle \overline{P^0} |{\cal H}_{\rm cp}| \overline{P^0} \rangle,
\ea
where
\be
{\cal H}_{\rm cp} \equiv ({\cal CP})\, {\cal H}\, ({\cal CP})^\dagger.
\ee
Therefore,
if CP is conserved, 
${\cal H}={\cal H}_{\rm cp}$,
\be
H_{12} = e^{-2i\xi} H_{21} \ \ \ \mbox{and}\ \ \ 
H_{11} = H_{22}.
\label{cp_conservation}
\ee
Because $\xi$ is a spurious phase without physical significance,
we conclude that the phases of $H_{12}$ and $H_{21}$ also lack meaning.
(This is clearly understood by noting that these matrix elements
change their phase under independent rephasings of
$|P^0 \rangle$ and $|\overline{P^0} \rangle$.)
As a result,
the conclusion with physical significance contained
in the first implication of CP conservation
in Eq.~(\ref{cp_conservation}) is $|H_{12}|=|H_{21}|$. 
A similar study can be made for the other discrete symmetries \cite{BLS},
leading to:
\begin{eqnarray}
{\cal CPT}\ \mbox{conservation}
& \Rightarrow & H_{11} = H_{22},
\nonumber\\ 
{\cal T}\ \mbox{conservation}
& \Rightarrow & |H_{12}| = |H_{21}|,
\nonumber\\ 
{\cal CP}\ \mbox{conservation}
& \Rightarrow & H_{11} = H_{22}\ 
\mbox{and}\ |H_{12}| = |H_{21}|.
\label{effect_of_discrete_symmetries}
\end{eqnarray}
In the most general case, these symmetries are broken and
the matrix $\mbox{\boldmath $H$}$ is completely arbitrary.

In the rest of this main text we will assume that CPT is conserved
and $H_{11} = H_{22}$.
As a result,
all CP violating observables occurring
in $P^0 - \overline{P^0}$ mixing must be proportional to
\be
\delta \equiv \frac{|H_{12}|-|H_{21}|}{|H_{12}|+|H_{21}|}.
\label{delta_definition}
\ee
For completeness,
the general case is discussed in appendix \ref{appendix-CPT}.

\subsection{Neutral meson mixing: the mass basis}

The time evolution in Eq.~(\ref{time_evolution}) becomes
trivial in the mass basis which diagonalizes the Hamiltonian
$\mbox{\boldmath $H$}$.
We denote the (complex) eigenvalues of $\mbox{\boldmath $H$}$
by 
\ba
\mu_H &=& m_H -i/2 \Gamma_H,
\nonumber\\
\mu_L &=& m_L -i/2 \Gamma_L,
\label{mass_eigenvalues}
\ea
corresponding to the
eigenvectors\footnote{A choice on the relative phase between
$| P_H \rangle$ and $| P_L \rangle$ was implicitly 
made in Eq.~(\ref{PaPb}).
Indeed, we chose $\langle P^0 | P_H \rangle$
to have the same phase as $\langle P^0 | P_L \rangle$.
Whenever using Eq.~(\ref{PaPb}) one should be careful
not to attribute physical significance to any phase which would vary
if the phases of $| P_H \rangle$ and of $| P_L \rangle$
were to be independently changed.
A similar phase choice affects Eq.~(\ref{K_pm}).
If one forgets that these phase choices have been made,
one can easily reach fantastic (and wrong!) ``new discoveries''.
}
\begin{eqnarray}
\left(
\begin{array}{c}
| P_H \rangle \\ | P_L \rangle
\end{array}
\right)
=
\left(
\begin{array}{cc}
p & - q\\ p & q
\end{array}
\right)
\
\left(
\begin{array}{c}
| P^0 \rangle \\ | \overline{P^0} \rangle
\end{array}
\right)
= 
\mbox{\boldmath $X$}^T\ 
\left(
\begin{array}{c}
| P^0 \rangle \\ | \overline{P^0} \rangle
\end{array}
\right).
\label{PaPb}
\end{eqnarray}
Although not strictly necessary, the labels $H$ and $L$ used
here stand for the ``heavy'' and ``light'' eigenstates
respectively. This means that we are using a convention in
which $\Delta m = m_H - m_L > 0$.
We should also be careful with the explicit choice
of $-q$ ($+q$) in the first (second) line of Eq.~(\ref{PaPb});
the opposite choice has been made in references \cite{BLS,reciprocal}.
Remember: in the end, minus signs \textit{do matter\/}!!

It is convenient to define
\ba
m - i \Gamma/2
\equiv
\mu
&\equiv&
\frac{\mu_H + \mu_L}{2}
\\
\Delta m - i \Delta \Gamma/2
\equiv 
\Delta \mu
&\equiv&
\mu_H - \mu_L
\ea
The relation between these parameters and the matrix elements
of $\mbox{\boldmath $H$}$ written in the flavor basis
is obtained through the diagonalization
\begin{equation}
\mbox{\boldmath $X$}^{-1} \mbox{\boldmath $H$} \mbox{\boldmath $X$} = 
\left(
\begin{array}{cc}
\mu_H & 0\\ 0 & \mu_L
\end{array}
\right),
\label{diagonalization}
\end{equation}
where \textbf{(Ex-2)}
\begin{equation}
\mbox{\boldmath $X$}^{-1} = \frac{1}{2 p q}
\left(
\begin{array}{cc}
q & - p\\ q & p
\end{array}
\right).
\label{X-1}
\end{equation}
We find
\ba
\mu 
& = &
H_{11}=H_{22},
\\
\Delta \mu 
& = &
2 \sqrt{H_{12} H_{21}},
\label{delta_mu}
\\
\frac{q}{p}
& = &
- \sqrt{\frac{H_{21}}{H_{12}}}
=
- \frac{2\, H_{21}}{\Delta \mu}.
\label{calc:q/p}
\ea
It is easier to obtain these equations by inverting
Eq.~(\ref{diagonalization}) \textbf{(Ex-3)}:
\be
\mbox{\boldmath $H$}  = 
\mbox{\boldmath $X$}
\left(
\begin{array}{cc}
\mu_H & 0\\*[2mm]
0 & \mu_L
\end{array}
\right)
\mbox{\boldmath $X$}^{-1}
=
\left(
\begin{array}{cc}
\mu & - \frac{\Delta \mu}{2} \frac{p}{q}\\*[2mm]
- \frac{\Delta \mu}{2} \frac{q}{p} & \mu
\end{array}
\right).
\label{diagonalization_inverted}
\ee
Eq.~(\ref{diagonalization_inverted}) is interesting because
it expresses the quantities which are calculated in a given
theory, $H_{ij}$,
in terms of the physical observables.
Recall that the phase of $H_{12}$ and, thus,
of $q/p$,
is unphysical,
because it can be changed through independent rephasings
of $|P^0 \rangle$ and $| \overline{P^0} \rangle$.

Eq.~(\ref{delta_mu}) can be cast in a more familiar form by
squaring it and separating the real and imaginary parts,
to obtain
\ba
(\Delta m)^2 - \frac{1}{4} (\Delta \Gamma)^2
& = &
4 |M_{12}|^2 -|\Gamma_{12}|^2,
\label{real_dmu_2}
\nonumber\\
(\Delta m) (\Delta \Gamma)
& = &
4 \mbox{Re}(M_{12}^\ast \Gamma_{12}).
\label{imag_dmu_2}
\ea
On the other hand,
it is easy to show that
\be
\delta
= 
\frac{ \left| \frac{p}{q} \right| - \left| \frac{q}{p} \right| }{
\left| \frac{p}{q} \right| + \left| \frac{q}{p} \right|}
=
\frac{2 \mbox{Im} (M_{12}^\ast \Gamma_{12})}{(\Delta m)^2
+|\Gamma_{12}|^2}.
\label{delta_new_expressions}
\ee
Eqs.~(\ref{real_dmu_2})--(\ref{delta_new_expressions}) can be
rearranged as in \textbf{(Ex-4)}.

\subsubsection{Mixing in the neutral kaon sector}

By accident,
the neutral kaon sector satisfies
$\Delta m_K \approx - \frac{1}{2} \Delta \Gamma_K$.
On the other hand,
$\delta_K$ is of order $10^{-3}$.
Combining these informations leads to
\be
\Delta m \approx 2 |M_{12}| \approx - \frac{1}{2} \Delta \Gamma
\approx |\Gamma_{12}|,
\label{relations_kaon_mixing}
\ee
and, thus,
\be
\delta_K 
\approx \frac{\mbox{Im}
(M_{12}^\ast\ \Gamma_{12}/|\Gamma_{12}|)}{\Delta m}
\approx
\frac{1}{4} \mbox{Im}\left( \frac{\Gamma_{12}}{M_{12}} \right).
\label{approx_kaon_delta}
\ee
Some authors describe this as the imaginary part of $M_{12}$
because they use a specific phase convention under which $\Gamma_{12}$
is real.

\subsubsection{\label{subsec:delta_B}Mixing in the neutral
$B_d$ and $B_s$ systems}

In both the $B_d$ and $B_s$ systems,
it can be argued
(as we will see below)
that $|\Gamma_{12}| \ll |M_{12}|$.
As a result,
\ba
\Delta m_B & = & 2 |M_{12}|,
\nonumber\\
\Delta \Gamma_B & = & 2 \mbox{Re}(M_{12}^\ast \Gamma_{12})/|M_{12}|,
\nonumber\\
\frac{q}{p} & = &
- \frac{M_{12}^\ast}{|M_{12}|}
\left[ 
1 - \frac{1}{2}
\mbox{Im}\left( \frac{\Gamma_{12}}{M_{12}} \right)
\right],
\ea
where the last expression has been expanded to next-to-leading
order in $|\Gamma_{12}/M_{12}|$ \cite{Nir01},
so that both the first and last equality in
Eq.~(\ref{delta_new_expressions}) lead consistently to
\be
\delta_B \approx
\frac{1}{2} \mbox{Im}\left( \frac{\Gamma_{12}}{M_{12}} \right).
\label{approx_kaon_B}
\ee

We now turn to an intuitive explanation of
why $|\Gamma_{12}|$ should be much smaller than $|M_{12}|$
\cite{Nir92,Nir01}.
The idea is the following:
one starts from
\be
\left| \frac{\Gamma_{12}}{M_{12}} \right|
=
2 \frac{ |\Gamma_{12}|/\Gamma}{\Delta m/\Gamma};
\ee
one argues that
\be
\Gamma_{12} = \sum_f \langle f | T | P^0 \rangle^\ast
\langle f | T | \overline{P^0} \rangle
\label{calc:Gamma_12}
\ee
should be dominated by the Standard Model tree-level diagrams;
one estimates what this contribution might be;
and, finally,
one uses a measurement of
$x_d = (\Delta m/\Gamma)_{B_d} = 0.771 \pm 0.012$ 
and an upper bound on
$x_s = (\Delta m/\Gamma)_{B_s} > 20.6$ with C.L.$= 95\%$
from experiment \cite{PDG}.

Clearly,
Eq.~(\ref{calc:Gamma_12}) only involves channels common
to $P^0$ and $\overline{P^0}$.
In the $B_d$ system,
such channels are CKM-suppressed and their
branching ratios are at or below the level of $10^{-3}$.
Moreover,
they come into Eq.~(\ref{calc:Gamma_12}) with opposite signs.
Therefore,
one expects that the sum does not exceed the individual level,
leading to $|\Gamma_{12}|/\Gamma < 10^{-2}$ as a rather
safe bound.
Combined with $x_d$,
we obtain $|\delta_{B_d}| < {\cal O} \left( 10^{-2} \right)$
for the $B_d$ system.

The situation in the $B_s$ system is rather different because
the dominant decays common to $B^0_s$ and
$\overline{B^0_s}$ are due to the tree-level transitions
$b \rightarrow c \bar c s$.
Therefore,
$\Gamma_{12}/\Gamma$ is expected to be large.
One estimate by Beneke, Buchalla and Dunietz
yields \cite{BBD},
\be
\left| \frac{\Gamma_{12}}{\Gamma} \right|
= \frac{1}{2} \left( 0.16^{+0.11}_{-0.09} \right).
\ee
Fortunately,
this large value is offset by the strong
lower bound on $x_s$,
leading, again,
to $|\delta_{B_s}| < {\cal O} \left( 10^{-2} \right)$.

These arguments are rather general and should hold in a variety of new
physics models.
Precise calculations within the SM lead to 
$\delta_{B_d} \sim - 2.5 \times 10^{-4}$  and 
$\delta_{B_s} \sim 0.1 \times 10^{-4}$ \cite{seemore}.

Incidentally,
the analysis discussed here
means that $\Delta \Gamma$ can be set to zero in the
$B_d$ system but that it must be taken into account
in the time evolution of the $B_s$ system \cite{Dunietz95}.

\subsection{\label{sec:reciprocal}The need for the reciprocal basis}

We now come to a problem frequently overlooked.
Is the matrix $\mbox{\boldmath $X$}$ in Eq.~(\ref{diagonalization})
a unitary matrix or not?
The answer comes from introductory algebra:
matrices satisfying 
$[ \mbox{\boldmath $H$}, \mbox{\boldmath $H$}^\dagger] = 0$
are called ``normal'' matrices.
Equivalent definitions are
(that is to say that $\mbox{\boldmath $H$}$ is normal
if and only if):
i) $\mbox{\boldmath $X$}$ is a unitary matrix;
ii) the left-eingenvectors and the right-eigenvectors
of $\mbox{\boldmath $H$}$ coincide;
iii) $[ \mbox{\boldmath $M$}, \mbox{\boldmath $\Gamma$} ] = 0$;
among many other possible equivalent statements.

Now,
the 1964 experiment mentioned above implies that
$|H_{12}| \neq |H_{21}|$
holds in the neutral kaon system,
thus establishing CP and T violation
in the $K^0 - \overline{K^0}$ mixing.
But,
this also has an important implication for the matrix $\mbox{\boldmath $X$}$.
Indeed,
the $(1,1)$ entry in the matrix 
$[\mbox{\boldmath $H$} , \mbox{\boldmath $H$}^\dagger]$
is given by $|H_{12}|^2 - |H_{21}|^2$.
Therefore,
that experimental result also implies that the matrix $\mbox{\boldmath $H$}$
is not normal and,
thus,
that we are forced to deal with non-unitary matrices in the neutral kaon
system \textbf{(Ex-5)}.
As for the other neutral meson systems,
$|H_{12}| \neq |H_{21}|$ has not yet been established experimentally.
Nevertheless,
the Standard Model predicts that,
albeit the difference is small,
$|H_{12}| \neq |H_{21}|$ does indeed hold.
As before,
this implies CP violation in the mixing and forces the use of 
a non-unitary mixing matrix $\mbox{\boldmath $X$}$ \cite{reciprocal}.

So, why do (most) people worry about performing non-unitary transformations?
The reason is that one would like the mass basis
$\{| P_H \rangle, | P_L \rangle \}$ 
to retain a number of the nice (orthonormality) features of the
$\{| P^0 \rangle, | \overline{P^0} \rangle \}$ flavor
basis;
Eqs.~(\ref{flavour:orthogonal})--(\ref{decomposition}).
%
The problem is that,
when $\mbox{\boldmath $H$}$ is not normal,
we \textit{cannot} find similar relations involving the basis of mass eigenkets
$\left\{| P_H \rangle, | P_L \rangle \right\}$
and the basis of the corresponding bras,
$\left\{ \langle P_H |, \langle P_L | \right\}$.
Indeed,
substituting Eq.~(\ref{diagonalization_inverted}) into
Eq.~(\ref{decomposition})
we find
\begin{eqnarray}
{\cal H}
&=&
\left(
\begin{array}{cc}
| P^0 \rangle, & | \overline{P^0} \rangle
\end{array}
\right)
\,
\mbox{\boldmath $X$}
\left(
\begin{array}{cc}
\mu_H & 0\\*[2mm]
0 & \mu_L
\end{array}
\right)
\mbox{\boldmath $X$}^{-1}
\,
\left(
\begin{array}{c}
\langle P^0 | \\
\langle \overline{P^0}|
\end{array}
\right)
\nonumber\\*[3mm]
&=&
\left(
\begin{array}{cc}
| P_H \rangle, & | P_L \rangle
\end{array}
\right)
\left(
\begin{array}{cc}
\mu_H & 0\\
0 & \mu_L
\end{array}
\right)
\left(
\begin{array}{c}
\langle \tilde P_H | \\
\langle \tilde P_L |
\end{array}
\right)
\nonumber\\*[3mm]
&=&
| P_H \rangle \mu_H \langle \tilde P_H|
+
| P_L \rangle \mu_L \langle \tilde P_L|
\label{decomposition-mass}
\end{eqnarray}
This does not involve the bras $\langle P_H |$ and $\langle P_L|$,
\begin{equation}
\left(
\begin{array}{c}
\langle  P_H | \\ \langle  P_L | 
\end{array}
\right)
=
\mbox{\boldmath $X$}^{\dagger}
\left(
\begin{array}{c}
\langle P^0 | \\ \langle \overline{P^0} |
\end{array}
\right),
\label{wrong-basis}
\end{equation}
but rather the so called `reciprocal basis'
\begin{equation}
\left(
\begin{array}{c}
\langle \tilde P_H | \\ \langle \tilde P_L | 
\end{array}
\right)
=
\mbox{\boldmath $X$}^{-1}
\left(
\begin{array}{c}
\langle P^0 | \\ \langle \overline{P^0} |
\end{array}
\right).
\label{PtildeaPtildeb}
\end{equation}
The reciprocal basis may also be defined by the orthonormality
conditions
\begin{eqnarray}
\langle \tilde P_H | P_L \rangle = \langle \tilde P_L | P_H \rangle &=& 0,
\nonumber\\
\langle \tilde P_H | P_H \rangle = \langle \tilde P_L | P_L \rangle &=& 1.
\label{definition-reciprocal}
\end{eqnarray}
Moreover,
$|P_H \rangle \langle \tilde P_H|$ and
$|P_L \rangle \langle \tilde P_L|$ are projection operators,
and  the partition of unity becomes
\begin{equation}
|P_H \rangle \langle \tilde P_H| + |P_L \rangle \langle \tilde P_L|=1.
\label{partition-of-unity}
\end{equation}
If $\mbox{\boldmath $H$}$ is not normal,
then $\mbox{\boldmath $X$}$ is not unitary,
and
$\left\{ \langle P_H |, \langle P_L | \right\}$
in Eq.~(\ref{wrong-basis}) do not coincide with
$\left\{ \langle \tilde P_H |, \langle \tilde P_L | \right\}$
in Eq.~(\ref{PtildeaPtildeb}).
Another way to state this fact is to note that 
$\mbox{\boldmath $H$}$ is normal
($\mbox{\boldmath $X$}$ is unitary)
if and only if its right-eigenvectors 
coincide with its left-eigenvectors.

That these features have an impact on the $K^0 - \overline{K^0}$
system,
was pointed out long ago by Sachs \cite{Sac63,Sac64},
by Enz and Lewis \cite{Enz65},
and by Wolfenstein \cite{WOLF}.
More recently,
they have been stressed by Beuthe, L\'{o}pez-Castro and Pestieu \cite{BLP},
by Alvarez-Gaum\'{e} {\it et al.} \cite{Alv99},
by Branco, Lavoura and Silva in their book ``CP violation'' \cite{BLS},
and expanded by Silva in \cite{reciprocal}.

We stress that this is not a side issue.
For example,
if we wish to describe a final state containing a $K_S$,
as we will do when discussing the extremely important
$B_d \rightarrow J/\psi K_S$ decay,
we will need to know that the correct ``bra'' to describe
a $K_S$ in the final state is 
$\langle \tilde K_S|$, and not $\langle K_S|$.

\subsection{Time evolution}

As mentioned,
the time evolution is trivial in the mass basis:
\ba
|P_H(t)\rangle &=& e^{-i \mu_H\, t} |P_H\rangle\ ,
\nonumber\\
|P_L(t)\rangle &=& e^{-i \mu_L\, t} |P_L\rangle\ .
\label{mass-basis-time-evolution}
\ea
This can be used to study the time evolution in the flavor basis.
Let us suppose that we have identified a state as $P^0$ at
time $t=0$.
Inverting Eq.~(\ref{PaPb}),
we may write the initial state as
\be
| P^0 \rangle
=
\frac{1}{2\, p}
\left(
| P_H \rangle + | P_L \rangle
\right).
\ee
From Eq.~(\ref{mass-basis-time-evolution})
we know that,
at a later time $t$,
this state will have evolved into
\be
| P^0 (t) \rangle
=
\frac{1}{2\, p}
\left(
e^{-i \mu_H\, t} | P_H \rangle + 
e^{-i \mu_L\, t} | P_L \rangle
\right),
\ee
which, using again Eq.~(\ref{PaPb}),
may be rewritten in the flavor basis as
\begin{equation}
| P^0 (t) \rangle 
= 
\frac{1}{2}
\left(
e^{-i \mu_H\, t} + e^{-i \mu_L\, t}
\right)
| P^0 \rangle 
- \frac{q}{p}\,
\frac{1}{2}
\left(
e^{-i \mu_H\, t} - e^{-i \mu_L\, t}
\right)
| \overline{P^0} \rangle .
\end{equation}
It is easy to repeat this exercise in order
to describe the time evolution
$| \overline{P^0} (t) \rangle $ of a state identified
as $\overline{P^0}$ at time $t=0$.
Introducing the auxiliary functions
\be
g_\pm(t) \equiv
\pm
\frac{1}{2}
\left(
e^{-i \mu_H\, t} \pm e^{-i \mu_L\, t}
\right)
=
e^{-i m\, t}\ e^{- \Gamma\, t/2}
\left\{
\begin{array}{l}
\cos{\left(\frac{\Delta \mu t}{2}\right)}\\
i \sin{\left(\frac{\Delta \mu t}{2}\right)}
\end{array}
\right.
.
\label{g+-}
\ee
we can combine both results into
\ba
|P^0(t)\rangle &=&
g_+(t) |P^0\rangle + \frac{q}{p} g_-(t) |\overline{P^0}\rangle\ ,
\nonumber\\
|\overline{P^0}(t)\rangle &=&
\frac{p}{q} g_-(t) |P^0\rangle + g_+(t) |\overline{P^0}\rangle\ ,
\label{standard-time-evolution-1}
\ea
These results may also be obtained making full use
of the matrix notation introduced in the preceding sections
\textbf{(Ex-6)}.

Eq.~(\ref{g+-}) contains another expression for which there are many
choices in the literature.
The explicit $-$ sign we have chosen here for the definition
of $g_-(t)$ is \textit{not} universal.
For instances,
the $+$ sign has been chosen for the definition of $g_-(t)$
in references \cite{BLS,reciprocal}.
This goes unnoticed in all expressions involving the
product of $q$ and $g_-(t)$,
because the minus signs introduced in both definitions cancel.
This is also the notation used in the recent PDG review
by Schneider on $B^0 - \overline{B^0}$ mixing.
Another notation is used in the recent PDG review of
CP violation by Kirby and Nir \cite{PDG};
they use the sign of $q$ in Eq.~(\ref{PaPb}),
but define $g_-(t)$ without the explicit minus sign
in Eq.~(\ref{g+-}). 
I cannot stress this enough:
when comparing different articles you should check all
definitions first.

\section{\label{ch:producao}Phenomenology of the 
production and decay of neutral mesons}

\subsection{\label{sec:relevant_parameters}Identifying
the relevant parameters}

\begin{figure}[htb]
\centerline{\includegraphics*[height=2in]{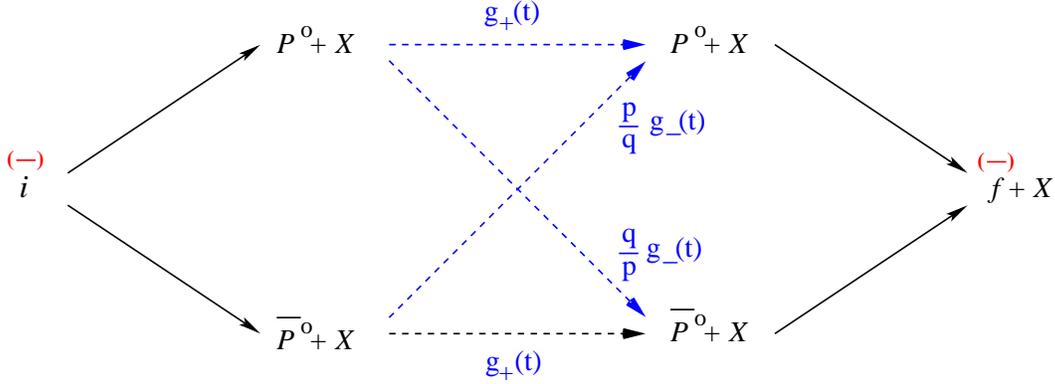}}
\caption{\label{figcasc}Schematic description of the decay
chain $i \rightarrow P + X \rightarrow f + X$.
}
\end{figure}

Let us consider the chain
$i \rightarrow P + X \rightarrow f + X$
shown in FIG.~\ref{figcasc},
in which the initial state $i$ originates the production of
a neutral meson $P$,
which evolves in time,
decaying later into a final state $f$.\footnote{This case,
as well as the case in which $i$ can also belong to a neutral meson
system, was first described 
in \cite{Santos}.
It allows us, for instance,
to provide a complete description for decays of the type
$B_d \rightarrow D X \rightarrow f X$,
even in the presence of $D^0 - \overline{D^0}$ mixing;
the so-called ``cascade decays''.}
In what follows we will skip the explicit reference
to the set of particles $X$ which is produced in
association with the neutral meson $P$,
except in appendix~\ref{appendix-rephasing}, 
where the explicit reference to $X$ will become necessary.

The amplitude for this decay chain (and its CP conjugated)
depends on the amplitudes for the initial process
\begin{eqnarray}
A_{i \rightarrow P^0} \equiv
\langle P^0 | T | i \rangle\ ,
& \hspace{5mm} &
A_{\bar i \rightarrow P^0} \equiv
\langle P^0 | T | \bar i \rangle\ ,
\nonumber\\
A_{i \rightarrow \overline{P^0}} \equiv
\langle \overline{P^0} | T | i \rangle\ ,
& \hspace{5mm} &
A_{\bar i \rightarrow \overline{P^0}} \equiv
\langle \overline{P^0} | T | \bar i \rangle\ ;
\end{eqnarray}
it depends on the parameters describing the time-evolution
of the neutral $P$ system, including $q/p$;
and it also depends on the amplitudes for the decay into the final state,
\begin{eqnarray}
A_f \equiv
\langle f | T | P^0 \rangle\ ,
& \hspace{5mm} &
A_{\bar f} \equiv
\langle \bar f | T | P^0 \rangle\ ,
\nonumber\\
\bar A_f \equiv
\langle f | T | \overline{P^0} \rangle\ ,
& \hspace{5mm} &
\bar A_{\bar f} \equiv
\langle \bar f | T | \overline{P^0} \rangle\ .
\end{eqnarray}

As mentioned,
all states may be redefined by an arbitrary
phase transformation \cite{Dirac}.
Such transformations change the mixing parameters and
the transition amplitudes\footnote{These issues are described
in detail in appendix~\ref{appendix-rephasing},
which contains discussions on these phase transformations;
the quantities which are invariant under those transformations;
the definition of CP transformations;
and the identification of those CP violating quantities which
are invariant under arbitrary phase redefinitions of the states.}.
Clearly,
the magnitudes of the transition amplitudes and the magnitude $|q/p|$
are all invariant under those transformations.
Besides these magnitudes,
there are quantities which are invariant under those arbitrary
phase redefinitions and which arise from the
``interference'' between the parameters describing the mixing and
the parameters describing the transitions:
\begin{eqnarray}
\lambda_f \equiv \frac{q}{p}
\frac{\bar A_f}{A_f}\ ,
& \hspace{5mm} &
\lambda_{\bar f} \equiv \frac{q}{p}
\frac{\bar A_{\bar f}}{A_{\bar f}}\ ,
\label{def:int-2}
\\ 
\xi_{i \rightarrow P} \equiv
\frac{A_{i \rightarrow \overline{P^0}}}{A_{i \rightarrow P^0}}
\frac{p}{q}\ ,
& \hspace{5mm} &
\xi_{\bar i \rightarrow P} \equiv
\frac{A_{\bar i \rightarrow \overline{P^0}}}{
A_{\bar i \rightarrow P^0}}
\frac{p}{q}\ .
\label{def:int-3}
\end{eqnarray}
The parameters in Eq.~(\ref{def:int-2})
describe the interference between the mixing in the
$P^0 - \overline{P^0}$ system and the subsequent
\textit{decay from that system}
into the final states $f$ and $\bar f$,
respectively.
In contrast,
the parameters in Eq.~(\ref{def:int-3})
describe the interference between the
\textit{production of the system} 
$P^0 - \overline{P^0}$ and the mixture in that
system\footnote{Please notice that the observables
$\xi_{i \rightarrow P}$ and $\xi_{\bar i \rightarrow P}$ 
bear no relation whatsoever to the spurious phases
$\xi$ which show up in the definition of the CP
transformations, as in Eq.~(\ref{CPonkaon}).}.

\subsubsection{\label{sec:usual_three}The usual three
types of CP violation}

With a simple analysis described in appendix~\ref{appendix-rephasing},
we can identify those observables which signal
CP violation:
\begin{enumerate}
\item $|q/p| - 1$ describes CP violation in the
mixing of the neutral meson system;
\item $|A_{i \rightarrow P^0}| - |A_{\bar i \rightarrow \overline{P^0}}|$
and $|A_{i \rightarrow \overline{P^0}}| - |A_{\bar i \rightarrow P^0}|$,
on the one hand,
and $|A_{f}| - |\bar A_{\bar f}|$ and $|A_{\bar f}| - |\bar A_{f}|$,
on the other hand,
describe the CP violation present directly in the
production of the neutral meson system and in its decay,
respectively;
\item $\arg \lambda_{f} + \arg \lambda_{\bar f}$
measures the CP violation arising from the interference
between mixing in the neutral meson system and
\textit{its subsequent decay} into the final states $f$ and $\bar f$.
We call this the 
\textit{``interference CP violation: first mix, then decay''}. 
When $f=f_{cp}$ is an CP eigenstate,
this CP violating observable $\arg \lambda_{f} + \arg \lambda_{\bar f}$,
becomes proportional to $\mbox{Im} \lambda_{f}$.
\end{enumerate}
These are the types of CP violation discussed in the usual
presentations of CP violation, since they are the ones
involved in the evolution and decay of the neutral meson
system (\textit{cf.\/} section~\ref{sec:decays_from}).

These three types of CP violation have been measured.
And, due to the rephasing freedom
$|P^0 \rangle \rightarrow e^{i \gamma} |P^0 \rangle$,
these must arise from the clash between two phases.
More information about these types of CP violation
will be discussed in later sections.
The combination of all this information
may be summarized schematically 
as\footnote{See \cite{complicacao} for the
relation with
$\epsilon_K$ and $\epsilon^\prime_K$.}:
\begin{enumerate}
\item Clash mixing $M_{12}$ with $\Gamma_{12}$: $|q/p|-1$
\begin{itemize}
\item CPV in mixing
\item measured in kaon system through $\epsilon_K$
\end{itemize}
\begin{figure}[htb]
\centerline{\includegraphics*[height=1in]{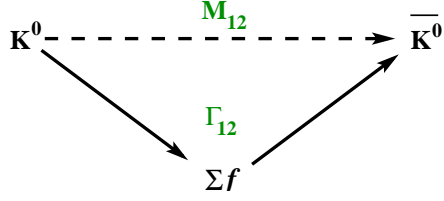}}
\caption{\label{fig:CPV_mixing}Schematic mixing CP violation
in neutral kaon mixing.
}
\end{figure}
\item Clash two direct decay paths:\footnote{Given two
distinct final states which are eigenstates of CP,
$f_{\rm cp}$ and $g_{\rm cp}$,
the difference $\lambda_f - \lambda_g$ also measures CPV
in the decays;
and it does so without the need for strong phases.
See appendix~\ref{appendix-rephasing} for details.} $|\bar A/A|-1$
\begin{itemize}
\item CPV in decay
\item measured in kaon system through $\epsilon^\prime_K$
\end{itemize}
\begin{figure}[htb]
\centerline{\includegraphics*[height=1in]{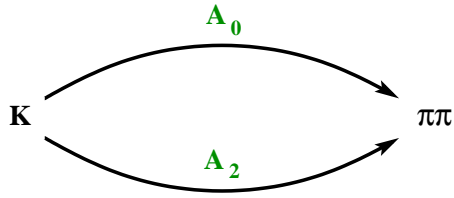}}
\caption{\label{fig:CPV_direct}Schematic direct CP violation
in neutral kaon decays into two pions.
}
\end{figure}
\item Clash direct path with mixing path; first mix--then decay: 
$\lambda_f = q_B/p_B\ \bar A_f/A_f$
\begin{itemize}
\item CPV in interference; first mix--then decay
\item measured in $B_d$ system through $\sin{2 \beta}$
\end{itemize}
\begin{figure}[htb]
\centerline{\includegraphics*[height=1in]{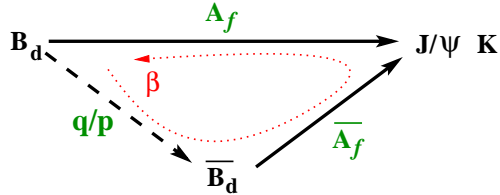}}
\caption{\label{fig:CPV_interference_1}Schematic interference
CP violation in the decay $B_d \rightarrow J/\psi K_S$.
}
\end{figure}
\end{enumerate}

\subsubsection{A fourth type of CP violation}

However,
in considering the production mechanism of the neutral meson system
we are lead to consider a novel observable,
which also signals CP violation,
\be
\arg \xi_{i \rightarrow P} + \arg \xi_{\bar i \rightarrow P}.
\ee
This observable measures the CP violation arising from the
interference between the \textit{production of the neutral meson
system} and the mixing in that system.
We call this the \textit{``interference CP violation:
first produce, then mix''}.
This was first identified in 1998 by Meca and Silva \cite{Meca},
when studying the effect of
$D^0 - \overline{D^0}$ mixing on the decay chain
\be
B^{\pm}
\rightarrow \{ D^0, \overline{D^0} \} K^{\pm} 
\rightarrow [f]_D K^{\pm}.
\ee
Later,
Amorim, Santos and Silva showed that adding
these new parameters $\xi_{i \rightarrow P}$
and $\xi_{\bar i \rightarrow P}$  is enough to describe fully
any decay chain involving a neutral meson system as an intermediate
step \cite{Santos}.

This information may be summarized schematically as:
\begin{itemize}
\item[4.] Clash direct path with mixing path; first produce--then decay: 
$\xi_i = A_{i \rightarrow D}/ A_{i \rightarrow \bar D}\ 
p_D/q_D$
\begin{itemize}
\item CPV in interference; first mix--then decay
\item Never measured
\item It can affect the determination of $\gamma$ from
$B \rightarrow D$ decays
\end{itemize}
\begin{figure}[htb]
\centerline{\includegraphics*[height=0.9in]{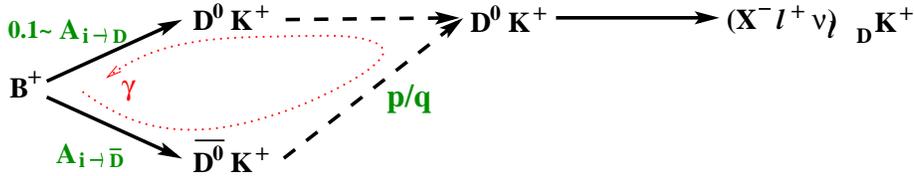}}
\caption{\label{fig:CPV_interference_2}Schematic interference
CP violation in the decay chain
$B^+ \rightarrow D K^+ \rightarrow D^0 K^+$.
}
\end{figure}
\end{itemize}
It is easy to see from the decay chain in FIG.~\ref{fig:CPV_interference_2}
that none of the types of CP violation described in the
previous subsection is involved here:
because we are using a charged $B$ meson,
we are not sensitive to the mixing or interference CP violation
involved in the decays of neutral $B$ mesons;
there is also no direct CP violation in $B \rightarrow D$ decays.
Furthermore,
this effect is  present already within the Standard Model,
and it involves the weak phase $\gamma$ to be discussed
later.\footnote{Should there be also a new physics,
CP violating contribution to $D^0 - \overline{D^0}$ mixing,
its effect would add to this one \cite{Meca}.}

Of course,
a non-zero mixing in the $D^0 - \overline{D^0}$ system is required.
A mixing of order $10^{-2}$ is still allowed by experiment,
which competes against
$A_{i \rightarrow D}/A_{i \rightarrow \bar D} \sim 10^{-1}$.
Therefore,
the lower decay path may give a correction of order
$10\%$ to the upper decay path.
It would be very interesting to measure this new type of
CP violation,
specially because it flies in the face of popular wisdom.
In any case,
this effect has to be taken into account as a source
of systematic uncertainty in the extraction of
$\gamma$ from $B \rightarrow D$ decays,
such as $B^\pm \rightarrow D K^\pm$ \cite{SilSof}.

\subsection{\label{sec:decays_from}Decays from a neutral meson system}

Henceforth,
we will ignore the production mechanism 
and concentrate on the time-dependent decay
rates from a neutral meson system into a final state $f$.
These formulae are used in the description of the 
CP violating asymmetries in section~\ref{sec:Bd}.

Let us consider the decay of a state $P^0$ or $\overline{P^0}$
into the final state $f$.
These decays depend on two decay amplitudes
\ba
A_f &\equiv& \langle f | T | P^0 \rangle,
\nonumber\\
\bar A_f &\equiv& \langle f | T | \overline{P^0} \rangle.
\ea
A state identified as the eigenvector of the strong
interaction (flavor eigenstate)
$P^0$ at time $t=0$,
will evolve in time according to
Eq.~(\ref{standard-time-evolution-1}) and,
thus,
decay into the final state $f$ at time $t$ with an amplitude
\be
A\left[P^0(t) \rightarrow f\right] =
\langle f | T | P^0(t) \rangle
=
g_+(t) A_f + \frac{q}{p} g_-(t) \bar A_f.
\ee
Similarly,
the decay amplitude for a state identified at time $t=0$ as
$\overline{P^0}$ is given by
\be
A\left[\overline{P^0}(t) \rightarrow f\right] =
\langle f | T | \overline{P^0}(t) \rangle
=
\frac{p}{q} g_-(t) A_f + g_+(t) \bar A_f.
\ee

The corresponding decay probabilities into the
CP conjugated states $f$ and $\bar f$ are given by
\ba
\Gamma [ P^0 (t) \ra f ]
& = &
\left| A_f \right|^2
\left\{
\left| g_+(t) \right|^2 + \left| \lambda_f \right|^2 \left| g_-(t) \right|^2
+ 2 \mbox{Re} \left[ \lambda_f g_+^\ast(t) g_-(t) \right]
\right\}
\nonumber\\*[3mm]
\Gamma [ P^0 (t) \ra \bar f ]
& = &
\left| \bar A_{\bar f} \right|^2 \left| \frac{q}{p} \right|^2
\left\{
\left| g_-(t) \right|^2 + \left| \bar \lambda_{\bar f} \right|^2
\left| g_+(t) \right|^2
+  2 \mbox{Re} \left[ \bar \lambda_{\bar f}\, g_+(t) g_-^\ast(t) \right]
\right\},
\nonumber\\*[3mm]
\Gamma [ \overline{P^0} (t) \ra f ]
& = &
\left| A_f \right|^2 \left| \frac{p}{q} \right|^2
\left\{
\left| g_-(t) \right|^2 + \left| \lambda_f \right|^2 \left| g_+(t) \right|^2
+ 2 \mbox{Re} \left[ \lambda_f g_+(t) g_-^\ast(t) \right]
\right\},
\nonumber\\*[5mm]
\Gamma [ \overline{P^0} (t) \ra \bar f ] &=&
\left| \bar A_{\bar f} \right|^2
\left\{
\left| g_+(t) \right|^2 + \left| \bar \lambda_{\bar f} \right|^2
\left| g_-(t) \right|^2 + 
2 \mbox{Re} \left[ \bar \lambda_{\bar f} g_+^\ast(t) g_-(t) \right]
\right\},
\label{master}
\ea
where we have used the definitions\footnote{Because the definition
of $\lambda_f$ in Eq.~(\ref{def:int-2})
is universally adopted,
Eqs.~(\ref{master}) remain the same under a simultaneous
change in the definitions of the sign of $q$ and of the sign
of $g_-(t)$.}
of $\lambda_f$ in Eq.~(\ref{def:int-2})
and $\bar \lambda_{\bar f} \equiv 1/ \lambda_{\bar f}$,
while the functions governing the time evolution
are given by \textbf{(Ex-7)}
\ba
|g_\pm(t)|^2
& = &
\frac{1}{4} \left[
e^{- \Gamma_H t} + e^{- \Gamma_L t} \pm 2 e^{- \Gamma t} \cos(\Delta m\, t)
\right]
\nonumber\\*[2mm]
             & = &
\frac{e^{- \Gamma t}}{2}
\left[
\cosh \frac{\Delta \Gamma t}{2} \pm \cos \left( \Delta m t \right)
\right],
\nonumber\\*[3mm]
g_+^\ast(t) g_-(t)
& = &
\frac{1}{4} \left[
- e^{- \Gamma_H t} + e^{- \Gamma_L t} + 2 i e^{- \Gamma t} \sin(\Delta m\, t)
\right]
\nonumber\\*[2mm]
                   & = &
\frac{e^{- \Gamma t}}{2} \left[
\sinh \frac{\Delta \Gamma t}{2} + i \sin \left( \Delta m t \right)
\right].
\label{g+-quad}
\ea
Eqs.~(\ref{master}) give us the probability,
divided by $dt$,
that the state identified as $P^0$
(or $\overline{P^0}$)
decays into the final state $f$ (or $\bar f$)
during the time-interval $[t, t+dt]$.
The time-integrated expressions are identical to these,
with the substitution of
$|g_+(t)|^2$, $|g_-(t)|^2$,
and $g_+^\ast(t) g_-(t)$ by \textbf{(Ex-8)}
\ba
G_\pm
& \equiv &
\displaystyle \int_0^{+ \infty} \!|g_\pm(t)|^2 dt
=
\frac{1}{2 \Gamma} \left( \frac{1}{1 - y^2} \pm \frac{1}{1 + x^2} \right),
\nonumber\\*[3mm]
G_{+-}
& \equiv &
\int_0^{+ \infty} \!g_+^\ast(t) g_-(t) dt
=
\frac{1}{2 \Gamma} \left( \frac{y}{1 - y^2} + \frac{ix}{1 + x^2} \right),
\label{GpmG+-}
\ea
where,
\be
x \equiv \frac{\Delta m}{\Gamma}\ \ \ \mbox{and}\ \ \ 
y \equiv \frac{\Delta \Gamma}{2 \Gamma}.
\label{eq:xandy}
\ee

\subsection{Flavor-specific decays and CP violation
in mixing}

Let us denote by $o$ a final state to which only $P^0$ may decay,
and by $\bar o$ its CP conjugated state, to which only
$\overline{P^0}$ can decay.
For example,
$o$ could be a semileptonic final state such as in
\be
K^0 \rightarrow \pi^- l^+ \nu_l
\ \ \ \ \ \mbox{or}
\ \ \ \ \ 
B^0_d \rightarrow h^- l^+ \nu_l,
\label{semileptonic_decays}
\ee
where $h^-$ is a negatively charged hadron,
$l^+$ a charged anti-lepton ($e^+$, $\mu^+$, or $\tau^+$),
and $\nu_l$ the corresponding neutrino.
Thus,
$A_{\bar o} = \langle \bar o | T | P^0 \rangle = 0$,
$\bar A_{o} = \langle o | T | \overline{P^0} \rangle = 0$,
and Eqs.~(\ref{master}) become
\ba
\Gamma [ P^0 (t) \ra o ]
& = &
\left| A_o \right|^2
\left| g_+(t) \right|^2 
\nonumber\\*[3mm]
\Gamma [ P^0 (t) \ra \bar o ]
& = &
\left| \bar A_{\bar o} \right|^2 \left| \frac{q}{p} \right|^2
\left| g_-(t) \right|^2 
\nonumber\\*[3mm]
\Gamma [ \overline{P^0} (t) \ra o ]
& = &
\left| A_o \right|^2 \left| \frac{p}{q} \right|^2
\left| g_-(t) \right|^2
\nonumber\\*[5mm]
\Gamma [ \overline{P^0} (t) \ra \bar o ] &=&
\left| \bar A_{\bar o} \right|^2
\left| g_+(t) \right|^2
\ea
Clearly,
$\Gamma [ P^0 (t) \ra \bar o ]$ and
$\Gamma [ \overline{P^0} (t) \ra o ]$ vanish at $t=0$,
but they are non-zero at $t \neq 0$ due to the mixing of the
neutral mesons.

We may test for CP violation through the asymmetry
\textbf{(Ex-9)}
\ba
A_M &=& 
\frac{\Gamma [ \overline{P^0} (t) \ra o ] -
\Gamma [ P^0 (t) \ra \bar o ]}{\Gamma [ \overline{P^0} (t) \ra o ] +
\Gamma [ P^0 (t) \ra \bar o ]
}
\nonumber\\
&=&
\frac{\left| p/q \right|^2 - \left| q/p \right|^2}{
\left| p/q \right|^2 + \left| q/p \right|^2}
=
\frac{2 \delta}{1 + \delta^2}
\nonumber\\
&=&
\frac{\left|  H_{12} \right|^2 - \left| H_{21} \right|^2}{
\left| H_{12} \right|^2 + \left| H_{21} \right|^2}
=
\frac{4 \mbox{Im}\left( M_{12}^\ast \Gamma_{12} \right)}{
4 \left| M_{12} \right|^2 + \left| \Gamma_{12} \right|^2},
\label{eq:A_M}
\ea
where we have used $|A_o|=|\bar A_{\bar o}|$.
Notice that this asymmetry does not depend on $t$.
This measures $\delta$,
\textit{i.e.}, it probes CP violation in mixing.
Because it is usually performed with the semileptonic decays
in Eq.~(\ref{semileptonic_decays}),
this is also known as the semileptonic decay asymmetry $a_{SL}$.

In the kaon system, 
we can use the approximate experimental equalities in
Eq.~(\ref{relations_kaon_mixing}) in order to find
\be
A_M \approx \frac{1}{2} \mbox{Im}\left(
\Gamma_{12}/M_{12} \right),
\ee
which, of course,
agrees with Eq.~(\ref{approx_kaon_delta}).
In fact CP violation in mixing has been measured in
the kaon system both through the $K_L \rightarrow \pi \pi$
decays and through the semileptonic decays. 

As discussed in subsection \ref{subsec:delta_B},
$\delta$ is expected to be very small for 
the $B_d$ and $B_s$ systems.
Because we will be looking for other CP
violating effects of order one,
we will neglect mixing CP violation in our ensuing
discussion of the $B$ meson systems.

\subsection{\label{sec:Bd}Approximations and notation for
$B$ decays}

In the next few years we will gain further information about
CP violation from the BABAR and Belle experiments,
concerning mainly $B^\pm$ and $B_d$ decays,
conjugated from results from CDF and D\O
(and, later, BTeV and LHCb),
which also detect $B_s$.

We will use the following approximations discussed in
subsection~\ref{subsec:delta_B}:
\ba
\mbox{both}\ \ B_d\ \ \mbox{and}\ \ B_s\ \ \mbox{systems}
& \Longrightarrow & 
\left| \frac{q}{p} \right| = 1
\ \ \Longrightarrow \ \
|\lambda_f| = \left| \frac{\bar A_f}{A_f} \right| ,
\label{approx_both}
\\*[2mm]
\mbox{only}\ \ B_d\ \ \mbox{system}
& \Longrightarrow & 
\Delta \Gamma = 0
\label{approx_no_width}
\ea
The first approximation leads to
\be
\frac{q}{p}
=
- \frac{2\, M_{21}}{\Delta m}
=
- \frac{M_{12}^\ast}{|M_{12}|},
\label{calc:q/p_Bsystems}
\ee
which will later be used to calculate $q/p$ in the
Standard Model.
However,
we know from Eqs.~(\ref{cp_conservation}) and (\ref{calc:q/p})
that CP conservation in the mixing implies that
\be
\frac{q}{p}
=
- \eta_P e^{i \xi},
\label{q/p_cp_conservation}
\ee
where $\xi$ is the arbitrary CP transformation phase
in Eq.~(\ref{CPonkaon}).
The sign $\eta_P = \pm 1$ arises from the square
root in Eq.~(\ref{calc:q/p}) and,
according to Eq.~(\ref{PaPb}),
it leads to ${\cal CP} | P_H \rangle = \eta_P | P_H \rangle$.
That is, 
$\eta_P$,
defined in the limit of CP conservation in the mixing,
is measurable and it determines whether the
heavier eigenstate is CP even ($\eta_P=1$) or CP odd ($\eta_P=-1$),
in that limit.
(See also appendix~\ref{appendix-rephasing}.)
Expressions (\ref{calc:q/p_Bsystems}) and (\ref{q/p_cp_conservation})
are often mishandled,
a fact we will come back to in section~\ref{sec:good_q/p}.

For the $B_s$ system,
we use the first approximation to transform the time-dependent 
decay probabilities of Eq.~(\ref{master}) into \textbf{(Ex-10)}
\ba
\Gamma [ B^0_s (t) \ra f ]
& = &
\frac{\left| A_f \right|^2 + \left| \bar A_f \right|^2}{2}\ 
e^{- \Gamma\, t}\,
\left\{
\cosh{\left( \frac{\Delta \Gamma\, t}{2} \right)}
+
D_f
\sinh{\left( \frac{\Delta \Gamma\, t}{2} \right)}
\right.
\nonumber\\*[2mm]
& &
\hspace{3.5cm}
\left.
+\;
C_f
\cos{(\Delta m\, t)}
-
S_f
\sin{(\Delta m\, t)}
\right\},
\nonumber\\*[4mm]
\Gamma [ \overline{B^0_s} (t) \ra f ]
& = &
\frac{\left| A_f \right|^2 + \left| \bar A_f \right|^2}{2}\ 
e^{- \Gamma\, t}\,
\left\{
\cosh{\left( \frac{\Delta \Gamma\, t}{2} \right)}
+
D_f
\sinh{\left( \frac{\Delta \Gamma\, t}{2} \right)}
\right.
\nonumber\\*[2mm]
& &
\hspace{3.5cm}
\left.
-\;
C_f
\cos{(\Delta m\, t)}
+
S_f
\sin{(\Delta m\, t)}
\right\},
\label{master_Bs}
\ea
where
\ba
D_f & \equiv &
\frac{2 \mbox{Re}( \lambda_f )}{1 + |\lambda_f|^2}
\label{D_f}
\\
C_f & \equiv &
\frac{1 - |\lambda_f|^2}{1 + |\lambda_f|^2}
\label{C_f}
\\
S_f & \equiv &
\frac{2 \mbox{Im}( \lambda_f )}{1 + |\lambda_f|^2}.
\label{S_f}
\ea
Clearly \textbf{(Ex-11)},
\be
\lambda_f =
\frac{1}{1+C_f} \left( D_f + i S_f \right)
\label{Lf_from_DCS}
\ee
is a physical observable, and
\be
D_f^2 + C_f^2 + S_f^2 = 1.
\label{D_C_S}
\ee
Therefore,
$C_f^2 + S_f^2 \leq 1$,
with the equality holding if and only if $\lambda_f$
is purely imaginary.
The importance of $\Delta \Gamma$ on the $B_s$ system
in order to provide a separate handle on $D_f$,
and in order to enable the use of untagged decays was first
pointed out by Dunietz\footnote{I recommend this article
very strongly to anyone wishing to learn about the $B_s$ system.}
\cite{Dunietz95}.

The expressions for the $B_d$ system are simplified
by setting $\Delta \Gamma = 0$, to obtain
\ba
\Gamma [ B^0_d (t) \ra f ]
& = &
\frac{\left| A_f \right|^2 + \left| \bar A_f \right|^2}{2}\ 
e^{- \Gamma\, t}\,
\left\{
1
+
C_f
\cos{(\Delta m\, t)}
-
S_f
\sin{(\Delta m\, t)}
\right\},
\nonumber\\*[3mm]
\Gamma [ \overline{B^0_d} (t) \ra f ]
& = &
\frac{\left| A_f \right|^2 + \left| \bar A_ f \right|^2}{2}\ 
e^{- \Gamma\, t}\,
\left\{
1
-
C_f
\cos{(\Delta m\, t)}
+
S_f
\sin{(\Delta m\, t)}
\right\}.
\label{master_Bd}
\ea
Notice that,
in this approximation of $\Delta \Gamma = 0$,
$D_f$ is not measured.
It can be inferred from 
Eq.~(\ref{D_C_S}) with a twofold ambiguity,
meaning that $\lambda_f$ is determined from
Eq.~(\ref{Lf_from_DCS}) with that twofold ambiguity.
In Eq.~(\ref{C_f}) we used $C_f$ as defined by BABAR.
When comparing results,
you should note that Belle uses a different notation
\be
{\cal A}_f \left( \mbox{Belle} \right)
=
- C_f \left( \mbox{BABAR} \right).
\label{A_Belle}
\ee
Here is another place where competing definitions
abound in the literature.
For example,
reference \cite{BLS} uses
$a^{\rm dir} = C_f$ and,
because of the sign change in the definition of $q$,
$a^{\rm int} = -S_f$.

In order to test CP,
we must compare $B^0 (t) \ra f$ with
$\overline{B^0}(t) \ra \bar f$,
or $B^0 (t) \ra \bar f$ with $\overline{B^0}(t) \ra f$.
To simplify the discussion we will henceforth
concentrate on decays into final states $f_{\rm cp}$ which
are CP eigenstates:
\be
{\cal CP} |f_{\rm cp} \rangle = \eta_f |f_{\rm cp} \rangle,
\ee
where $\eta_f = \pm 1$.
For these,
we define the CP asymmetry
\ba
A_{\rm CP}(t)
&\equiv&
\frac{\Gamma[\overline{B^0}(t) \rightarrow f_{\rm cp}]
- \Gamma[B^0(t) \rightarrow f_{\rm cp}]
}{\Gamma[\overline{B^0}(t) \rightarrow f_{\rm cp}]
+ \Gamma[B^0(t) \rightarrow f_{\rm cp}]}
\label{ACP_1}
\\
&=&
- C_f \cos{\Delta m t}
+ S_f \sin{\Delta m t}.
\label{ACP_2}
\ea
This is another place where two possibilities exist in the
literature.
Although this seems to be the most common choice
nowadays,
some authors define $A_{\rm CP}(t)$ to have the opposite sign,
specially in their older articles.

Since we have assumed that $|q/p|=1$,
$1 - |\lambda_f|^2 \propto |A_f|^2 - |\bar A_f|^2$,
and $C_f$ measures CP violation in the decay amplitudes.
On the other had,
$S_f \propto \mbox{Im} \lambda_f$ measures CP violation
in the interference between the mixing in the
$B^0_d - \overline{B^0_d}$ system and its decay
into the final state $f_{\rm cp}$.

There is a similar CP violating
asymmetry defined for charged $B$ decays.
However,
since there is no mixing (of course),
it only detects direct CP violation
\be
A_D \equiv
\frac{\Gamma[B^+ \rightarrow f^+]
- \Gamma[B^- \rightarrow f^-]
}{\Gamma[B^+ \rightarrow f^+]
+ \Gamma[B^- \rightarrow f^-]}
=
\frac{|A_+|^2 - |A_-|^2}{|A_+|^2 + |A_-|^2},
\ee
where the notation is self-explanatory.

\subsection{\label{sec:checklist}Checklist of crucial notational signs}

There are countless reviews and articles on CP violation,
each with its own notational hazards.
When reading any given article,
there are a few signs whose definition is crucial.
We have mentioned them when they arose,
and we collect them here for ease of reference.
One should check:
\begin{enumerate}
\item the sign of $q$ in the definition of $| P_H \rangle$
in terms of the flavor eigenstates 
-- \textit{c.f.\/} Eq.~(\ref{PaPb});
\item the sign choice, if any, for $\Delta m$;
\item the definitions of the functions $g_\pm(t)$,
in particular the sign of $g_-(t)$
-- \textit{c.f.\/} Eq.~(\ref{g+-});
\item the definitions of the coefficients of the various
time-dependent functions in the decay rates;
$D_f$, $C_f$ and $S_f$, or any others defined in their place
-- \textit{c.f.\/} Eqs.~(\ref{master_Bs})--(\ref{master_Bd});
\item the order in which the decay rates of
$B^0_d$ and $\overline{B^0_d}$ appear in the definition
of $A_{\rm CP}(t)$,
and its relation with the time-dependent functions
-- \textit{c.f.\/} Eqs.~(\ref{ACP_1}) and (\ref{ACP_2}).
\end{enumerate}
To be extra careful,
check also the definition of $\lambda_f$.

In addition,
one should also check whether specific conventions
for the CP transformation phases are used.

This completes our discussion of the phenomenology
of CP violation at the hadronic (experimental) level,
which was concentrated on the ``bras'' and ``kets'' 
in Eq.~(\ref{quarkVShadron}).
We will now turn to a specific theory of the electroweak
interactions, which will enable us to
discuss CP violation at the level of the quark-field
operators in Eq.~(\ref{quarkVShadron}).
These two analysis will later be combined into specific
predictions for observable quantities.

\section{\label{ch:SM}CP violation in the Standard Model}

\subsection{Some general features of the SM}

Since the Standard Model (SM) of electroweak interactions 
\cite{GWS},
and its parametrization of CP violation through the
Cabibbo-Kobayashi-Maskawa (CKM)  mechanism \cite{CKM},
are well discussed in virtually every book of
particle physics,
we will only review here some of its main features.

The SM can be characterized by its gauge group,
\be
SU(3)_C \otimes SU(2)_L \otimes U(1)_Y,
\ee
with the associated gauge fields,
\begin{itemize}
\item gluons:\ \ \ \ 
$G_{\mu \nu}^k\ \ \ \ k=1\dots8$,
\item $SU(2)_L$ gauge bosons:\ \ \ \ 
$W_\mu^a\ \ \ \ a=1,2,3$,
\item $U(1)_Y$ gauge boson:\ \ \ \ 
$B_\mu$;
\end{itemize}
by its non-gauge field content,
\begin{itemize}
\item quarks:\ \ \ \ \ 
$q_L =
\left(
\begin{array}{c}
p_L\\
n_L
\end{array}
\right)\ \ [1/2,1/6]$,
\ \ \ \ $p_R\ \  [0,2/3]$,
\ \ \ \ $n_R\ \  [0,-1/3]$,
\item leptons:\ \ \ 
$L_L =
\left(
\begin{array}{c}
\nu_L\\
C_L
\end{array}
\right)\ \ [1/2,-1/2]$,
\hspace{3cm}
$C_R\ \  [0,-1]$,
\item Higgs Boson:\ \ \ \ 
$\Phi =
\left(
\begin{array}{c}
\phi^+\\
\phi^0
\end{array}
\right)\ \ [1/2,1/2]$,
\end{itemize}
where the square parenthesis show the 
electroweak quantum numbers
$[T,Y]$, with $Q=T_3+Y$;
and by the symmetry breaking scheme
\be
SU(3)_C \otimes SU(2)_L \otimes U(1)_Y
\longrightarrow
SU(3)_C \otimes U(1)_{\rm elmg}\ ,
\ee
induced by the potential
\be
V(\Phi^\dagger \Phi) = 
- \mu^2 (\Phi^\dagger \Phi) + \lambda (\Phi^\dagger \Phi)^2
= - {\cal L}_{\rm Higgs},
\ee
with minimum at
\be
\langle \Phi^\dagger \Phi \rangle = \frac{v^2}{2} = \frac{\mu^2}{2 \lambda}.
\ee

The electroweak part of the Standard Model lagrangian may
be written as
\be
{\cal L}_{EW}
=
{\cal L}_{\rm pure\ gauge}
+
{\cal L}_{\rm Higgs}
+
{\cal L}_{\rm kinetic}
+
{\cal L}_{\rm Yukawa},
\ee
where the first term involves only the gauge bosons,
and
\ba
{\cal L}_{\rm kinetic} &=&
i \bar q_L\, \gamma^\mu\, D_\mu^{q_L}\, q_L
+
i \bar p_R\, \gamma^\mu\, D_\mu^{p_R}\, p_R
+
i \bar n_R\, \gamma^\mu\, D_\mu^{n_R}\, n_R
\nonumber\\
& &
+
i \bar L_L\, \gamma^\mu\, D_\mu^{L_L}\, L_L
+
i \bar C_R\, \gamma^\mu\, D_\mu^{C_R}\, C_R
+
\left|
\left(
i \partial_\mu - \frac{g}{2} \vec{\tau}.\vec{W}_\mu - 
\frac{g^\prime}{2} B_\mu
\right) \Phi
\right|^2,
\label{L_kinetic}
\ea
with
\ba
i D_\mu &=&
i \partial_\mu  - \frac{g}{2} \vec{\tau}.\vec{W}_\mu
- g^\prime Y B_ \mu,
\nonumber\\
i D_\mu &=&
i \partial_\mu  - g^\prime Y B_ \mu,
\label{cov_der}
\ea
for the $SU(2)_L$ doublets and singlets, respectively.
The vector $\vec{\tau}$ is made out of the three Pauli matrices.

The Yukawa interactions are given by
\be
{\cal L}_{\rm Yukawa}
=
- \bar q_L Y_d n_R \Phi - \bar q_L Y_u p_R (i \tau_2 \Phi^\ast)
- \bar L_L Y_l C_R \Phi + h.c.,
\label{L_Yukawa}
\ee
where $Y_u$, $Y_d$, and $Y_l$ are complex
$3 \times 3$ Yukawa coupling matrices. We are using a 
very compact (and, at first, confusing) matrix convention
in which the fields $q_L$, $n_R$, etc. are $3 \times 1$
vectors in generation (family) space.
Expanding things out would read
\be
q_L
=
\left[
\begin{array}{c}
\left(
\begin{array}{c}
{p_{L}}_{1}\\
{n_{L}}_{1}
\end{array}
\right)
\\*[6mm]
\left(
\begin{array}{c}
{p_{L}}_{2}\\
{n_{L}}_{2}
\end{array}
\right)
\\*[6mm]
\left(
\begin{array}{c}
{p_{L}}_{3}\\
{n_{L}}_{3}
\end{array}
\right)
\end{array}
\right],
\ \ \ \ \ \ \ \ 
n_R =
\left[
\begin{array}{c}
{n_{R}}_{1}\\
{n_{R}}_{2}\\
{n_{R}}_{3}
\end{array}
\right]
\ee
and
\be
Y_u
=
\left[
\begin{array}{ccc}
{Y_u}_{11} & {Y_u}_{12} & {Y_u}_{13}\\
{Y_u}_{21} & {Y_u}_{22} & {Y_u}_{23}\\
{Y_u}_{31} & {Y_u}_{32} & {Y_u}_{33}
\end{array}
\right],
\ \ \ \ \ \ \ \ 
i \tau_2 \Phi^\ast 
=
\left(
\begin{array}{c}
{\phi^0}^\ast\\
- \phi^-
\end{array}
\right),
\ee
where we have used regular parenthesis for the $SU(2)_L$ space
and square brackets for the family space.
These spaces appear in addition to the usual spinor space
\textbf{(Ex-12)}.

The fact that no right-handed neutrino field was introduced above
leads to the nonexistence of neutrino masses and to the
conservation of individual lepton flavors.
Those who viewed this as the ``amputated SM'',
were not surprised to learn from experiment that neutrino masses
do exist and, thus, that a more complex neutrino sector
is called for.
We will not comment further on neutrinos in these lectures,
and the reader is referred to one of the many excellent reviews
on that subject \cite{neutrinos}.

After spontaneous symmetry breaking, the Higgs field may be parametrized
conveniently by
\be
\Phi =
\left(
\begin{array}{c}
\phi^+\\
\phi^0
\end{array}
\right)
\longrightarrow
\left(
\begin{array}{c}
G^+\\
\frac{1}{\sqrt{2}} (v + H^0 + i G^0)
\end{array}
\right),
\label{eq:Higgs_basis}
\ee
where $H^0$ is the Higgs particle and $G^+$ and $G^0$
are the Goldstone bosons that, in the unitary gauge,
become the longitudinal components of the $W^+$
and $Z$ bosons, respectively.
The charged gauge bosons acquire a tree-level mass $M_W = g v/2$.
The $U(1)_Y$ gauge boson and the neutral $SU(2)_L$ gauge boson
are mixed into the massless $U(1)_{\rm elmg}$ gauge boson and
another neutral gauge boson $Z$,
with tree-level mass $M_Z = \sqrt{g^2+g^{\prime 2}} v/2$.
This rotation is characterized by the weak mixing angle
$\tan{\theta_W}=g^\prime/g$:
\be
\left(
\begin{array}{c}
B_\mu \\
{W_3}_\mu
\end{array}
\right)
=
\left(
\begin{array}{cc}
\cos{\theta_W} & - \sin{\theta_W}\\
\sin{\theta_W} & \cos{\theta_W}
\end{array}
\right)
\left(
\begin{array}{c}
A_\mu \\
Z_\mu
\end{array}
\right).
\label{BW_from_AZ}
\ee
The gauge bosons have interactions with the quarks given,
in a weak basis, by \textbf{(Ex-13)}
\ba
- {\cal L}_W &=&
\frac{g}{\sqrt{2}}\, \bar p_L\, \gamma^\mu\, n_L\, W_\mu^+ + h.c.,
\label{LW_weak}
\\
- {\cal L}_Z &=&
\frac{g}{\cos{\theta_W}} Z_\mu
\left\{
c_L^{\rm up}\, \bar p_L\, \gamma^\mu\, p_L 
+ c_L^{\rm down}\, \bar n_L\, \gamma^\mu\, n_L
+
\left( L \leftrightarrow R \right)
 \right\},
\label{LZ_weak}
\ea
where $c = T_3 - Q \sin^2{\theta_W}$.
We have designated by a ``weak basis'' any basis choice for
$q_L$, $p_R$, and $n_R$ which leaves 
${\cal L}_{EW} - {\cal L}_{\rm Yukawa}$
invariant.
Two such basis are related by a ``Weak Basis Transformation'' (WBT)
\ba
\left( 
\begin{array}{c}
p_L^\prime\\
n_L^\prime
\end{array}
\right)
=
q_L^\prime = W_L\, q_L
=
W_L\,
\left( 
\begin{array}{c}
p_L\\
n_L
\end{array}
\right),
\nonumber\\*[3mm]
p_R^\prime = W_{pR}\; p_R, \ \ \ \ \ \ 
n_R^\prime = W_{nR}\; n_R.
\label{WBT}
\ea
This corresponds to a global flavor symmetry
\be
F = U(3)_{qL} \otimes U(3)_{pR} \otimes U(3)_{nR}
\ee
of ${\cal L}_{EW} - {\cal L}_{\rm Yukawa}$,
which is broken by ${\cal L}_{\rm Yukawa}$ down to
\be
F^\prime = U(1)_{\rm B},
\ee
corresponding to baryon number conservation.

Unfortunately,
for any weak basis,
the interactions with the Higgs are not diagonal.
We may solve this problem by taking the quarks to the mass
basis $u_L$, $u_R$, $d_L$, $d_R$,
through
\ba
\bar p_L = \bar u_L\, U_{u_L}^\dagger,
&\ \ \ \ \ \ \ &
\bar n_L = \bar d_L\, U_{d_L}^\dagger,
\nonumber\\
p_R = U_{u_R}\, u_R,
&\ \ \ \ \ \ \ &
n_R = U_{d_R}\, d_R,
\label{mass_basis}
\ea
where the unitary matrices $U$ have been chosen in order to
diagonalize the Yukawa couplings,
\ba
M_U &\equiv& 
diag (m_u, m_c, m_t) =
\frac{v}{\sqrt{2}}\, U_{u_L}^\dagger\, Y_u\, U_{u_R},
\nonumber\\
M_D &\equiv& 
diag (m_d, m_s, m_b) =
\frac{v}{\sqrt{2}}\, U_{d_L}^\dagger\, Y_d\, U_{d_R}.
\label{mass_diagonalization}
\ea
In this new basis \textbf{(Ex-14, Ex-15)}
\ba
- {\cal L}_H &=&
\left( 1 + \frac{H^0}{v} \right)
\left\{
\bar u\, M_U\, u + \bar d\, M_D\, d 
\right\},
\label{LH_mass}
\\
- {\cal L}_W &=&
\frac{g}{\sqrt{2}}
\bar u_L\;
\left( U_{u_L}^\dagger U_{d_L} \right)
\; \gamma^\mu d_L\, W_\mu^+ + h.c.,
\label{LW_mass}
\\
- {\cal L}_Z &=&
\frac{g}{\cos{\theta_W}} Z_\mu
\left\{
c_L^{\rm up} \bar u_L\;
(V V^\dagger)\; \gamma^\mu  u_L
+
c_L^{\rm down} \bar d_L\  
(V^\dagger V)\ \gamma^\mu  u_L
+
\left( L \leftrightarrow R\right)
\right\},
\label{LZ_mass}
\ea
where
\be
V \equiv U_{u_L}^\dagger U_{d_L},
\ee
is the Cabibbo-Kobayashi-Maskawa matrix \cite{CKM}.
Notice that the charged $W$ interactions are purely left-handed.
Also the lack of flavor changing neutral $Z$ interactions 
is due to the unitarity of the CKM matrix.
Indeed,
since $U_{u_L}$ and $U_{d_L}$ are unitary,
so is $V$,
implying that
\be
V V^\dagger = 1 = V^\dagger V,
\ee
which renders the interactions in
Eq.~(\ref{LZ_mass}) flavor diagonal.
This is no longer the case for theories
containing extra quarks in exotic representations
of the $SU(2)_L$ group \cite{many}.

A further complication arises from the fact that
the matrices $U$ are not uniquely determined by
Eqs.~(\ref{mass_diagonalization}).
And, thus,
the mass basis definition in Eqs.~(\ref{mass_basis})
is not well defined.
Indeed,
introducing the diagonal matrices
\be
\Theta_u = diag (e^{i \theta_{u1}}, e^{i \theta_{u2}}, e^{i \theta_{u3}})
\ \ \ \mbox{and}\ \ \ 
\Theta_d = diag (e^{i \theta_{d1}}, e^{i \theta_{d2}}, e^{i \theta_{d3}})
\ee
and redefining the mass eigenstates by
\ba
\bar u_L^\prime = \bar u_L\, \Theta_u^\dagger,
&\ \ \ \ \ \ \ &
\bar d_L^\prime = \bar d_L\, \Theta_d^\dagger,
\nonumber\\
u_R^\prime = \Theta_u\, u_R,
&\ \ \ \ \ \ \ &
d_R^\prime = \Theta_d\, d_R,
\label{mass_basis_rephasing}
\ea
leaves
Eqs.~(\ref{LH_mass}) and (\ref{LZ_mass})
unchanged.
This is just the standard rephasing of the quark field
operators.

We are now ready to compute the number of parameters
in the CKM matrix $V$ in two different ways.
In the first procedure,
we note that any $3 \times 3$ unitary matrix $V$
has 3 angles and 6 phases.
However,
the quark rephasings in Eq.~(\ref{mass_basis_rephasing})
leave ${\cal L}_W$  in Eq.~(\ref{LW_mass}) invariant as long
as we change $V$ simultaneously,
according to
\be
V^\prime = \Theta_u V \Theta_d^\dagger.
\ee
This allows us to remove 5 relative phases from $V$.
Notice that a global rephasing,
redefining all quarks by the same phase,
leaves $V$ unchanged.
As a result,
such a (sixth) transformation cannot be used to remove
a (sixth) phase from $V$.
Thus, we are left with three real parameters (angles) and one
(CP violating) phase in the CKM matrix.
In the second procedure,
we note that there are $N_{\rm Yuk}=18$ magnitudes plus 
$N_{\rm Yuk}=18$ phases for a total of 36
parameters in the two Yukawa matrices $Y_u$ and $Y_d$.
When these are turned on,
they reduce the global flavor symmetry $F$ of
${\cal L}_{EW} - {\cal L}_{\rm Yukawa}$ into $F^\prime$.
Therefore,
we are left with \cite{San93}
\be
N = N_{\rm Yuk} - N_F + N_{F^\prime}
\ee
parameters,
where $N_F$ and $N_{F^\prime}$ are the number of parameters
in $F$ and $F^\prime$, respectively.
This equation holds in a very general class of models,
and is valid separately for the magnitudes and
for the phases \cite{San93}.
Applying it to the SM model we find that the Yukawa
couplings lead to
$9 = 2 \times 9 - 3 \times 3$ real parameters
(which are the 6 masses and the three mixing angles
in $V$),
and to only $1 = 2 \times 9 - 3 \times 6 + 1$
phase (which is the CP violating phase in the
CKM matrix $V$).
It is easy to understand,
using either method,
that there would be no CP violating phase if we had
only one or two generations of quarks.

The fact that there is only one CP violating
phase in the CKM matrix $V$ has an immensely
important implication:
within the SM,
any two CP violating observables are proportional
to each other.
In general,
the proportionality will involve CP conserving
quantities,
such as mixing angles and hadronic matrix elements.
If it involves only mixing angles,
it can be used for a clean test of the SM;
if it also involves hadronic matrix elements,
the test is less precise.
We will come back to this when we discuss
the $\rho - \eta$ plane in section~\ref{sec:rho-eta}.

A few points are worth emphasizing:
\begin{itemize}
\item the existence of CP violation is connected with the Yukawa
couplings which appear in the interaction with the scalar and,
thus, it is intimately related with the sector which provides
the spontaneous symmetry breaking;
\item because the masses have the same origin,
it is also related to the flavor problem;
\item the existence of (no less than) three generations is
crucial for CP violation,
which relates this with the problem of the number of generations;
\item although in the SM CP is violated explicitly by
the Lagrangian,
it is also possible to construct theories which break CP spontaneously;
\item at tree-level, CP violation arises in the SM only through
flavor changing transitions involving the charged currents.
Hence,
flavour diagonal CP violation is, at best, loop suppressed;
\item the fact that the SM exhibits a single CP violating
phase makes it a very predictive theory and, thus,
testable/falsifiable;
\item CP violation is a crucial ingredient for bariogenesis and,
thus to our presence here to discuss it.
\end{itemize}
These are some of the theoretical reasons behind the
excitement over CP violation,
which should be added to the experimental reasons discussed
in the introduction.

\subsection{\label{sec:def_CP_transf}Defining the CP transformation}

Here we come to two other of those 
(perhaps best to be ignored in a first reading) 
fine points which plague the study of CP violation.
Both were stressed as early as 1966 by Lee and Wick \cite{Lee66}.

The first point concerns the consistency of describing
P, CP or T in theories in which these symmetries are violated.
For example, the geometrical transformation
$\vec{r} \rightarrow - \vec{r},\ \ t \rightarrow + t$,
corresponding to parity P (or CP),
should commute with a time translation.
In terms of the infinitesimal generators ${\cal P}$ and ${\cal H}$,
this translates into $[{\cal P},{\cal H}]=0$,
which one recognizes as the correct commutation relation for
the corresponding generators of the Poincar\'{e} group.
Thus, for a theory in which parity is violated,
one cannot define parity in a way consistent with this basic
geometrical requirements.
A similar reasoning applies to the other discrete symmetries
discussed here.
The correct procedure is to define the discrete symmetries
in some limit of the Lagrangian in which they hold.
This is particularly useful if one wishes to understand
which (clashes of) terms of the Lagrangian are generating the violation
of these symmetries.

The second point concerns the ambiguity in this procedure.
First,
because we can break the Lagrangian in a variety of ways.
Second,
and most importantly,
because there is great ambiguity in defining the discrete
symmetries when the theory possesses extra internal symmetries.
To be specific,
suppose that the Lagrangian is invariant under
some group of unitary internal symmetry operators $\{F\}$.
Then,
if ${\cal P}$ is a space inversion operator,
then $F {\cal P}$ is an equally good space inversion operator.
A particularly useful case arises when we take this group
to correspond to basis transformations,
since then one can build easily basis independent quantities
violating the discrete symmetry in question.

These observations provide us with a way to construct
basis-invariant quantities that signal CP violation,
which is applicable to any theory with an arbitrary gauge group,
arbitrary fermion and scalar content,
renormalizable or not \cite{Bot95}.
The basic idea is the following:
\begin{enumerate}
\item Divide the Lagrangian into two pieces,
${\cal L} = {\cal L}_{\rm invariant} + {\cal L}_{\rm break}$,
such that the first piece is invariant under a CP
transformation.
\item Find the most general set of basis transformations $\{F\}$
which leaves ${\cal L}_{\rm invariant}$ invariant. 
\item Define the generalized CP transformations at the level
of ${\cal L}_{\rm invariant}$,
including the basis transformations $\{F\}$ in that definition.
These are called the ``spurious matrices'' (``spurious phases'' if only
rephasings of the fields are included) brought about by the CP
transformations.
\item Inspired by perturbation theory,
search for expressions involving the couplings in ${\cal L}_{\rm break}$,
which are invariant under the generalized CP transformations.
Such expressions are a sign of CP conservation;
their violation is a sign of CP violation.
\end{enumerate}
And, because the basis transformations $\{F\}$ have already been
included in the definition of the generalized CP transformations,
the signs of CP violation constructed in this way do not depend
on the basis transformations;
essentially, those basis transformations have been traced over.
This is the method which we will follow below to get $J_{\rm CKM}$.
After seeing it in action a few times,
one realizes that steps 3 and 4 may be substituted by \cite{Bot95}:
\begin{itemize}
\item[3$^\prime$.] Inspired by perturbation theory,
build products of the coupling matrices in ${\cal L}_{\rm break}$,
of increasing complexity,
taking traces over all the (scalar or fermion) internal flavor spaces.
Since traces have been taken,
these expressions are invariant under the transformations $\{F\}$.
\item[4$^\prime$.] Those traces with an imaginary part signal
CP violation.\footnote{Of course,
we are excluding the artificial option of introducing by hand
some phases in the definition of quantities which would
otherwise be real and invariant under a WBT.}
\end{itemize}
Below,
we will use this second route,
in order to get $J$.
This method can be extended to provide invariant
quantities which signal the breaking of other discrete symmetries,
such as $R$-parity breaking in supersymmetric theories \cite{Dav97}.
Next, we will apply these ideas to the SM.

\subsection{\label{sec:def_CPV_J}Defining the CP 
violating quantity in the SM}

In studying a particular experiment,
we are faced with hadronic matrix elements like
those in Eq.~(\ref{quarkVShadron}).
We have stressed that any observable has to be
invariant under a rephasing of the ``kets'' and ``bras'',
and we have used this property in chapter~\ref{ch:producao} and in
appendix~\ref{appendix-rephasing} in order to identify CP violating
observables, invariant under such rephasings.
But a CP violating observable must also be
invariant under the rephasings of the quark field
operators in Eq.~(\ref{quarkVShadron}).
This ties into the previous sections.

We start with the CP conserving Lagrangian
${\cal L}_{EW} - {\cal L}_W$,
written in the mass basis.
This is invariant under the quark rephasings in
Eq.~(\ref{mass_basis_rephasing}),
which may be included as spurious phases $\xi$ in the general definition
of CP violation:
\ba
\left( \cp \right) W_\mu^+ \left( \cp \right)^\dagger
&=&
- e^{i \xi_W} W^{\mu -},
\label{CP_on_W}
\\
\left( \cp \right) \overline{u_\alpha} \left( \cp \right)^\dagger
&=&
- e^{- i \xi_\alpha} u_\alpha^T C^{-1} \gamma^0,
\nonumber\\
\left( \cp \right) d_k \left( \cp \right)^\dagger
&=&
e^{i \xi_k} \gamma^0 C \overline{d_k}^T.
\label{CP_with_rephasing}
\ea
It is easy to check \textbf{(Ex-16, Ex-17)}
that the Lagrangian describing the
interactions of the charged current
\be
\frac{g}{2 \sqrt{2}} \sum_{\alpha = u, c, t}
\sum_{k = d, s, b} \left[ W_\mu^+ V_{\alpha k} \overline{u_\alpha} \gamma^\mu
\left( 1 - \gamma_5 \right) d_k
+ W_\mu^- V_{\alpha k}^\ast \overline{d_k} \gamma^\mu
\left( 1 - \gamma_5 \right) u_\alpha \right]
\label{LW_expanded}
\ee
is invariant under CP if and only if
\be
V_{\alpha k} = e^{i \left( - \xi_W + \xi_\alpha - \xi_k \right)}
V_{\alpha k}^\ast.
\label{afinal}
\ee
For any given matrix element $V_{\alpha k}$,
it is always possible to choose the spurious phases
$\xi_W + \xi_\alpha - \xi_k$ in such a way that Eq.~(\ref{afinal})
holds.
However,
the same reasoning tells us that CP conservation
implies
\be
V_{\alpha i} V_{\beta j} 
V_{\alpha j}^\ast V_{\beta i}^\ast
=
\left(
V_{\alpha i} V_{\beta j} 
V_{\alpha j}^\ast V_{\beta i}^\ast
\right)^\ast,
\label{afinal2}
\ee
where $\alpha \neq \beta$, $i \neq j$.
And,
this equality no longer involves the spurious phases
brought about by the CP transformations,
\textit{i.e.\/},
it is invariant under a rephasing of the quarks in
Eq.~(\ref{mass_basis_rephasing}).
Therefore, a nonzero
\be
J_{\rm CKM} = \left|
\mbox{Im} \left( V_{\alpha i} V_{\beta j} 
V_{\alpha j}^\ast V_{\beta i}^\ast \right) \right|
\label{J_CKM}
\ee
constitutes an unequivocal sign of
CP violation.\footnote{The magnitude
is only introduced here because
the sign of $\mbox{Im} \left( V_{\alpha i} V_{\beta j} 
V_{\alpha j}^\ast V_{\beta i}^\ast \right)$ changes for some
re-orderings of the flavor indexes \textbf{(Ex-18, Ex-19)}.}
We can build more complex combinations of
$ V_{\alpha i}$ which signal CP violation but they
are all proportional to $J_{\rm CKM}$ \cite{BLS}.
This is a simple consequence of the fact that there is
only one CP violating phase in the SM.
We learn that CP violation requires all the
CKM matrix elements to be non-zero.

But, we might equally well have performed all calculations
before changing into the mass basis,
and start with the
CP conserving Lagrangian ${\cal L}_{EW} - {\cal L}_{\rm Yukawa}$.
This is invariant under the matrix redefinitions in
Eq.~(\ref{WBT}),
which can be included in a more general definition of
the CP transformations \cite{Ber86,Bot95}:
\ba
\left( \cp \right) \Phi \left( \cp \right)^\dagger
&=&
\Phi^\ast \equiv \left( \Phi^\dagger \right)^T
\label{CP_on_Phi}
\\
\left( \cp \right) \bar q_L \left( \cp \right)^\dagger
&=&
- q_L^T\, C^{-1}\, \gamma^0\, K_L^\dagger,
\nonumber\\
\left( \cp \right) n_R \left( \cp \right)^\dagger
&=&
K_{nR}\, \gamma^0\, C\, {\bar{n}}_R^T,
\nonumber\\
\left( \cp \right) p_R \left( \cp \right)^\dagger
&=&
K_{pR}\, \gamma^0\, C\, {\bar{p}}_R^T,
\label{CP_with_WBT}
\ea
where $K$ are unitary matrices acting in the respective flavor
spaces.\footnote{Due to the vev,
dealing with a phase in the CP transformation of $\Phi$
is very unfamiliar,
and we have not included it.
It requires one to study the transformation properties
of the vevs under a redefinition of the scalar fields,
as is explained in reference \cite{Bot95}.}
It is easy to check \textbf{(Ex-20)} that ${\cal L}_{\rm Yukawa}$
in Eq.~(\ref{L_Yukawa})
would be invariant under CP if and only if
matrices $K$ were to exist such that
\ba
K_L^\dagger Y_u K_{pR} &=& Y_u^\ast,
\nonumber\\
K_L^\dagger Y_d K_{pR} &=& Y_d^\ast.
\label{afinal3}
\ea
The crucial point is the presence of $K_L$ in both conditions,
which is a consequence of the fact that the left-handed
up and down quarks belong to the same $SU(2)_L$ doublet.
As for $J_{\rm CKM}$,
we could now use these generalized CP transformations to
identify signs of CP violation.

Instead,
we will follow the second route presented at the end of the
previous section,
because it is easier to apply to any model \cite{Bot95}.
As mentioned,
we can build CP violating quantities by
tracing over the basis transformations in the family spaces,
and looking for imaginary parts which remain.
To start,
we ``trace over'' the right-handed spaces with
\ba
H_u &\equiv& \frac{v^2}{2} Y_u Y_u^\dagger
=
U_{u_L} M_U^2 U_{u_L}^\dagger,
\nonumber\\
H_d &\equiv& \frac{v^2}{2} Y_d Y_d^\dagger
=
U_{d_L} M_D^2 U_{d_L}^\dagger,
\ea
where the second equalities express the matrices
in terms of the parameters written in the mass basis,
as in Eq.~(\ref{mass_diagonalization}).
Working with $H_u$ and $H_d$,
and inspired by perturbation theory,
we seek combinations of these couplings of
increasing complexity
(which appear at higher order in perturbation theory),
such as
\ba
H_u H_d
&=&
U_{u_L} M_U^2 V M_D^2 U_{d_L}^\dagger,
\nonumber\\
H_u^2 H_d^2
&=&
U_{u_L} M_U^4 V M_D^4 U_{d_L}^\dagger,
\nonumber\\
H_u H_d H_u^2 H_d^2
&=&
U_{u_L} M_U^2 V M_D^2 V^\dagger M_U^4 V M_D^4 U_{d_L}^\dagger,
\label{H_combinations}
\ea
and so on\dots
We can now `trace over'' the left-handed space.
Taking traces over the first two combinations,
it is easy to see that they are real \textbf{(Ex-21)}.
However \textbf{(Ex-22)},
\ba
J
&=&
\mbox{Im}
\left\{
\mbox{Tr} \left( H_u H_d H_u^2 H_d^2 \right)
\right\}
\nonumber\\
&=&
\mbox{Im}
\left\{
\mbox{Tr} \left(
V^\dagger M_U^2 V M_D^2 V^\dagger M_U^4 V M_D^4
\right)
\right\}
\nonumber\\
&=&
(m_t^2 - m_c^2)(m_t^2 - m_u^2)(m_c^2 - m_t^2)
(m_b^2 - m_s^2)(m_b^2 - m_d^2)(m_s^2 - m_d^2)
J_{\rm CKM},
\label{J}
\ea
is not zero and,
since we have already traced over basis transformations,
this imaginary part
is a signal of CP violation \cite{Rol91,Bot95,BLS}.

Historically,
many alternatives for this quantity have been derived,
with different motivations \cite{Ber86,Jarlskog,Dun85}:
\be
\mbox{Tr} \left[ H_u, H_d\right]^3
=
3\, \mbox{det} \left[ H_u , H_d \right]
=
6 i\, \mbox{Im}
\left\{ \mbox{Tr} \left( H_u H_d H_u^2 H_d^2 \right)\right\},
\ee
but the technique used here to arrive at Eq.~(\ref{J})
has the advantage that it
can be generalized to an arbitrary theory \cite{Bot95}.

As expected $J$ is proportional to $J_{\rm CKM}$,
with the proportionality coefficient involving only CP conserving
quantities.
But we learn more from Eq.~(\ref{J}) than we do from
Eq.~(\ref{J_CKM});
we learn that CP violation can only occur because all the up quarks
are non-degenerate and all the down quarks are non-degenerate.

You may now be confused. In order to reach $J_{\rm CKM}$ we defined
the CP transformations at the ${\cal L}_{EW} - {\cal L}_W$ level.
CP violation seems to arise from ${\cal L}_W$,
which seems to arise from a term in ${\cal L}_{\rm kinetic}$.
In contrast,
we have reached $J$ by defining
the CP transformations at the ${\cal L}_{EW} - {\cal L}_{\rm Yukawa}$ level.
CP violation seems to arise from ${\cal L}_{\rm Yukawa}$
(with the implicit utilization of CP conservation in
${\cal L}_{\rm kinetic}$).
Now, which is it?
Is CP violation in ${\cal L}_W$, or in ${\cal L}_{\rm Yukawa}$?
The answer is\dots: the question does not make sense!!!
CP violation has to do with a phase.
But phases can be brought in an out of the various
terms in the Lagrangian through rephasings or
more general basis transformations.
Thus,
asking about the specific origin of a phase makes no sense.
CP violation must always arise from a clash of two different
terms in the Lagrangian.
One way to state what happens in the SM,
is to say that CP violation arises from a clash between the Yukawa terms
and the charged current interactions,
as seen clearly on the second line of Eq.~(\ref{J}).

Having developed $J$ as a basis invariant quantity
signaling CP violation,
we might expect it to appear in every calculation of
a CP violating observable.
This is not the case
because $J$ may appear multiplied or divided by
some combination of CP conserving quantities,
such as hadronic matrix elements, functions of masses,
or even mixing angles.
In addition,
some of the information encapsulated into $J$ may even
be contained in the setup of the experiment.
For example,
if we try do describe a decay such as $B_d \rightarrow K \pi$,
we use implicitly the fact that the experiment
can distinguish between a $B_d$, a $K$, and a $\pi$.
That is,
the fact that $d$, $s$, and $b$ are non-degenerate is
already included in the experimental setup itself;
thus,
the non-degeneracy constraint contained in the term
$(m_b^2 - m_s^2)(m_b^2 - m_d^2)(m_s^2 - m_d^2)$ has
been taken into account from the start.
In fact,
even $J_{\rm CKM}$ may appear truncated in a given
observable.
Questions such as these may be dealt with through appropriately
defined projection operators \cite{Bot04}.

\subsection{Parametrizations of the CKM matrix and beyond}

\subsubsection{The standard parametrizations of the CKM matrix}

The CKM matrix has four quantities with physical significance:
three mixing angles and one CP violating phase.
These may be parametrized in a variety of ways.
The particle data group \cite{PDG} uses the Chau--Keung
parametrization \cite{Cha84}
\be
V =
\left(
\begin{array}{ccc}
  c_{12} c_{13} & s_{12} c_{13} & s_{13} e^{-i \delta_{13}}\\
  - s_{12} c_{23} - c_{12} s_{23} s_{13} e^{i \delta_{13}} &
         c_{12} c_{23} - s_{12} s_{23} s_{13} e^{i \delta_{13}} &
                s_{23} c_{13}\\
  s_{12} s_{23} - c_{12} c_{23} s_{13} e^{i \delta_{13}} &
         - c_{12} s_{23} - s_{12} c_{23} s_{13} e^{i \delta_{13}} &
                c_{23} c_{13}
\label{Chau-Keung}
\end{array}
\right),
\ee
where $c_{ij} = \cos{\theta_{ij}}$ and
$s_{ij} = \sin{\theta_{ij}}$ control the mixing between
the $ij$ families \textbf{(Ex-23)},
while $\delta_{13}$ is the CP violating phase.

Perhaps the most useful parametrization
is the one developed by Wolfenstein \cite{Wol83},
based on the experimental result
\be
|V_{us}|^3 \approx |V_{cb}|^{3/2} \approx |V_{ub}|,
\ee
and on unitarity,
to obtain the matrix elements as series expansions in 
$\lambda \equiv |V_{us}| \approx 0.22$.
Choosing a phase convention in which
$V_{ud}$,
$V_{us}$,
$V_{cd}$,
$V_{ts}$,
and $V_{tb}$ are real,
Wolfenstein found
\be
V =
\left(
\begin{array}{ccc}
1 - \frac{1}{2}\lambda^2 &
	\lambda &
	A \lambda^3 (\rho - i \eta)
\\*[2mm]
- \lambda &
	1 - \frac{1}{2}\lambda^2 &
	A \lambda^2
\\*[2mm]
A \lambda^3 (1 - \rho - i \eta) &
	- A \lambda^2 &
	1
\end{array}
\right)
\ \
+
\ \
{\cal O} \left( \lambda^4 \right),
\label{Wolfenstein_parametrization}
\ee
where the series expansions are truncated at order $\lambda^3$.
Here,
$\eta$ is CP violating,
while the other three parameters are CP conserving.
The experimental result on $|V_{cb}|$ corresponds to $A \approx 0.8$.
So, it remains to discuss the experimental bounds
on $\rho$ and $\eta$ in section~\ref{sec:rho-eta}.

In terms of these parametrizations,
\be
J_{\rm CKM} = c_{12} c_{23} c_{13}^2
s_{12} s_{23} s_{13}\; | \sin{\delta_{13}} |
\sim
A^2 \lambda^6\;  | \eta |.
\ee
We recognize immediately that $J_{\rm CKM}$ is necessarily
small,
even if $\delta_{13}$ and $\eta$
turn out to be of order one (as will be the case),
because of the smallness of mixing angles.
As a result,
large CP violating asymmetries should only be found in channels
with small branching ratios;
conversely, channels with large branching ratios are likely
to display small CP violating asymmetries.
This fact drives the need for large statistics and,
thus, for experiments producing large numbers of $B$ mesons.

\subsubsection{Bounds on the magnitudes of CKM matrix elements}

This is a whole area of research in itself.
It involves a precise control of many advanced topics,
such as radiative corrections,
Heavy Quark Effective Theory,
estimates of the theoretical errors involved in
quantities used to parameterize certain hadronic matrix elements,
the precise way used to combine the various experimental and
theoretical errors, etc\dots \ 
For recent reviews see \cite{CKMfitter-04,UTfit-04}.

Schematically,
\begin{itemize}
\item[$|V_{ud}|$:] This is obtained from three independent methods:
i) superallowed Fermi transitions, which are beta decays connecting two
$J^P = 0^+$ nuclides in the same isospin multiplet;
ii) neutron $\beta$-decay;
and iii) the pion $\beta$-decay $\pi^+ \rightarrow \pi^0 e^+ \nu_e$.
\item[$|V_{us}|$:] This is obtained from kaon semileptonic decays,
$K^+ \rightarrow \pi^0 e^+ \nu_e$ and 
$K_L \rightarrow \pi^- e^+ \nu_e$.
Less precise are the values obtained from
semileptonic decays of hyperons,
such as $\Lambda \rightarrow p e^- \bar \nu_e$.
This matrix element determines the Wolfenstein 
expansion parameter $\lambda$.
\item[$|V_{cd}|$ \& $|V_{cs}|$:] The direct determination of these
matrix elements is plagued with theoretical uncertainties.
They are obtained from deep inelastic neutrino excitation
of charm, in reactions such as
$\nu_\mu d \rightarrow \mu^- c$ for $|V_{cd}|$,
and $\nu_\mu s \rightarrow \mu^- c$ for $|V_{cs}|$.
They may also be obtained from semileptonic $D$ decays;
$D^0 \rightarrow \pi^- e^+ \nu_e$, for $|V_{cd}|$,
or 
$D^0 \rightarrow K^- e^+ \nu_e$ and
$D^+ \rightarrow \overline{K^0} e^+ \nu_e$, 
for $|V_{cs}|$.
Better bounds are obtained through CKM unitarity.
\item[$|V_{cb}|$:] This is obtained from exclusive decays
$B \rightarrow D^{(\ast)} l \bar \nu_l$,
and from inclusive decays $B \rightarrow X_c l^- \bar \nu_l$.
This matrix element determines the Wolfenstein parameter
$A$.
\item[$|V_{ub}|$:] This is obtained from exclusive decays
such as $B \rightarrow \left\{ \pi, \rho, \dots \right\} l \bar \nu_l$,
and from inclusive decays $B \rightarrow X_u l^- \bar \nu_l$.
For a compilation of results for $|V_{ub}|$ and
$|V_{cb}|$ containing recent developments,
see,
for example, references \cite{Lig03,HFAG,Bau04}.
\end{itemize}

Using these and other constraints together with CKM unitarity,
the Particle Data Group obtains the following $90\% \mbox{C.L.}$
limits on the magnitudes of the CKM matrix elements \cite{PDG}
\be
\left(
\begin{array}{ccc}
0.9739-0.9751 & 0.221-0.227 & 0.0029-0.0045\\
0.221-0.227 & 0.9730-0.9744 & 0.039-0.044\\
0.0048-0.014 & 0.037-0.043 & 0.9990-0.9992
\end{array}
\right).
\ee

We should be aware that all the techniques
mentioned here are subject to many
(sometimes hot) debates.
Not surprisingly,
the major points of contention involve the assessment of
the theoretical errors and the precise procedure to include
those in overall constraints on the CKM mechanism.
Nevertheless,
everyone agrees that $\lambda$ and $A$ are rather well determined.
For example,
the CKMfitter Group finds \cite{CKMfitter-04}
\ba
\lambda &=& 0.2265^{\; +0.0025}_{\; -0.0023}\ ,
\\
A &=& 0.801^{\; +0.029}_{\; -0.020}\ ,
\ea
through a fit to the currently available data.

\subsubsection{Further comments on parametrizations of the CKM matrix}

Sometimes it is useful to parametrize the CKM matrix
with a variety of rephasing invariant 
combinations \cite{CKM_parametrizations},
including the magnitudes
\ba
R_b &=& \left| \frac{V_{ud} V_{ub}}{V_{cd} V_{cb}} \right| ,
\label{R_b}
\\
R_t &=& \left| \frac{V_{td} V_{tb}}{V_{cd} V_{cb}} \right| ;
\label{R_t}
\ea
the large CP violating phases
\ba
\alpha \equiv \phi_2 &\equiv&
\arg{\left( - \frac{V_{td} V_{tb}^\ast}{V_{ud} V_{ub}^\ast} \right)},
\label{alpha}
\\
\beta \equiv \phi_1 &\equiv&
\arg{\left( - \frac{V_{cd} V_{cb}^\ast}{V_{td} V_{tb}^\ast} \right)},
\label{beta}
\\
\gamma \equiv \phi_3 &\equiv&
\arg{\left( - \frac{V_{ud} V_{ub}^\ast}{V_{cd} V_{cb}^\ast} \right)};
\label{gamma}
\ea
or the small CP violating phases\footnote{The history of these
small phases is actually quite controverted.
They were first introduced as $\epsilon$ and $\epsilon^\prime$
by Aleksan, Kayser and London in \cite{AKL},
but this lead to confusion with the phenomenological
CP violating parameters in use for decades in the
kaon system ($\epsilon_K$ and $\epsilon^\prime_K$).
So,
the authors changed into the notation used here.
Later,
in a series of excellent and very influential lecture notes,
Nir \cite{Nir01}, changed the notation into
$\beta_s = \chi$ and $\beta_K = - \chi^\prime$.}
\ba
\chi &\equiv&
\arg{\left( - \frac{V_{cb} V_{cs}^\ast}{V_{tb} V_{ts}^\ast} \right)}\ ,
\nonumber\\
\chi^\prime &\equiv&
\arg{\left( - \frac{V_{us} V_{ud}^\ast}{V_{cs} V_{cd}^\ast} \right)}\ .
\ea
Notice that, \textit{by definition},
\be
\alpha + \beta + \gamma = \pi\ \ \  (\mbox{mod } 2 \pi)\ .
\label{a+b+g}
\ee
This arises directly from the definition of the angles,
regardless of whether $V$ is unitary or not \textbf{(Ex-24)}.
One interesting feature of such parametrizations,
is the observation that the unitarity of the CKM matrix
relates magnitudes with CP violating phases.
For example,
one can show that \cite{BLS} \textbf{(Ex-25)}
\ba
R_b &=&
\frac{\sin{\beta}}{\sin{(\beta + \gamma)}},
\nonumber\\
R_t &=&
\frac{\sin{\gamma}}{\sin{(\beta + \gamma)}}.
\label{Rb_Rt_and_sines}
\ea
Thus,
there is nothing forcing us to parametrize the CKM
matrix with three angles and one phase,
as done in the two parametrizations discussed above.
One could equally well parametrize the CKM matrix
exclusively with the four magnitudes $|V_{us}|$,
$|V_{ub}|$, $|V_{cb}|$, and $|V_{td}|$
\cite{Bra-Lav-88};
or with the four CP violating phases
$\beta$, $\gamma$, $\chi$, and $\chi^\prime$
\cite{AKL}.

Within the SM, these CP violating phases are all related
to the single CP violating parameter $\eta$ through
\ba
R_t e^{- i \beta} &\approx& 1 - \rho - i \eta,
\label{R_t_Wolf}
\\
R_b e^{- i \gamma} &\approx& \rho - i \eta,
\label{R_b_Wolf}
\\
\chi &\approx& \lambda^2 \eta,
\label{chi_Wolf}
\\
\chi^\prime &\approx& A^2 \lambda^4 \eta.
\label{chi_prime_Wolf}
\ea
What is interesting is that these four phases are also
useful in models beyond the SM \cite{BLS}.
Indeed,
even if there are extra generations of quarks,
and even if they have exotic $SU(2)_L$ quantum numbers,
the charged current interactions involving exclusively
the three first families may still be parametrized
by some $3 \times 3$ matrix $V$.
In general,
this matrix will cease to be unitary,
but one may still use the rephasing freedom of the
first three families in order to remove five phases.
Thus, this generalized matrix depends on 9 independent magnitudes
and 4 phases.
Therefore,
any experiment testing exclusively
(or almost exclusively)
CP violation in the interactions of the quarks 
$u$, $c$, $t$,
$d$, $s$ and $b$
with $W^\pm$,
should depend on only four phases.
Branco, Lavoura, and Silva have put forth this argument and
shown that
we may parametrize the CP violating phase structure of such a
generalized CKM matrix with \cite{BLS}
\be
\arg V =
\left(
\begin{array}{ccc}
	0 & \chi^\prime & - \gamma\\
	\pi & 0 & 0\\
	- \beta& \pi + \chi & 0 
\end{array}
\right),
\label{arg_V}
\ee
where a convenient phase convention has been chosen.
Of course,
since such a generalized matrix is no longer unitary,
Eqs.~(\ref{Rb_Rt_and_sines}) cease to be valid.
The fact that CP violation in the neutral kaon system
is small implies that $\chi^\prime$ is small in 
a very wide class of models \cite{BLS}.
In contrast,
$\chi$ might be larger than in the SM
\cite{BotBra04},
a fact which may eventually be probed by experiment \cite{SW2}.
The impact of $\chi$ may be accessed straightforwardly
through back of the envelope calculations,
with\footnote{I find Eq.~(\ref{crazy_V}) more useful than I claim here.
The point is that $R_b$ and $\beta$ are usually measured with
observables which involve the mixing, while $R_t$ and $\gamma$
are measured with processes involving only the decays.
The (approximate) redundant parametrization in Eq.~(\ref{crazy_V})
shows where that comes in.}
\be
V \approx
\left(
\begin{array}{ccc}
1 - \frac{1}{2}\lambda^2 &
	\lambda e^{i \chi^\prime}&
	A \lambda^3 R_b e^{- i \gamma}
\\*[2mm]
- \lambda &
	1 - \frac{1}{2}\lambda^2 &
	A \lambda^2
\\*[2mm]
A \lambda^3 R_t e^{- i \beta} &
	- A \lambda^2 e^{i \chi}&
	1
\end{array}
\right).
\label{crazy_V}
\ee

\subsection{\label{sec:unit_triangle}The unitarity triangle}

The fact that the SM CKM matrix is unitary,
$V V^\dagger = 1 = V^\dagger V$,
leads to six relations among the magnitudes.
They express the normalization  to unity of the three columns
and of the three rows of the CKM matrix.
It also leads to six relations involving both
magnitudes and phases,
\ba
V_{ud} V_{us}^\ast + V_{cd} V_{cs}^\ast + V_{td} V_{ts}^\ast &=& 0,
\label{triangle_ds}
\\
V_{us} V_{ub}^\ast + V_{cs} V_{cb}^\ast + V_{ts} V_{tb}^\ast &=& 0,
\label{triangle_sb}
\\
V_{ud} V_{ub}^\ast + V_{cd} V_{cb}^\ast + V_{td} V_{tb}^\ast &=& 0,
\label{triangle_db}
\\
V_{ud} V_{cd}^\ast + V_{us} V_{cs}^\ast + V_{ub} V_{cb}^\ast &=& 0,
\label{triangle_uc}
\\
V_{cd} V_{td}^\ast + V_{cs} V_{ts}^\ast + V_{cb} V_{tb}^\ast &=& 0,
\label{triangle_ct}
\\
V_{ud} V_{td}^\ast + V_{us} V_{ts}^\ast + V_{ub} V_{tb}^\ast &=& 0,
\label{triangle_ut}
\ea
where the first three relations express the orthogonality
of two different columns,
and the last three express the orthogonality of two different rows.
These relations may be represented as triangles in the complex plane,
with rather different shapes.
Indeed,
using the Wolfenstein expansion,
we see that the sides of the triangles in
Eqs.~(\ref{triangle_db}) and (\ref{triangle_ut}) are all of order
$\lambda^3$.
The triangles in Eqs.~(\ref{triangle_sb}) and (\ref{triangle_ct})
have two sides of order $\lambda^2$ and one side of order $\lambda^4$;
while those in Eqs.~(\ref{triangle_ds}) and (\ref{triangle_uc})
have two sides of order $\lambda$ and one side of order $\lambda^5$.
Remarkably,
all have the same area $J_{\rm CKM}/2$ which is,
thus,
a sign of CP violation \textbf{(Ex-26)}.

\begin{figure}[htb]
\centerline{\includegraphics*[height=3cm]{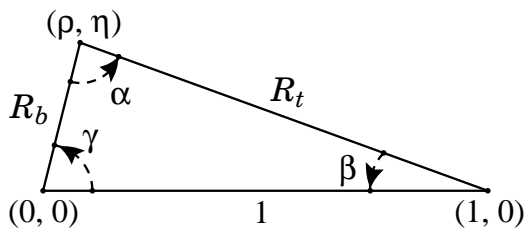}}
\caption{\label{unit_triangle}The unitarity triangle.
}
\end{figure}

The name ``unitarity triangle'' is usually
reserved for the orthogonality between the first ($d$) and third ($b$)
column,
shown in Eq.~(\ref{triangle_db}).
Aligning $V_{cd} V_{cb}^\ast$ with the real axis;
dividing all sides by its magnitude $|V_{cd} V_{cb}|$;
and using the definitions in Eqs.~(\ref{R_b})--(\ref{gamma}),
leads to
\be
R_b e^{i \gamma} + R_t e^{- i \beta} = 1.
\label{rescalled_triangle}
\ee
This is shown in 
FIG.~\ref{unit_triangle}.
In terms of the Wolfenstein parameters,
this triangle has an apex at coordinates
$(\rho, \eta)$ and area $|\eta|/2$ \textbf{(Ex-27)}.

In the SM,
$R_t$ and $\beta$ are determined in processes which
involve $B - \overline B$ mixing,
while $R_b$ and $\gamma$ are determined from processes
which do not and, thus, come purely from decay.
As a result,
the unitarity triangle checks for the consistency of
the information obtained from mixing with the
information obtained from decay \cite{mix_decay}.

There is one further feature of the CKM picture of CP violation
which is graphically seen in the unitarity triangle.
Imagine that we measure the magnitudes of two
sides of the triangle
(say, $R_b$ and $R_t$),
and that they add up to more than the third one
($R_b + R_t > 1$).
Then,
the triangle cannot be completely flat and
it will have a nonzero area.
But,
since this area is proportional to $J_{\rm CKM}$,
we would have identified CP violation by measuring
only three CP conserving magnitudes of sides.
We can now understand how the full CKM matrix,
including CP violation,
may be parametrized exclusively with the (CP conserving)
magnitudes of four matrix elements \cite{Bra-Lav-88}.

Sometimes it is claimed that the unitarity triangle
tests the relation $\alpha + \beta + \gamma = \pi$.
This is poor wording.
We have seen that a generalized CKM matrix has only four
independent phases,
which we may choose to be $\beta$, $\gamma$, $\chi$,
and $\chi^\prime$.
It follows directly from the definitions of $\alpha$,
$\beta$,
and $\gamma$ that these phases satisfy Eq.~(\ref{a+b+g}),
regardless of whether $V$ is unitary or not.
In the SM,
$\beta$ and $\gamma$ are large,
while $\chi$ and $\chi^\prime$ are small and smaller.
This is easily seen to leading order in the Wolfenstein approximation,
\textit{c.f.\/} Eqs.~(\ref{R_t_Wolf})--(\ref{chi_prime_Wolf}).
So, there are only two large independent phases in
the CKM matrix.
To put it bluntly: ``there is no such thing as $\alpha$''.
So what is meant by a ``test the relation
$\alpha + \beta + \gamma = \pi$''?
Imagine that one has measured $\beta$ and $\gamma$ with two
separate experiments.
Now imagine that a third experiment allows the determination
of the combination $\beta + \gamma$ 
(which, because $\alpha = \pi - \beta - \gamma$, is sometimes referred
to as a measurement of $\alpha$).
Clearly,
one may now probe whether the value of $\beta + \gamma$
obtained from the third experiment is consistent with the
results obtained previously for $\beta$ and $\gamma$.
This is what is meant by a ``test of the relation
$\alpha + \beta + \gamma = \pi$''.
But rewording it as we do here,
highlights the fact that there is nothing really fundamental about
the unitarity triangle;
we may equally well ``test the relation $\beta = \beta$'' by
measuring the angle $\beta$ with two independent processes;
or we may probe $\chi$, or \dots

\subsection{\label{sec:rho-eta}The $\rho - \eta$ plane}

What is true is that,
within the SM,
and given that $\lambda$ and $A$ are known rather well,
all other experiments which probe the CKM matrix will
depend only on $\rho$ and $\eta$.
This means that any experimental constraint of this
type may be plotted as some allowed region on
the $\rho - \eta$ plane.
Therefore,
a very expeditious way to search for new physics consists
in plotting the constraints from those experiments
in the $\rho - \eta$ plane,
looking for inconsistencies.
If any two regions fail to overlap,
we will have uncovered physics beyond the SM.

\begin{figure}[htb]
\centerline{\includegraphics*[height=3in]
{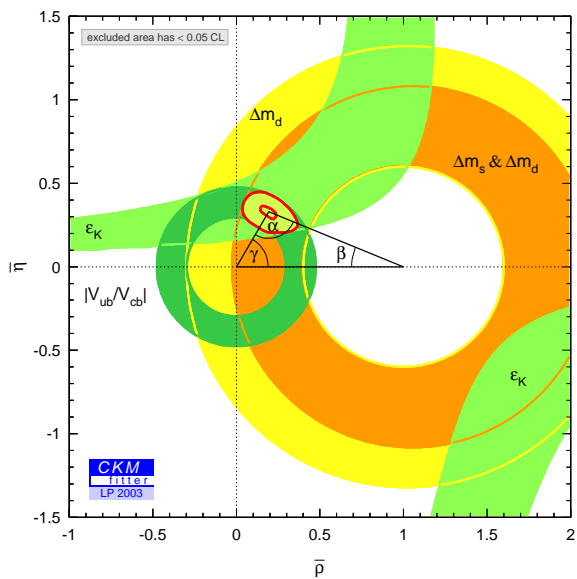}}
\caption{\label{figura rho-eta} ``Classic'' experimental constraints
on the $\rho - \eta$ plane.
The circles centered
at $(0,0)$ come from Eq.~(\ref{1}).
The circles centered at 
$(1,0)$ come from Eq.~(\ref{2}):
yellow for $\Delta m_d$,
brown for the improvement due to $\Delta m_s$.
The hyperbolic curves in green arise from
Eq.~(\ref{3}).
The intersection of all constraints (red curves)
determines the region in the
$\rho - \eta$ plane consistent with these experiments.
}
\end{figure}

There are three classic experiments which
constrain the SM parameters $\rho$ and $\eta$.
The SM formulae for these observables are well
described elsewhere \cite{elsewhere},
and we will only summarize some of the resulting features:
\begin{itemize}
\item The experimental determination of $|V_{ub}/(\lambda V_{cb})|$
from $b \rightarrow u$ and $b \rightarrow c$ decays
constrains
\be
R_b = \sqrt{\rho^2 + \eta^2}.
\label{1}
\ee
These bounds correspond to circles centered at $( \rho, \eta ) = (0,0)$. 
\item The mass difference in the
$B^0_d - \overline{B^0_d}$ system
is dominated by the box diagram with intermediate
top quarks,
being proportional to $|V_{tb} V_{td}|^2$.
This leads to a constraint on
\be
R_t = \sqrt{(1-\rho)^2 + \eta^2},
\label{2}
\ee
whose upper bound is improved by using also the
lower bound on the mass difference in the 
$B^0_s - \overline{B^0_s}$ system.
These bounds correspond to circles centered at $(\rho, \eta) = (1,0)$. 
\item The parameter $\delta_K$ measuring CP violation
in $K^0 - \overline{K^0}$ mixing arises from
a box diagram involving all up quarks as intermediate
lines.
Using CKM unitarity,
the result may be written as a function of
the imaginary parts of
$(V_{cs}^\ast V_{cd})^2$,
$(V_{ts}^\ast V_{td})^2$
and $V_{cs}^\ast V_{cd} V_{ts}^\ast V_{td}$.
This leads to a constraint of the type
\be
\eta( a - \rho) = b\ ,
\label{3}
\ee
with suitable constants $a$ and $b$.
These bounds correspond to an hyperbola in the $\rho - \eta$ plane.
\end{itemize}
The combination of these results,
with the limits available at the time of the conference
LP2003,
is shown in FIG.~\ref{figura rho-eta}
taken from the CKMfitter group \cite{CKMfitter-site}.

A few points should be noticed:
\begin{itemize}
\item the tests implicit in the rescaled unitarity triangle
of Eq.~(\ref{rescalled_triangle}) are also
illustrated in FIG.~\ref{figura rho-eta};
\item the sizable improvement provided when we utilize the
lower bound on $\Delta m_s$;
\item the agreement of all the allowed regions into a single
overlap region means that these experiments by themselves
are not enough to uncover new physics;
\item the rather large allowed regions provided by each
experiment individually, which are mainly due to theoretical
uncertainties.
As mentioned in the introduction,
hadronic messy effects are our main enemy in the search for 
signals of new physics.
\end{itemize}
The improvement provided by $\Delta m_s$ is mostly due to the fact that
the theoretical errors involved in extracting
$|V_{tb} V_{tq}|$ from $\Delta m_q$ ($q$ = $d$, $s$) cancel
partly in the ratio \cite{Schneider-PDG}
\be
\frac{\Delta m_s}{\Delta m_d}
= \frac{m_{B_s}}{m_{B_d}}
\xi^2 
\left| \frac{V_{ts}}{V_{td}} \right|^2 .
\ee
Here, $\xi = 1.15 \pm 0.05^{+0.12}_{-0.00}$
is an SU(3) breaking parameter obtained from lattice QCD
calculations \cite{xi_for_deltams}.
Thus, a measurement of $\Delta m_s$, when it becomes available
from experiments at hadronic machines, will be very important in
reducing the (mostly theoretical) uncertainties in the extraction of
$R_t$.

The bounds discussed in this section carry somewhat large
theoretical errors.
As we will see shortly,
the CP violating asymmetry in
$B_d \rightarrow J/\psi K_S$ provides us
with a very clean measurement of the CKM phase
$\beta$.
This was the first real test of the SM to come out of
the $B$ factories.

\section{\label{ch:road}On the road to $\lambda_f$}

In chapter~\ref{ch:producao} we saw that CP violation in the decays of
neutral $B$ mesons may be described by a phenomenological
parameter $\lambda_f$.
In chapter~\ref{ch:SM} we reviewed the SM;
a specific theory of electroweak interactions.
This theory will now be tested by calculating $\lambda_f$ for
a variety of final states and confronting its parameters
(most notably $\rho$ and $\eta$) with those experiments.
A few of the following sections were designed to avoid the
common potholes on that road.

\subsection{\label{sec:good_q/p}Can we calculate $q/p$?}

As we have seen in subsection~\ref{subsec:delta_B}
and in section~\ref{sec:Bd},
when studying large CP violating effects in $B$ meson decays,
we may assume $|\Gamma_{12}| \ll |M_{12}|$ and
CP conservation in $B - \overline{B}$ mixing.
As a result,
\ba
\frac{q_B}{p_B}
&=&
- \eta_B e^{i \xi},
\label{good_q/p_1}
\\
&=&
- \sqrt{\frac{M_{12}^\ast}{M_{12}}},
\label{good_q/p_2}
\ea
where $\xi$ is the arbitrary CP transformation phase
in
\ba
{\cal CP} | B^0_q \rangle &=&
e^{i \xi} | \overline{B^0_q} \rangle,
\\
{\cal CP} | \overline{B^0_q} \rangle &=&
e^{- i \xi} | B^0_q \rangle.
\ea
The parameter $\eta_B = \pm 1$ appears in
${\cal CP} | B_H \rangle = \eta_B | B_H \rangle$,
consistently with the fact that,
if there is CP conservation in the mixing,
then the eigenstates of the Hamiltonian must also be
CP eigenstates.\footnote{This sign $\eta_B$ should not be
confused with the parameter $\eta_B$ introduced in the
calculation of $M_{12}$ as a result of QCD corrections
to the relevant box diagram.
It is unfortunate that,
historically,
the same symbol is used for these two quantities.}
In the SM,
and when neglecting CP violation,
one obtains $\eta_B = -1$,
meaning that the heavier state is CP odd in that limit.

Having reached this point,
it is tempting to
``parametrize'' the phase of $M_{12}$ within a given
model and with ``suitable phase choices'' to be
$e^{- 2 i \phi_M}$.
One then concludes from Eq.~(\ref{good_q/p_2}) that
\be
\frac{q_B}{p_B} = - \eta_B e^{2 i \phi_M},
\label{bad_q/p}
\ee
where $\phi_M$ would be some measurable phase.
For instances,
in the SM one would obtain $\frac{q_B}{p_B} = e^{- 2 i \beta}$.
Strictly speaking, this is wrong,
because it is at odds with Eq.~(\ref{good_q/p_1});
\textit{i.e.\/}, it contradicts the quantum mechanical rule that,
when CP is conserved in mixing, the eigenstates of the
Hamiltonian should coincide with the eigenstates of CP.

Let us use Eq.~(\ref{good_q/p_2}) to
perform a correct calculation of $q_B/p_B$ \cite{appendixA-BLS}.
The quantity $M_{12}$ is calculated from an effective Hamiltonian
having a weak (CP odd) phase $- 2 \phi_M$,
and a $\Delta B = 2$ operator ${\cal O}$:
\ba
M_{12} &=& e^{- 2 i \phi_M}
\langle B^0_q | {\cal O} | \overline{B^0_q} \rangle,
\nonumber\\*[1mm]
M_{12}^\ast &=& e^{2 i \phi_M}
\langle \overline{B^0_d} | {\cal O}^\dagger | B^0_q \rangle.
\label{I need m12}
\ea
The operator ${\cal O}$ and its Hermitian conjugate
are related by the CP transformation
\be
\left(  {\cal CP}\right) {\cal O}^\dagger 
\left( {\cal CP} \right)^\dagger
= e^{2 i \xi_M} {\cal O}.
\label{a cp do o}
\ee
We may use two insertions
of $\left( {\cal CP} \right)^\dagger \left( {\cal CP} \right) = 1$
in the second Eq.~(\ref{I need m12}) to derive
\ba
M_{12}^\ast &=& e ^{2 i \phi_M} \langle \overline{B^0_q} |
\left( {\cal CP} \right)^\dagger \left( {\cal CP} \right) 
{\cal O}^\dagger
\left(  {\cal CP}\right)^\dagger \left( {\cal CP} \right) | B^0_q \rangle
\nonumber\\ &=&
e ^{2 i \left( \phi_M + \xi + \xi_M \right)}
\langle B^0_q | {\cal O} | \overline{B^0_q} \rangle
\nonumber\\ &=&
e ^{2 i \left( 2 \phi_M + \xi + \xi_M \right)} M_{12}.
\ea
Then,
from Eq.~(\ref{good_q/p_2}),
\be
\label{good_q/p_3}
\frac{q_B}{p_B} = - \eta_B e^{i \left( 2 \phi_M + \xi + \xi_M \right)}.
\ee
This should be equal to $- \eta_B e^{i \xi}$,
as in Eq.~(\ref{good_q/p_1}).
The CP transformation phase $\xi_M$ must therefore be chosen
such that $2 \phi_M + \xi_M = 0$.

How does that come about?
Let us illustrate this point with the calculation of $q_B/p_B$ within the SM.
There,
\be
{\cal O} \propto \left[ \overline{q} \gamma^\mu
\left( 1 - \gamma_5 \right) b \right]
\left[ \overline{q} \gamma_\mu \left( 1 - \gamma_5 \right) b \right],
\label{o o}
\ee
and
\be
e^{- 2 i \phi_M} = \frac{V_{tb} V_{tq}^\ast}{V_{tb}^\ast V_{tq}}
=
\left\{
\begin{array}{llr}
e^{- 2 i \beta} & \ \ \mbox{for} & B_d,
\\
e^{2 i \chi} & \ \ \mbox{for} & B_s.
\end{array}
\right.
\label{o phim do sm}
\ee
Now,
in the mass basis,
the most general CP transformation of the quark fields $b$ and $q$ is,
according to Eqs.~(\ref{CP_with_rephasing}),
\ba
\left( \cp \right) b \left( \cp \right)^\dagger
&=&
e^{i \xi_b} \gamma^0 C \overline{b}^T,
\nonumber\\*[1mm]
\left( \cp \right) \overline{q} \left( \cp \right)^\dagger
&=&
- e^{- i \xi_q} q^T C^{-1} \gamma^0.
\ea
Then,
from Eqs.~(\ref{o o}) and (\ref{a cp do o}),
$\xi_M = \xi_q - \xi_b$ and
\be
\frac{q_B}{p_B} =
- \eta_B e^{i (\xi + \xi_q - \xi_b)}
\frac{V_{tb}^\ast V_{tq}}{V_{tb} V_{tq}^\ast}.
\label{q/p_B_SM}
\ee
The requirement that $2 \phi_M + \xi_M = 0$ is equivalent to
\be
V_{tb} V_{tq}^\ast = e^{i \left( \xi_q -\xi_b \right)} V_{tb}^\ast V_{tq}.
\label{a condicao do co}
\ee
It is clear that we may always choose $\xi_q$ and $\xi_b$
such that Eq.~(\ref{a condicao do co}) be verified,
thus obtaining CP invariance.
We recognize Eq.~(\ref{a condicao do co}) as resulting
from Eq.~(\ref{afinal}),
which expresses CP conservation in the SM.
We conclude from this particular example that,
when one discards the free phases in the CP transformation of the quark
fields,
one may occasionally run into contradictions.

But now we have another problem.
If Eq.~(\ref{good_q/p_1}) holds in any model leading to
CP conservation in mixing,
and since $\xi$ is an arbitrary phase,
what does it mean to calculate $\lambda_f$?

\subsection{\label{sec:cancellation}Cancellation of the 
CP transformation phases in $\lambda_f$}

Let us consider the decays of $B^0_q$ and 
$\overline{B^0_q}$ into a CP eigenstate $f_{\rm cp}$:
\be
{\cal CP} | f_{\rm cp} \rangle =  \eta_f | f_{\rm cp} \rangle,
\ee
with $\eta_f = \pm 1$.
We assume that the decay amplitudes have only one weak phase $\phi_A$,
with an operator ${\cal O}^\prime$ controlling the decay,
\ba
A_f &=& 
e^{i \phi_A} \langle f_{\rm cp} | {\cal O}^\prime | B^0_q \rangle,
\nonumber\\*[1mm]
\bar A_f &=&
e^{- i \phi_A} 
\langle f_{\rm cp} | {{\cal O}^\prime}^\dagger | \overline{B^0_q} \rangle.
\ea
The CP transformation rule for ${\cal O}^\prime$ is
\be
\left( {\cal CP} \right) 
{{\cal O}^\prime}^\dagger 
\left( {\cal CP} \right)^\dagger
= e^{- i \xi_D} {\cal O}^\prime.
\ee
Then,
\ba
\bar A_f &=& 
e^{- i \phi_A}
\langle f_{\rm cp} |
\left( {\cal CP} \right)^\dagger
\left( {\cal CP} \right) 
{{\cal O}^\prime}^\dagger
\left( {\cal CP} \right)^\dagger 
\left( {\cal CP} \right) 
| \overline{B^0_q} \rangle
\nonumber\\
&=&
\eta_f
e^{- i \left( \phi_A + \xi + \xi_D \right)}
\langle f_{\rm cp} | {\cal O}^\prime | B^0_q \rangle
\nonumber\\
&=&
\eta_f e^{- i \left( 2 \phi_A + \xi + \xi_D \right)} A_f.
\label{equacao da relacao}
\ea
Combining Eq.~(\ref{good_q/p_3}) and (\ref{equacao da relacao}),
we obtain
\be
\lambda_f \equiv \frac{q_B}{p_B} \frac{\bar A_f}{A_f} = - \eta_B \eta_f
e^{2 i (\phi_M - \phi_A)}
e^{i (\xi_M - \xi_D)}.
\label{ytoui}
\ee
We now state the following:
if the calculation has been done correctly,
then the phases $\xi_M$ and $\xi_D$,
which arise in the CP transformation of the mixing and decay operators,
are equal and cancel out.
This cancellation is due to the fact that,
because they involve the same quark fields,
the CP transformation properties of the $\Delta B = 2$ operators
describing the mixing are related to those of the $\Delta B = 1$ operators
describing the decay.
Thus,
\be
\label{aquihadronicuncertfall}
\lambda_f = - \eta_B \eta_f e^{2 i (\phi_M - \phi_A)}.
\ee
An explicit example of the cancellation of
the CP transformation phases
occurs in the SM computation of the parameter $\lambda_f$,
as shown in chapter 33 of reference \cite{BLS} for a variety
of final states.
Below we will check this cancellation explicitly
for the decay $B_d \rightarrow J/\psi K_S$.

There are two important points to note
in connection with Eq.~(\ref{aquihadronicuncertfall}):
\begin{itemize}
\item If we had set $\xi_M = \xi_D = 0$ from the very beginning
we would have obtained the correct result for $\lambda_f$.
This is what most authors do.
The price to pay is,
as pointed out above,
an inconsistency between Eqs.~(\ref{bad_q/p}) and (\ref{good_q/p_1}).
\item The $- \eta_B = \mp 1$ 
sign in Eq.~(\ref{aquihadronicuncertfall}) is important.
That sign comes from $q_B/p_B$ in Eq.~(\ref{good_q/p_1}).
And, to be precise,
the sign of $q_B/p_B$ is significant only when compared with
the sign of either $\Delta m$ or $\Delta \Gamma$.
Therefore,
it is not surprising to find that $\lambda_f$ always appears multiplied
by an odd function of either $\Delta m$
or $\Delta \Gamma$ in any experimental observable\footnote{It is
sometimes stated that the sign of $\mbox{Im} \lambda$ can be predicted.
The meaning of that statement should be clearly understood.
What can be predicted is the sign of $\Delta m\, \mbox{Im} \lambda_f$.
Indeed,
the interchange $P_H \leftrightarrow P_L$ makes $\Delta m$,
$\Delta \Gamma$,
$q/p$,
and $\lambda_f$ change sign.
If one chooses,
as we do,
$\Delta m > 0$,
then the sign of $\mbox{Im} \lambda_f$ becomes well defined
and can indeed be predicted,
at least in some models.},
\textit{c.f.\/} Eqs.~(\ref{master_Bs}) and (\ref{master_Bd}).
\end{itemize}

\subsection{\label{parametrization}A common parametrization
for mixing and decay within a given model}

Having realized where contradictions might (and do) arise
and that the calculations of $\lambda_f$ are safe,
we will now brutally simplify the discussion by
ignoring the ``spurious'' phases brought about by
CP transformations,

Let us consider the decay
$B^0_d \rightarrow f_{\rm cp}$,
mediated by two diagrams with magnitudes
$A_1$ and $A_2$, 
CP odd phases (weak-phases)
$\phi_{A1}$ and $\phi_{A2}$,
and CP even phases (strong phases)
$\delta_1$ and $\delta_2$.
Let us take $\phi_M$ as the CP odd phase
in $B^0_d - \overline{B^0_d}$ mixing.
Then,
\ba
\frac{q_B}{p_B}
&=&
- \eta_B e^{2i\phi_M}\ ,
\label{q/p_parametrization}
\\
A_f
&=&
A_1 e^{i \phi_{A1}} e^{i \delta_1} + A_2 e^{i \phi_{A2}} e^{i \delta_2}\ ,
\\
\bar A_f
&=&
\eta_f 
\left(
A_1 e^{- i \phi_{A1}} e^{i \delta_1} + A_2 e^{- i \phi_{A2}} e^{i \delta_2}
\right)\ ,
\ea
from which
\be
\lambda_f = - \eta_B \eta_f e^{-2i \phi_1}
\frac{1+r e^{i(\phi_1 - \phi_2)} e^{i \delta}}{
1+r e^{- i(\phi_1 - \phi_2)} e^{i \delta}}\ ,
\label{Lf_r_and_delta}
\ee
where $\eta_B = \pm 1$, $\eta_f = \pm 1$,
$\phi_1 \equiv \phi_{A1}-\phi_M$,
$\phi_2 \equiv \phi_{A2}-\phi_M$,
$\delta=\delta_2-\delta_1$
and $r=A_2/A_1$.

In a model,
such as the Standard Model,
the CP odd phases are determined by the weak interaction
and are easily read off from the fundamental Lagrangian.
In contrast,
the CP even phases are determined by the strong interactions
(and, on occasion, the electromagnetic interactions)
and involve the calculation of hadronic matrix elements,
including the final state interactions.
These are usually calculated within a model
of the hadronic interactions.
Naturally,
such calculations depend on the model used and,
therefore,
these quantities and the studies associated with them
suffer from the corresponding ``hadronic uncertainties''.
Therefore,
we are most interested in decays for which these
hadronic uncertainties are small, or,
in the limit, nonexistent.

\subsubsection{Decays mediated by a single weak phase}

This case includes those situations in which the decay
is mediated by a single diagram
($r=A_2/A_1=0$)
as well as those situations in which there are several
diagrams mediating the decay,
but all share the same weak phase
($\phi_1-\phi_2=\phi_{A1}-\phi_{A2}=0$).
Then
\be
\lambda_f = - \eta_B \eta_f e^{-2i \phi_1},
\ee
from which $|\lambda_f|=1$,
and Eqs.~(\ref{C_f}) and (\ref{S_f}) yield
\ba
C_f &=& 0\ ,
\label{best_C_f}
\\
S_f &=& \eta_B \eta_f \sin{2 \phi_1}\ .
\label{best_S_f}
\ea

These are clearly ideal decays,
because the corresponding CP asymmetry depends on a single
weak phase (which may be calculated in the Standard Model as
well as other models);
it does not depend on the strong phases, nor on the
magnitudes of the decay amplitudes
(meaning that these asymmetries do not depend on the
hadronic uncertainties.)
Therefore,
the search for CP violating asymmetries in decays
into final states which are eigenstates of CP and
whose decay involves only one weak phase constitutes
the Holy Grail of
CP violation in the $B$ system.

\subsubsection{Decays dominated by one weak phase}

Unfortunately,
most decays involve several diagrams,
with distinct weak phases.
To understand the devious effect that a second
weak phase has,
it is interesting to consider the case in which,
although there are two diagrams with different weak phases,
the magnitudes of the corresponding decay amplitudes
obey a steep hierarchy $r<<1$.
In that case,
Eqs.~(\ref{C_f}) and (\ref{S_f}) yield \textbf{(Ex-28)}
\ba
C_f & \approx & 2 r \sin{(\phi_1-\phi_2)} \sin{\delta}\ ,
\label{def:adir}
\\
S_f & \approx & \eta_B \eta_f
\left[
\sin{2 \phi_1}
- 2 r \cos{2 \phi_1} \sin{(\phi_1-\phi_2)} \cos{\delta}
\right]\ .
\label{def:aint}
\ea
These equations allow us to learn a few important lessons.

First,
the CP violation present in the decays (direct CP violation)
is only non-zero if
\begin{itemize}
\item there are at least two diagrams mediating the decay;
\item these two diagrams have different weak phases;
\item and these two diagrams also have different strong phases.
\end{itemize}
On the other hand,
since it depends on
$r$ and $\delta$, 
the calculation of the direct CP violation parameter
$C_f$ depends always on the hadronic uncertainties.
These features do not depend on the $r$ expansion
which we have used;
they are valid in all generality and hold also for the
direct CP violation probed with $B^\pm$ decays.

Second,
when we have two diagrams involving two distinct weak phases,
the interference CP violation also becomes dependent on
$r$ and $\delta$.
As a result,
the calculation of $S_f$ is also subject to hadronic
uncertainties.
Notice that,
for $S_f$,
this problem is worse than it seems.
Indeed,
even if the final state interactions are very small
(in which case $\delta \sim 0$,
and $C_f \sim 0$ does not warn us about the presence of a
second weak phase.),
$S_f$ will still depend on $r$ \cite{Gronau93}.
That is,
the presence of a second amplitude with a different
weak phase can destroy the measurement of
$\sin 2 \phi_1$,
even when the strong phase difference vanishes.
This problem occurs even for moderate values of $r$.

To simplify the discussion,
we could say that some $B$ decays we are interested in
have both a tree level diagram and a gluonic penguin diagram,
which is higher order in perturbation theory.
As such, we could expect that $r = A_2/A_1 < 1$.
However,
this might not be the case,
both because the tree level diagram might be suppressed
by CKM mixing angles,
and because the decay amplitudes
involve hadronic matrix elements which,
in some cases,
are difficult to estimate.
For this purpose,
it is convenient to write $r = r_{\rm ckm} r_h$,
where $r_{\rm ckm}$ is the ratio of the magnitudes of
the CKM matrix elements in the two diagrams.
We can now separate two possibilities,
according to the size of 
$r_{\rm ckm} \sin{(\phi_1 - \phi_2)}$ \cite{garbage}:
\begin{enumerate}
\item if $r_{\rm ckm} \sin{(\phi_1 - \phi_2)} \ll 1$,
then combining this
with some rough argument that $r_h$ is small
will allow us to conclude that
$r \sin{(\phi_1 - \phi_2)} \ll 1$,
and $S_f \approx  \eta_B \eta_f \sin{2 \phi_1}$; 
\item if $r_{\rm ckm} \sin{(\phi_1 - \phi_2)} \sim 1$,
then we must really take the second weak phase into account,
because any rough argument about the
``smallness'' of $r_h$,
by itself,
will not guarantee that the error introduced by
$r \sin{(\phi_1 - \phi_2)}$
will indeed be small.
\end{enumerate}
An exhaustive search shows that the best case occurs for
$B_d \rightarrow J/\psi K_S$.

\section{\label{ch:B_decays}$B$ decays as a test of the Standard Model}

The interest in $B$ decays had its origin in the seminal articles
by Carter and Sanda \cite{Carter-Sanda} and by Bigi and Sanda
\cite{Bigi-Sanda}.\footnote{It is rumored that when
Bigi, Carter, and Sanda started to give seminars suggesting the
search for CP violation in $B$ decays,
some audiences were less than enthusiastic (to put it politely).
Two decades later, we cannot thank them enough for their resilience.}
They identified early on the decay $B_d \rightarrow J/\psi K_S$
as a prime candidate in the search for CP violation outside
the kaon system.

\subsection{The decay $B_d \rightarrow J/\psi K_S$}

\subsubsection{\label{subsec:rephasing_invariance}Rephasing invariance,
reciprocal basis, and other details}

Although this is the most famous $B$ decay,
its calculation is fraught with hazardous details:
\begin{enumerate}
\item $J/\psi K_S$ is not a CP eigenstate, given that
$K_S$ itself is not a CP eigenstate.
However,
ignoring this detail does not affect our conclusions,
because $\delta_K \sim 10^{-3}$ may be neglected with respect to the
CP violating asymmetry of order unity present in 
$B_d \rightarrow J/\psi K_S$.
\item In some sense,
$K_S$ is not even a good final state,
since what we detect are its decay products $\pi^+ \pi^-$.
Now,
it is clear that by selecting those events in which the kaon lives
for proper times much greater than $\tau_S$,
the $K_S$ component will have decayed away,
and what one really measures is $K_L$.
Therefore,
the correct calculation must consider both paths
of the cascade decay:
$B_d \rightarrow J/\psi K_S \rightarrow J/\psi (\pi^+ \pi^-)$
and $B_d \rightarrow J/\psi K_L \rightarrow J/\psi (\pi^+ \pi^-)$
\cite{cascades}.
Of course,
this effect is negligible for the small kaon decay times
used in extracting $B_d \rightarrow J/\psi K_S$ 
data.\footnote{Nevertheless,
in $B \rightarrow D X \rightarrow [f]_D X$ cascade decays,
the fact that $D$ is really an intermediate state must be taken
into account,
even for vanishing $D^0 - \overline{D^0}$ mixing.
Otherwise,
the result will not be rephasing invariant.
See sections 34.4 and 34.5 of \cite{BLS}.}
\item In the spectator quark approximation,
the SM only allows the decays
$B^0_d \rightarrow J/\psi K^0$ and
$\overline{B^0_d} \rightarrow J/\psi \overline{K^0}$.
The decay $B_d \rightarrow J/\psi K_S$ is only possible
due to $K^0 - \overline{K^0}$ mixing,
which must be taken into account through \cite{BLS}
\be
\frac{q_K}{p_K} = - \eta_K
e^{i (\xi_K + \xi_d - \xi_s)}
\frac{V_{us}^\ast V_{ud}}{V_{us} V_{ud}^\ast}.
\label{q/p_K_SM}
\ee
\item Although we will neglect CP violation in $K^0 - \overline{K^0}$ mixing,
we \textit{must} use the reciprocal basis,
or some expressions will be wrong.
\item Because the vector meson $J/\psi$ and the kaon arise from
a $B$ decay,
they must be in a relative $l=1$ state,
which upon a CP (P) transformation yields an extra minus sign. 
\end{enumerate}
We may now combine the last three remarks into the calculation
of $\lambda_{B_d \rightarrow J/\psi K_S}$.

As we have mentioned in section~\ref{sec:reciprocal},
when dealing with a $K_S$ in the final state we \textit{must}
use the reciprocal basis
\be
\langle \tilde{K}_S |
=
\frac{1}{2 p_K} \langle K^0 |
+ \frac{1}{2 q_K} \langle \overline{K^0} |.
\ee
Therefore,
\ba
\langle J/\psi \tilde{K}_S | T | B^0_d \rangle
&=&
\frac{1}{2 p_K} \langle J/\psi K^0 | T | B^0_d \rangle,
\nonumber\\
\langle J/\psi\tilde{K}_S | T | \overline{B^0_d} \rangle
&=&
\frac{1}{2 q_K} \langle J/\psi \overline{K^0} | T | \overline{B^0_d} \rangle,
\ea
leading to
\be
\lambda_{B_d \rightarrow J/\psi K_S}
=
\frac{q_B}{p_B}
\frac{\bar A_{J/\psi K_S}}{A_{J/\psi K_S}}
=
\frac{q_B}{p_B}
\frac{\langle J/\psi \overline{K^0} | T | \overline{B^0_d} \rangle}{
\langle J/\psi K^0 | T | B^0_d \rangle}
\frac{p_K}{q_K}.
\label{psi_KS_1}
\ee
The presence of $p_K/q_K$ 
expresses the fact that the interference can only occur
due to $K^0 - \overline{K^0}$ mixing.
\footnote{An alternative formula which highlights this fact is
\be
\lambda_{B_d \rightarrow J/\psi K_S}
=
\frac{q_B}{p_B}
\frac{\langle J/\psi \overline{K^0} |T| \overline{B^0_d} \rangle}{
\langle J/\psi K^0 |T| B^0_d \rangle}
\frac{\langle \tilde{K}_S|\overline{K^0}\rangle}{
\langle \tilde{K}_S|K^0 \rangle}.
\ee
}

\begin{figure}[htb]
\centerline{\includegraphics*[height=1.5in]{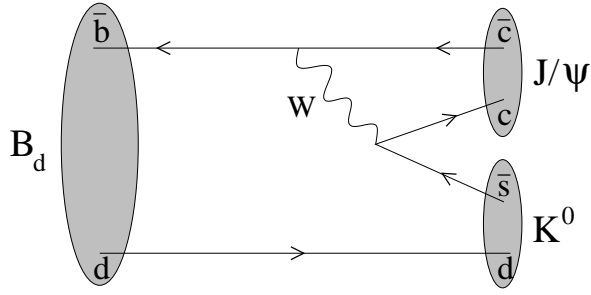}}
\caption{\label{figura treeJK}Tree level diagram for
$B_d \rightarrow J/\psi K_S$.
}
\end{figure}

The mixing parameters are shown in Eqs.~(\ref{q/p_B_SM}) and 
(\ref{q/p_K_SM}).
We must now turn to the decay amplitudes.
We start by assuming that the decay is mediated only by
the tree level diagram in FIG.~\ref{figura treeJK}.
Then
\be
\langle J/\psi K^0 | T | B^0_d \rangle
\propto
V_{cb}^\ast V_{cs}
\langle J/\psi K^0 |
\left[ \overline{b} \gamma^\mu
\left( 1 - \gamma_5 \right) c \right]
\left[ \overline{c} \gamma_\mu \left( 1 - \gamma_5 \right) s \right]
| B^0_d \rangle.
\ee
We now use multiple insertions of
$\left( {\cal CP} \right)^\dagger \left( {\cal CP} \right) = 1$.
Then
\be
\langle J/\psi K^0 | \left( {\cal CP} \right)^\dagger 
=
- e^{-i\xi_K} \langle J/\psi \overline{K^0} |,
\ee
where the minus sign appears because $J/\psi$ and $K$
are in a relative $l=1$ state.
Also
\ba
\left( {\cal CP} \right)
\left[ \overline{b} \gamma^\mu
\left( 1 - \gamma_5 \right) c \right]
\left( {\cal CP} \right)^\dagger
&=&
- e^{i(\xi_c -\xi_b)}
\left[ \overline{c} \gamma_\mu
\left( 1 - \gamma_5 \right) b \right],
\nonumber\\
\left( {\cal CP} \right)
\left[ \overline{c} \gamma_\mu
\left( 1 - \gamma_5 \right) s \right]
\left( {\cal CP} \right)^\dagger
&=&
- e^{i(\xi_s -\xi_c)}
\left[ \overline{s} \gamma^\mu \left( 1 - \gamma_5 \right) c \right],
\nonumber\\
\left( {\cal CP} \right)
| B^0_d \rangle
&=&
e^{i \xi_B}
| \overline{B^0_d} \rangle.
\ea
We obtain
\be
\frac{\langle J/\psi \overline{K^0} | T | \overline{B^0_d} \rangle}{
\langle J/\psi K^0 | T | B^0_d \rangle}
=
- e^{i (\xi_K - \xi_B + \xi_b - \xi_s)}
\frac{V_{cb} V_{cs}^\ast}{V_{cb}^\ast V_{cs}}.
\label{Abar/A_JpsiKS_M}
\ee

Substituting Eqs.~(\ref{q/p_B_SM}),
(\ref{q/p_K_SM}),
and (\ref{Abar/A_JpsiKS_M}) into Eq.~(\ref{psi_KS_1}),
we find,
\ba
\lambda_{B_d \rightarrow J/\psi K_S}
&=&
- \eta_B e^{i (\xi_B + \xi_d - \xi_b)}
\frac{V_{tb}^\ast V_{td}}{V_{tb} V_{td}^\ast}\;
(- \eta_K)
e^{-i (\xi_K + \xi_d - \xi_s)}
\frac{V_{us} V_{ud}^\ast}{V_{us}^\ast V_{ud}}\;
(-) e^{i (\xi_K - \xi_B + \xi_b - \xi_s)}
\frac{V_{cb} V_{cs}^\ast}{V_{cb}^\ast V_{cs}}
\nonumber\\
&=&
- \eta_B \eta_K
\frac{V_{tb}^\ast V_{td}}{V_{tb} V_{td}^\ast}\;
\frac{V_{us} V_{ud}^\ast}{V_{us}^\ast V_{ud}}\;
\frac{V_{cb} V_{cs}^\ast}{V_{cb}^\ast V_{cs}}
\nonumber\\
&=&
- e^{- 2 i(\beta - \chi^\prime)}.
\label{lambda_psiKs_SM}
\ea
where Eq.~(\ref{arg_V}) and $\eta_B=\eta_K=-1$
were used to obtain the last line.
Notice the cancellation of the various spurious phases
$\xi$ brought about by the CP transformations in going
from the first to the second line;
the result is manifestly rephasing invariant.
The cancellation occurs both for the spurious phases
involving the kets and bras, $\xi_B$ and $\xi_K$,
and for the spurious phases involving the fields in the quark field
operators, $\xi_q$, \textit{c.f.\/} Eq.~(\ref{quarkVShadron}).
The last cancellation involves the balance between
the CP transformation properties of the $\Delta B=2$ mixing
and the $\Delta B=1$ decay operators and provides an explicit
example of the cancellation $\xi_M - \xi_D = 0$
mentioned in section~\ref{sec:cancellation}.

\subsubsection{\label{sec:simplified_q/p}Simplified $q/p$ and new physics}

These lengthy calculation involving the phases $\xi$ and fractions
with $V^\ast V$ bilinears are very reassuring,
but they are a very hefty price to pay for
consistency between Eqs.~(\ref{good_q/p_1}) and (\ref{bad_q/p}).
It would be much easier to ignore the 
CP transformation phases in Eqs.~(\ref{q/p_B_SM}) and
(\ref{q/p_K_SM}) and to substitute the phases of
all CKM matrix elements by Eq.~(\ref{arg_V}).
We would obtain
\ba
\frac{q_B}{p_B} &=& e^{-2 i \tilde{\beta}},
\label{simple_q/p_B}
\\
\frac{q_K}{p_K} &=& e^{- 2 i \chi^\prime}
\sim 1.
\label{simple_q/p_K}
\ea
The phase $\tilde{\beta}$ in Eq.~(\ref{simple_q/p_B}) includes
the possibility that there might be new physics contributions
to the relevant phase in $B^0_d - \overline{B^0_d}$ mixing
\cite{new_phys_mix}.
In the SM,
$\tilde{\beta}$ coincides with the CKM phase $\beta$.
Because $\delta_K \sim 10^{-3}$,
$\chi^\prime$ is likely to be small in almost any
new physics model \cite{BLS},
and we will ignore it.
Similarly,
we may take
\be
\frac{q_{Bs}}{p_{Bs}} = e^{2 i \tilde{\chi}},
\label{simple_q/p_Bs}
\ee
where $\tilde{\chi}$ allows for new physics contributions
to the relevant phase in $B^0_s - \overline{B^0_s}$ mixing.
In the SM $\tilde{\chi}$ coincides with the CKM phase $\chi$.

Because we know that,
in the end,
the expression for $\lambda_f$ yields the same result,
and because we know where the pitfall is,
we will henceforth use
Eqs.~(\ref{simple_q/p_B})--(\ref{simple_q/p_Bs}).
It is trivial to reproduce the last line of
Eq.~(\ref{lambda_psiKs_SM}) with this \textbf{(Ex-29)}.

\subsubsection{$B_d \rightarrow J/\psi K_S$ involves one weak phase}

\begin{figure}[htb]
\centerline{\includegraphics*[height=1.5in]{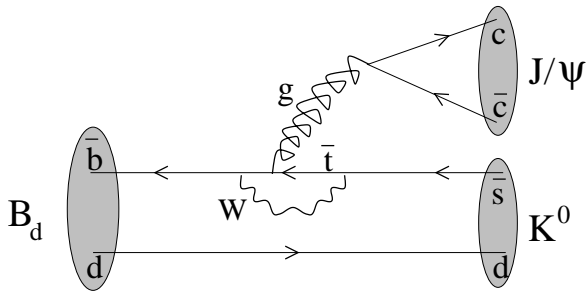}}
\caption{\label{figura pengJK}Penguin diagram with a
virtual top quark which mediates the decay
$B_d \rightarrow J/\psi K_S$.
The gluonic line represents two or more gluons.
}
\end{figure}

The decay $B_d \rightarrow J/\psi K_S$ is mediated by the
tree level diagram in FIG.~\ref{figura treeJK}.
But it also gets a contribution from
FIG.~\ref{figura pengJK}.
The two diagrams are proportional to
\ba
V_{cb}^\ast V_{cs}
&\sim&
(A \lambda^2) (1),
\nonumber\\
V_{tb}^\ast V_{ts}
&\sim&
(1) (-A \lambda^2 e^{i \chi}),
\label{CKM_coef_JK}
\ea
respectively,
where we have used the parametrization in Eq.~(\ref{crazy_V}).
The difference between the two weak phases is
$\phi_1 - \phi_2 = \chi$ which,
in the SM,
is proportional to $\lambda^2$.
On the other hand,
since the penguin diagram is higher order in the
weak interactions,
we expect $r$ to be suppressed also.
Therefore
$r \sin{(\phi_1 - \phi_2)} \sim \lambda^2 r$,
the decay is overwhelmingly dominated by one weak phase,
and Eq.~(\ref{lambda_psiKs_SM}) remains valid -- 
possibly with $\tilde{\beta}$ in the place of $\beta$,
to allow for the possibility of a new physics contribution to
the phase in $B^0_d - \overline{B^0_d}$ mixing.

As a result,
we conclude from Eqs.~(\ref{C_f}) -- or (\ref{best_C_f}) --
that there is no direct CP violation in this decay,
and that the interference CP violation term in
Eqs.~(\ref{S_f}) -- or (\ref{best_S_f}) -- is simply
$S_{J/\psi K_S} = \sin{2 \tilde{\beta}}$.
The measurement of this parameter by BABAR and Belle
constituted the first observation of CP violation
outside the kaon system.
It had to wait over 35 years!
The PDG2004 world average is \cite{PDG}
\ba
|\lambda_{B_d \rightarrow (c \bar c) K}| &=& 0.949 \pm 0.045\ ,
\label{current_bound_|L|_psiKS}
\\
\sin{2 \tilde{\beta}} &=& 0.731 \pm 0.056\ ,
\label{current_bound_sin2beta}
\ea
which provides an extremely precise constraint on a CKM parameter
of the SM.
The HFAG group has updated this result after the conferences
of the summer of 2004,
including all the charmonium states,
obtaining
$|\lambda_{B_d \rightarrow (c \bar c) K}| = 0.969 \pm 0.028$
and 
$\sin{2 \tilde{\beta}} = 0.725 \pm 0.037$ \cite{HFAG}.
Recalling that $|\lambda_f| = |q_B/p_B| |\bar A_f/A_F|$,
Eq.~(\ref{current_bound_|L|_psiKS}) is consistent with very
small or vanishing CP violation in both $B_d$ mixing and
the $b \rightarrow c \bar c s$ decay amplitudes.

There are other diagrams contributing to the decay
$B_d \rightarrow J/\psi K_S$,
besides those in FIGs.~\ref{figura treeJK}
and \ref{figura pengJK}.
For example,
we could use a virtual up quark instead of the virtual top quark
in FIG.~\ref{figura pengJK},
which would seem to bring with it a third CKM
combination $V_{ub}^\ast V_{us}$.
However,
due to the CKM unitarity relation in Eq.~(\ref{triangle_sb}),
we are still left with only two independent weak phases.
Strictly speaking,
one should refer to the operators multiplying
each of the (chosen) two weak phases relevant for any 
particular decay,
rather than to specific diagrams.
This is elegantly included in the effective 
Hamiltonian approach \cite{BBL96}.
However,
the pictorial description of figures like
FIGs.~\ref{figura treeJK}
and \ref{figura pengJK} provides a very intuitive idea
of the mechanisms at hand in each decay.

\subsubsection{Setting $\sin{2 \beta}$ on the $\rho - \eta$ plane}

\begin{figure}[htb]
\centerline{\includegraphics*[height=3in]
{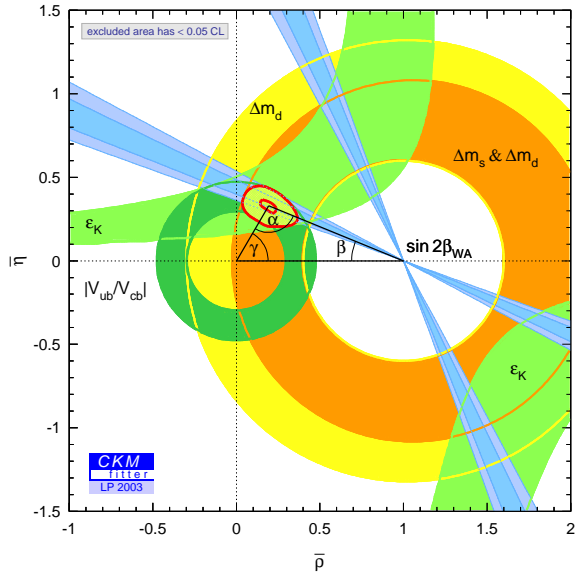}}
\caption{\label{figura rho-eta with 2b}Constraints on
the $\rho - \eta$ plane, with the results from
$\sin{2 \beta}$ overlaid.
}
\end{figure}

In the SM,
$\tilde{\beta} = \beta$ is related to the Wolfenstein
parameters $\rho$ and $\eta$ through
\be
\frac{1 - \rho + i \eta}{\sqrt{(1 - \rho)^2 + \eta^2}}
\approx
e^{i \beta}.
\ee
Therefore, Eq.~(\ref{current_bound_sin2beta})
corresponds to the area between two lines
passing through $(\rho, \eta) = (1,0)$.
Overlaying this constraint in FIG.~\ref{figura rho-eta},
the CKMfitter group obtained,
at the time of the conference LP2003,
the result in FIG.~\ref{figura rho-eta with 2b}
\cite{CKMfitter-site}.
The various blue areas shown correspond to a discrete ambiguity arising
when one extracts $\beta$ from $\sin{2 \beta}$.

Notice the perfect agreement of this measurement of
$\sin{2 \beta}$,
in blue,
with the results of FIG.~\ref{figura rho-eta} known previously.
This is a major success for the SM,
and it might well mean that the leading CP violating effects
are dominated by the CKM mechanism.
If so,
we will need to combine a number of different experiments
(preferably, with small theoretical uncertainties) in
order to uncover new physics effects.

\subsection{The penguin decay $B_d \rightarrow \phi K_S$
and related channels}

The angle $\beta$ can be probed in a variety of decay channels.
Performing those experiments allows us to
``test the relation $\beta=\beta$''.\footnote{This terminology
is used here to parallel the usual claim that measuring
$\alpha$ allows one to ``test the relation 
$\alpha + \beta + \gamma = \pi$;
a relation which,
as stressed in the last paragraph of
section~\ref{sec:unit_triangle},
holds by definition.}
Several candidates include:
\begin{itemize}
\item $b \rightarrow s \bar s s$ decays,
such as $B_d \rightarrow \phi K_S$, $B_d \rightarrow \eta^\prime K_S$,
and $B_d \rightarrow K^+ K^- K_S$;
\item $b \rightarrow c \bar c d$ decays, such as
$B_d \rightarrow \psi \pi^0$,
and $B_d \rightarrow D^{(\ast)\, +}  D^{(\ast)\, -}$.
Since the tree level amplitudes for these decays are suppressed
by $\lambda$ with respect to the  $b \rightarrow c \bar c s$,
these will be more sensitive to new physics in
$b \rightarrow d$ penguins than
$b \rightarrow c \bar c s$ are to new physics in
$b \rightarrow s$ penguins;
\item $B_d \rightarrow A K_S$, where $A = \chi_1, \eta_c, \dots$ is
some axial vector $c \bar c$ state. Comparing this with
$B_d \rightarrow J/\psi K_S$ tests models which break P and CP
\cite{AH03};
\item $B_d \rightarrow J/\psi K_L$. Comparing this with
$B_d \rightarrow J/\psi K_S$,
instead of including it in an overall $b \rightarrow c \bar c s$
analysis of $\tilde{\beta}$,
allows for tests of CPT or of exotic
$B_d^0 \rightarrow \overline{K^0}$ decays 
\cite{Lav00,Gro02}.
\end{itemize}

\begin{figure}[h]
\centerline{\includegraphics*[height=1.5in]{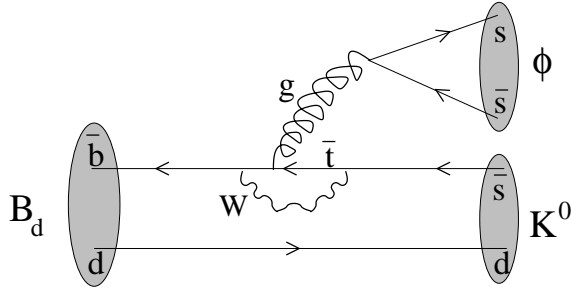}}
\caption{\label{figura pengPK}Penguin diagram with a
virtual top quark which mediates the decay
$B_d \rightarrow \phi K_S$.
The gluonic line represents two or more gluons.
}
\end{figure}

The decay $B_d \rightarrow \phi K_S$ is mediated by the penguin
diagram in FIG.~\ref{figura pengPK},
which should be compared with that involved in the
decay $B_d \rightarrow J/\psi K_S$,
shown in FIG.~\ref{figura pengJK}.
Clearly,
they share the CKM structure,
shown on the second line of Eq.~(\ref{CKM_coef_JK}).
As before,
there is also a contribution from the penguin
diagram with an intermediate charm quark,
which involves the CKM structure shown on the first line of
Eq.~(\ref{CKM_coef_JK});
and, using the CKM unitarity relation in Eq.~(\ref{triangle_sb}),
any other contribution may be written as a linear
combination of these two.
In contrast to the situation in the decay $B_d \rightarrow J/\psi K_S$,
here there is no tree level contribution;
this is a penguin decay.
However,
the relative phase between the two contributions is
the same as in $B_d \rightarrow J/\psi K_S$;
in the SM it is $\chi \sim \lambda^2$.
Thus,
four qualitative predictions are possible:
\begin{enumerate}
\item we also expect this decay to measure $\tilde{\beta}$, \textit{i.e.\/},
\be
\tilde{\beta} (\mbox{in}\ b \rightarrow s\ \mbox{penguin})
=
\tilde{\beta} (\mbox{in}\ b \rightarrow c \bar c s);
\ee
\item but, these $b \rightarrow s$ penguin decays are likely
to be more affected by new physics than the tree level
$b \rightarrow c \bar c s$ decays;
\item these new effects may both alter the
interference CP violation, $S_f \neq \pm \sin{2 \tilde{\beta}}$, 
and introduce CP violation in the decay,
$C_f \neq 0$;
\item and, due to the different hadronic matrix elements involved,
such new physics may have a different impact in different decays,
such as $B_d \rightarrow \phi K_S$ and
$B_d \rightarrow \eta^\prime K_S$.
\end{enumerate}

The results at the time of the conference ICHEP2004 were \cite{HFAG}
\ba
C_{\phi K} &=& 
\left\{
\begin{array}{ll}
0.00 \pm 0.23 \pm 0.05 & \mbox{BABAR}\\
-0.08 \pm 0.22 \pm 0.09 & \mbox{Belle}\\
-0.04 \pm 0.17 & \mbox{HFAG average}
\end{array}\ ,
\right.
\label{ICHEP04_bound_CPK}
\\
S_{\phi K} &=&
\left\{
\begin{array}{ll}
+0.50 \pm 0.25^{+0.07}_{-0.04} & \mbox{BABAR}\\
+0.06 \pm 0.33 \pm 0.09 & \mbox{Belle}\\
+0.34 \pm 0.20 & \mbox{HFAG average}
\end{array}\ .
\right.
\label{ICHEP04_bound_SPK}
\ea
Recall that a sizable difference 
between $\sin{2 \tilde{\beta}}$ extracted
from $B_d \rightarrow J/\psi K_S$ and from
$B_d \rightarrow \phi K_S$ is a problem for the SM.
In 2003 $S_{\phi K}=-0.14 \pm 0.33$ posed a 
serious problem
(which now seems to be much reduced in this channel).
This spurred a renewed interest in new physics
contributions to this decay \cite{various_PK}.
In general,
attempts to reconcile $\sin{2 \tilde{\beta}} \sim 0.73$ with
a much smaller value for $S_{\phi K_S} $ involve new physics 
in $b \rightarrow s$ penguins;
with amplitudes comparable to the SM;
and with a large relative CP violating phase.

\subsubsection{Enhanced electroweak penguins and small $S_{\phi K}$}

One possibility consists in introducing
non SM $s Z b$ couplings through
\be
{\cal L}_Z^{\rm new}
=
\frac{g^2}{4 \pi} \frac{g}{2 \cos{\theta_W}}
\left[
Z_{sb}\  \bar b_L \gamma_\mu s_L +
Z_{sb}^\prime\  \bar b_R \gamma_\mu s_R
\right]\,
Z^\mu + h.c.\ ,
\label{new_sZB}
\ee
where a loop-type suppression factor has been introduced
in the definition of the coefficients
$Z_{sb}$ and $Z_{sb}^\prime$.
These couplings are already constrained by a number of
observables, including the exclusive decay
$B_d \rightarrow X_s e^+ e^-$ \cite{sZb-review},
implying that the new $Z$ penguins are at most
two to three times larger than the SM contributions
to the decay $b \rightarrow s$.

\begin{figure}[htb]
\centerline{\includegraphics*[height=1.5in]{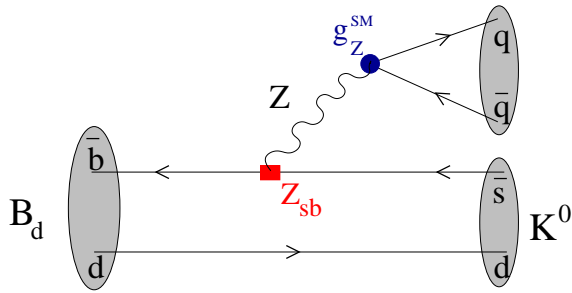}}
\caption{\label{figura Z_pengPK}New diagram generated by the
interactions in Eq.~(\ref{new_sZB}).
For $q \bar q = s \bar s$ this contributes to the
decay $B_d \rightarrow \phi K_S$;
for $q \bar q = c \bar c$,
this contributes to the decays $B_d \rightarrow (c \bar c) K_S$.
Notice the presence of the new $s Z b$ coupling
and of the SM coupling $g_Z^{\rm SM}$ in the $Z q q$ vertex.
}
\end{figure}

Atwood and Hiller have used this possibility to illustrate an
interesting point \cite{AH03}.
Consider the new physics diagram in FIG.~\ref{figura Z_pengPK},
which competes with the SM one in FIG.~\ref{figura pengJK}.
When $q \bar q = s \bar s$ this contributes to the
decay $B_d \rightarrow \phi K_S$,
which might explain the discrepancy between
the naive average of Eq.~(\ref{ICHEP04_bound_SPK}) and 
Eq.~(\ref{current_bound_sin2beta}).
But,
when $q \bar q = c \bar c$,
this contributes to the decays $B_d \rightarrow (c \bar c) K_S$,
thus altering the extraction of $\sin{2 \tilde{\beta}}$.
Now comes the important argument:
because the SM $Z q \bar q$ coupling is involved,
and because this coupling treats left and right handed quarks 
differently -- \textit{c.f.\/} Eq.~(\ref{LZ_weak}) --
it is crucial whether the $c \bar c$ quarks combine into
a vector (V) or an axial-vector (A) meson.
Indeed,
\be
g_Z^{{\rm SM}, V} (\psi, \psi^\prime, \dots)= +0.19\ ,
\hspace{2cm}
g_Z^{{\rm SM}, A} (\eta_c, \chi_1, \dots) = -0.5\ .
\ee
But this implies that comparing the value for
$\sin{2 \tilde{\beta}}$ extracted from 
$B_d \rightarrow (c \bar c)\, K_S$ decays in which
the $c \bar c$ quarks combine into a vector meson,
with those obtained when the $c \bar c$ combine into an
axial vector meson,
will allow us to probe the type and parameter space of
the new physics models proposed.
A recent analysis \cite{CKMfitter-04} finds that
the electroweak penguin explanation of an eventual
discrepancy in this channel may be disfavored with
respect to a modification of the gluonic penguins.
Nevertheless,
this important lesson remains:
given two measurements of some quantity
(say $\tilde{\beta}$) we should always ask what
features of new physics models would be probed by a discrepancy
in those measurements,
and where else such features would show up.
If the measurements show a clear sign of new physics,
we throw a party;
otherwise,
we have a constraint on that class of models.

\subsubsection{Tantalizing signals from $b \rightarrow s$ penguin decays}

\begin{figure}[h]
\centerline{\includegraphics*[height=2.8in]{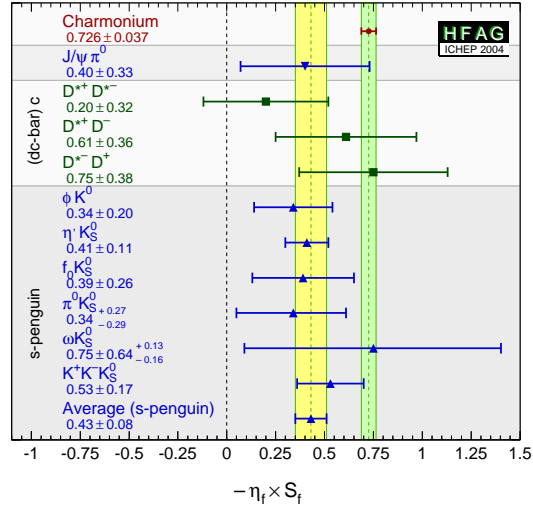}}
\caption{\label{HFAG S_btos}Experimental results
for the interference CP violation parameter $S$
extracted from $b \rightarrow s$ penguin decays,
compared with the results extracted with decays into charmonium states.
}
\end{figure}

Things certainly heat up when we compare the 
results in Eq.~(\ref{current_bound_sin2beta}),
obtained from $b \rightarrow c \bar c s$ transitions,
with the results obtained by comparing all
$b \rightarrow s$ penguin decays.
This is shown in HFAG's FIG.~\ref{HFAG S_btos} \cite{HFAG}.
\begin{figure}[h!]
\centerline{\includegraphics*[height=2.8in]{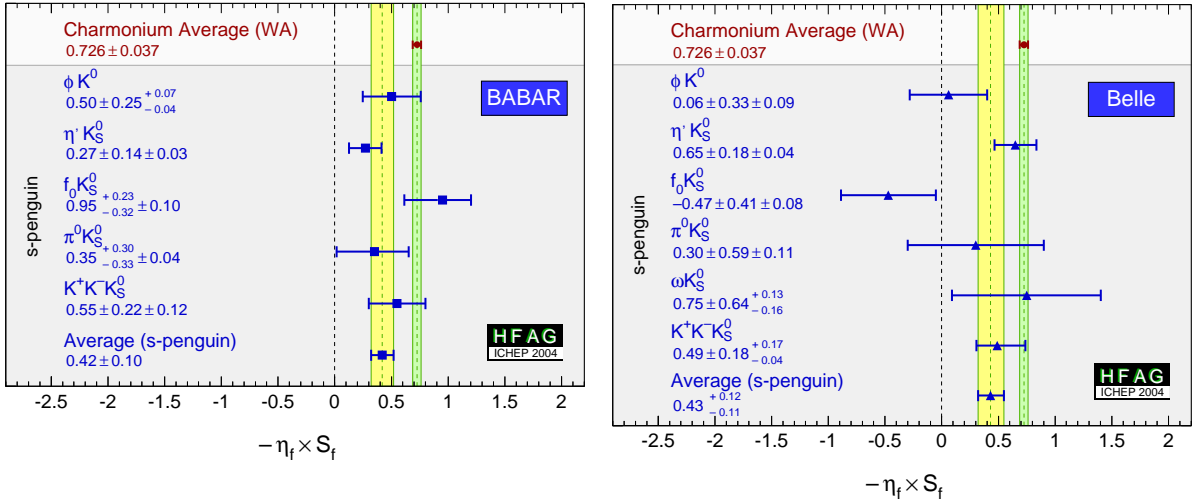}}
\caption{\label{HFAG S_btos Individual}Experimental results
for the interference CP violation parameter $S$
extracted from $b \rightarrow s$ penguin decays,
compared with the results extracted with decays into charmonium states,
as obtained by each experiment:
BABAR Collaboration (left); Belle Collaboration (right).
}
\end{figure}

The world averages after ICHEP2004 were \cite{HFAG} 
\ba
C_{b \rightarrow s} = 0.02 \pm 0.05\ ,
\label{current_bound_C_from_btos}
\\
S_{b \rightarrow s} = 0.43 \pm 0.08\ . 
\label{current_bound_sin2beta_from_btos}
\ea
Although no signal is seen for direct CP violation,
Eq.~(\ref{current_bound_sin2beta_from_btos}) differs
from Eq.~(\ref{current_bound_sin2beta})
by $3.6\sigma$. This is when we start paying attention.

This is all the more striking since this effect is clearly seen
by both BABAR and Belle independently,
as shown in HFAG's FIG.~\ref{HFAG S_btos Individual}.
It is clear that $b \rightarrow s$ decays will be under close scrutiny
during the next few years,
both theoretically and experimentally.

\subsection{The decay $B_d \rightarrow \pi^+ \pi^-$ and related channels}

\subsubsection{Penguin pollution}

\begin{figure}[htb]
\centerline{\includegraphics*[height=1.5in]{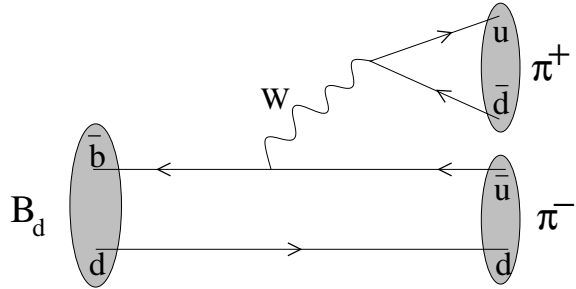}}
\caption{\label{figura treepp}Tree level diagram for
$B_d \rightarrow \pi^+ \pi^-$.
}
\end{figure}

\begin{figure}[htb]
\centerline{\includegraphics*[height=1.5in]{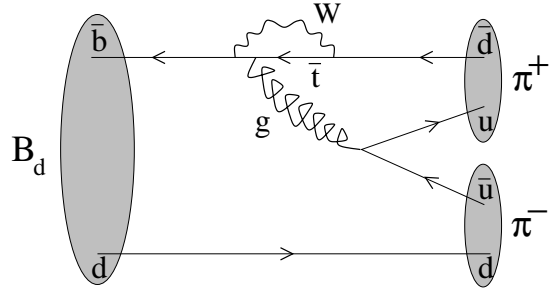}}
\caption{\label{figura pengpp}Penguin diagram with a
virtual top quark which mediates the decay
$B_d \rightarrow \pi^+ \pi^-$.
The gluonic line stands for any number of gluons.
}
\end{figure}

The tree level and penguin diagrams affecting the
decay $B_d \rightarrow \pi^+ \pi^-$
are represented in FIGs.~\ref{figura treepp} and
\ref{figura pengpp}.
These diagrams are proportional to
\ba
V_{ub}^\ast V_{ud} & \approx &
\left( A R_b \lambda^3 e^{i \gamma} \right)
(1),
\nonumber\\
V_{tb}^\ast V_{td} & \approx &
(1)
\left( A R_t \lambda^3 e^{- i \beta} \right),
\label{CKM_structure_pipi}
\ea
respectively.

To understand the interest behind this decay,
let us start by considering only the tree level diagram,
neglecting the penguin diagram.
If that were reasonable,
then
\be
\lambda_{B_d \rightarrow \pi^+ \pi^-} =
\frac{q_B}{p_B}
\frac{\langle \pi^+ \pi^- | T | \overline{B^0_d} \rangle}{
\langle \pi^+ \pi^- | T | B^0_d \rangle}
=
e^{- 2 i (\tilde{\beta} + \gamma)}.
\label{would_be_pipi}
\ee
Thus,
if there were only tree level diagrams,
the CP violating asymmetry in this decay would measure
$\tilde{\beta} + \gamma$.
Within the SM,
this coincides with $\beta+\gamma$ and provides another constraint
to be placed on the $\rho - \eta$ plane,
thus improving our search for new physics.
Since $\alpha = \pi - \beta - \gamma$ by definition,
this is sometimes referred to as a measurement of $\alpha$.
To highlight the fact that there are only two large phases in the CKM matrix,
it may be preferable to view this as a measurement
of $\tilde{\beta} + \gamma$.
In models were the new physics is only in 
$B - \overline{B}$ mixing,
$\tilde{\beta}$ is known from $B_d \rightarrow J/\psi K_S$,
and $B_d \rightarrow \pi^+ \pi^-$ provides a measurement
of $\gamma$.

When both diagrams are taken into account
\be
\lambda_{B_d \rightarrow \pi^+ \pi^-}
=
e^{- 2 i \tilde{\beta}}\
\frac{e^{-i \gamma} \langle t \rangle + e^{i \beta} \langle p \rangle}{
e^{i \gamma} \langle t \rangle + e^{-i \beta} \langle p \rangle}
=
e^{- 2 i (\tilde{\beta} + \gamma)}\
\frac{1 + r e^{i \delta} e^{i (\beta + \gamma)}}{
1 + r e^{i \delta} e^{- i (\beta + \gamma)}},
\ee
where $\langle t \rangle$ ($\langle p \rangle$)
contains the matrix element of the operator that appears
multiplied by the CKM coefficient 
$V_{ub}^\ast V_{ud}$ ($V_{tb}^\ast V_{td}$),
and the magnitude of that coefficient.
Clearly,
the terms proportional to 
\be
r e^{i \delta} = \frac{\langle p \rangle}{\langle t \rangle},
\ee
where $r$ is a positive real number and $\delta$ a relative
strong phase:
\begin{itemize}
\item destroy the simple relation in Eq.~(\ref{would_be_pipi});
\item imply that $\lambda_{B_d \rightarrow \pi^+ \pi^-}$ is not
a pure phase;
\item and, because $r e^{i \delta}$ depends crucially on the details 
of the hadronic matrix elements,
such terms introduce a theoretical uncertainty into the interpretation
of this experiment.
\end{itemize}

Indeed,
for this decay,
the difference between the two phases,
$\phi_1 - \phi_2 = \gamma + \beta$,
is large.
Therefore,
whatever convictions one might have about $r$,
are not enough to guarantee that
$C_{\pi^+ \pi^-}$  vanishes or that
$S_{\pi^+ \pi^-}$ measures $- \sin{(2 \tilde{\beta} + 2 \gamma)}$.
The situation is made worse by the fact that
$R_b \sim 0.4$ -- Buras and Fleischer
named this the $R_b$ suppression \cite{BF98},
-- thus enhancing $r$.
Gronau showed that this problem affects the measurement
of $\tilde{\beta} + \gamma$,
even for moderate values of $r$ \cite{Gronau93}.
This is known as ``penguin pollution''.

\subsubsection{\label{subsec:trapping}Methods for trapping the penguin}

The extraction of $\tilde{\beta} + \gamma$ from the CP asymmetry
in $B_d \rightarrow \pi^+ \pi^-$ would be straightforward
if we were able to know $r e^{i \delta}$;
sometimes known as ``trapping the penguin''.
This may be achieved via two different paths:
one may relate this decay to other decays invoking some
symmetry property;
or, one may try to calculate $r e^{i \delta}$ directly
within a given theoretical treatment of the hadronic
interactions.

Some possibilities are listed below,
where $\mbox{BR}_{\rm av}$ stands for the branching ratio
averaged over a particle and its antiparticle.
\begin{itemize}
\item Gronau and London advocated the use of isospin \cite{GL} to
relate the decay $B_d \rightarrow \pi^+ \pi^-$ with the
decays $B_d \rightarrow \pi^0 \pi^0$ and
$B^+ \rightarrow \pi^+ \pi^0$ through 
\be
\frac{1}{\sqrt{2}}
\langle \pi^+ \pi^- | T | B^0_d \rangle
+
\langle \pi^0 \pi^0 | T | B^0_d \rangle
=
\langle \pi^+ \pi^0 | T | B^+ \rangle.
\label{GL_isospin_triangle}
\ee
This method requires the measurement of
\be
\mbox{BR}_{\rm av}(\pi^+ \pi^-),\ \
C_{\pi^+ \pi^-},\ \ 
S_{\pi^+ \pi^-},\ \ 
\mbox{BR}_{\rm av}(\pi^0 \pi^0),\ \
C_{\pi^0 \pi^0},\ \ 
\mbox{BR}_{\rm av}(\pi^+ \pi^0).
\ee
Of these, all have been available for some time,
except for $C_{\pi^0 \pi^0}$,
which was only recently measured by BABAR to be
$C_{\pi^0 \pi^0} = -0.12 \pm 0.56 \pm 0.06$ \cite{BABAR-Cpi0pi0},
and by Belle to be
$C_{\pi^0 \pi^0} = -0.43 \pm 0.51 \pm 0.17$ \cite{Belle-Cpi0pi0}.
This method determines $\tilde{\beta}+\gamma$ with a 16-fold ambiguity.
If one were able to measure also $S_{\pi^0 \pi^0}$,
one would determine $\tilde{\beta}+\gamma$ with a 4-fold ambiguity.
\item Grossman and Quinn pointed out that one may get
some information even with a partial realization of
the isospin analysis \cite{GQ,Charles,GLSS}.
In its simplest form \cite{GQ},
this requires the measurement of
\be
C_{\pi^+ \pi^-},\ \ 
S_{\pi^+ \pi^-},\ \ 
\mbox{BR}_{\rm av}(\pi^+ \pi^0),\ \
\mbox{upper bound on }
\mbox{BR}_{\rm av}(\pi^0 \pi^0).
\ee
%
\item Silva and Wolfenstein proposed the use of flavor SU(3)
-- in fact, $U$-spin -- to relate the decay $B^0_d \rightarrow \pi^+ \pi^-$ 
with $B_d \rightarrow K^+ \pi^-$ \cite{SW1}.
Indeed,
the diagrams mediating the decay
$B_d \rightarrow K^+ \pi^-$
are obtained from those in FIGs.~\ref{figura treepp} and
\ref{figura pengpp},
which mediate the decay $B^0_d \rightarrow \pi^+ \pi^-$,
with the simple substitution
of $\bar d \rightarrow \bar s$,
leading to $\pi^+ \rightarrow K^+$.
The new diagrams are proportional to
\ba
V_{ub}^\ast V_{us} & \approx &
\left( A R_b \lambda^3 e^{i \gamma} \right)
(\lambda),
\nonumber\\
V_{tb}^\ast V_{ts} & \approx &
(1)
\left( - A \lambda^2 e^{- i \chi} \right),
\label{CKM_structure_Kpi}
\ea
respectively.
The crucial point behind this idea may be understood
by comparing Eq.~(\ref{CKM_structure_Kpi}) with
Eq.~(\ref{CKM_structure_pipi}):
the ratio $r$ in $B_d \rightarrow K^+ \pi^-$
is enhanced by $1/\lambda^2$ with respect to the
ratio $r$ in $B_d \rightarrow \pi^+ \pi^-$.
In fact,
the $B_d \rightarrow K \pi$ decays 
are predicted to be penguin dominated,
which makes them a very good source of information
on the penguin needed to extract $\tilde{\beta} + \gamma$
from $B_d \rightarrow \pi^+ \pi^-$ \cite{SW1}.

In the simplest approximation \cite{SW1},
one needs only the measurements of
\be
\mbox{BR}_{\rm av}(\pi^+ \pi^-),\ \
S_{\pi^+ \pi^-},\ \ 
\mbox{BR}_{\rm av}(K^+ \pi^-).
\ee 
Since then,
a variety of other methods using $U$-spin to
extract information from CP asymmetries
have been proposed \cite{SU3-group}.
%
\item One may also use some theoretical method in order
to \textit{calculate} the hadronic matrix elements required
\cite{QCDF1,QCDF2,QCDF-BN,pQCD1,pQCD2}.
\end{itemize}
The first two methods,
based on isospin,
will be discussed in detail in subsection~\ref{subsec:isospin}.

\subsubsection{The $C - S$ plane}

\begin{figure}[h]
\centerline{\includegraphics*[height=2.7in]
{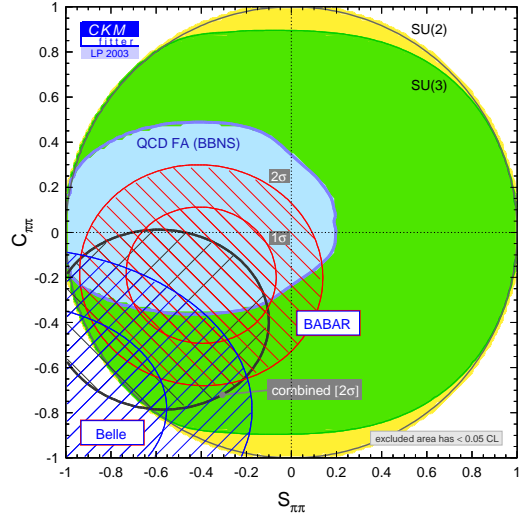}}
\caption{\label{Cpipi_Spipi_plane_04}Experimental results for
$B_d \rightarrow \pi^+ \pi^-$ at the time of LP2003,
compared with some theoretical analysis.
}
\end{figure}

The PDG2004 world averages for the CP violating observables
in the decay $B_d \rightarrow \pi^+ \pi^-$ were \cite{PDG}
\ba
C_{\pi^+ \pi^-} &=& -0.51 \pm 0.23\ ,
\nonumber\\
S_{\pi^+ \pi^-} &=& -0.5 \pm 0.6\ ,
\ea
which, after ICHEP2004, became \cite{HFAG}
\ba
C_{\pi^+ \pi^-} &=& -0.37 \pm 0.11\ (0.24)\ ,
\nonumber\\
S_{\pi^+ \pi^-} &=& -0.61 \pm 0.14\ (0.34)\ .
\ea
The inflated errors in between parenthesis result from
the fact that the measurements of BABAR and Belle are
inconsistent with each other.
It is clear that a naive average is not the best procedure to
combine inconsistent measurements,
but how exactly this should be implement is the subject of some debate 
\cite{average_incompatible}.
FIG.~\ref{Cpipi_Spipi_plane_04}
\begin{figure}[h!]
\centerline{\includegraphics*[height=2.7in]{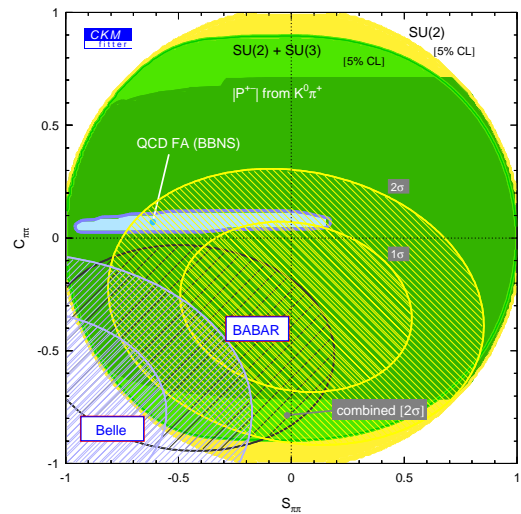}}
\caption{\label{Cpipi_Spipi_plane_02}Experimental results for
$B_d \rightarrow \pi^+ \pi^-$ from data available
in early 2003,
compared with some theoretical analysis.
Taken from \cite{CKMfitter-Durham}.}
\end{figure}
shows an analysis
made by the CKMfitter group \cite{CKMfitter-site}
at the time of LP2003,
where a later (rumored) result attributed to BABAR \cite{BABAR-later-pipi}
had also been taken into account.

It is interesting to compare this with a similar 
FIG.~\ref{Cpipi_Spipi_plane_02},
drawn by some members of the CKMfitter group 
at an earlier date \cite{CKMfitter-Durham}.
Clearly,
the experimental constraints have moved about within their
error bars,
as experimental results do.
But the most remarkable feature is the enormous enlargement
of the blue range obtained theoretically 
within the QCD factorization approach \cite{QCDF1,QCDF2,QCDF-BN}.
This is due to the fact that \cite{QCDF2,QCDF-BN} improved on
the earlier analysis by including hard scattering spectator
interactions $X_H$ and annihilation diagrams $X_A$,
which cannot be estimated in a model independent fashion.
As a result,
one obtains larger CP even strong phases,
thus enlarging considerably the $C_{\pi^+ \pi^-}$
allowed range.
Hadronic ``messy'' effects are really a nuisance.

\subsubsection{\label{subsec:Bpi0pi0}$B_d \rightarrow \pi^0 \pi^0$
and predictions from global fits}

The methods mentioned in subsection~\ref{subsec:trapping} can be used
in order to perform a global fit to all available data
on two-body charmless $B$ decays.
Because,
typically,
SU(2) introduces more parameters than there are data points,
we are left with two classes of analysis:
\begin{enumerate}
\item Analysis utilizing a diagrammatic decomposition 
\cite{SU3-group} based on SU(3) \cite{early-SU3},
in order to
\textit{parametrize}
unknown matrix elements for different channels.
Recent analysis may be found in
\cite{U3} and \cite{SU3-1,SU3-04}.
\item QCD based
\textit{calculations}
of the hadronic matrix elements,
in the context of perturbative QCD (pQCD) \cite{pQCD1,pQCD2},
QCD factorization (QCDF) \cite{QCDF1,QCDF2,QCDF-BN},
and soft collinear effective theory (SCET) \cite{SCET}.
\end{enumerate}

Before the experimental measurement of $\mbox{BR}_{\rm av}(\pi^0 \pi^0)$
was announced,
global fits were performed in the context of
U(3) \cite{U3}, SU(3) \cite{SU3-1}, pQCD \cite{pQCD2},
and QCDF \cite{QCDF-BN} with 
projections for this observable.
I will name these ``\textit{predictions\/}'',
because they appeared \textit{before\/} the
experimental results.
The results are compiled in Table~\ref{table:Bpi0pi0},
together with the values for this branching ratio
listed in PDG2004 \cite{PDG} and improved at ICHEP2004.
%
\begin{table}[hbt]
\begin{centering}
\begin{tabular}{|c|c|c|c|c|}
\hline
pQCD                                                     &
QCDF                                                     &
SU(3)                                                    &
U(3)                                                     &
$\mbox{BR}_{\rm av}^{\rm exp}(\pi^0 \pi^0)$              \\*[2mm]
\hline
$0.33 - 0.65$                                            &
$0.3
\begin{array}{llll}
+0.2 & + 0.2 & +0.3 & +0.2\\
-0.2 & - 0.1 & -0.1 & -0.1
\end{array}$                                             &
$0.4 - 1.6$                                              &
$1.2 - 2.7$                                              &
$\begin{array}{ll}
1.9 \pm 0.5 & \mbox{PDG2004}\\
1.5 \pm 0.3 & \mbox{ICHEP04}
\end{array}$                                              \\*[2mm]
\hline
\end{tabular}
\caption{\label{table:Bpi0pi0}Theoretical predictions
for $\mbox{BR}_{\rm av}(\pi^0 \pi^0)$,
compared with the experimental measurement,
in units of $10^{-6}$.}
\end{centering}
\end{table}
%
The observation of this value
for $\mbox{BR}_{\rm av}(\pi^0 \pi^0)$ has a few important
consequences:
\begin{enumerate}
\item it enables the partial isospin, Grossman-Quinn bound;
\item it implies that $C_{\pi^0 \pi^0}$ is within reach,
thus enabling the full isospin analysis;
\item it poses a challenge to the QCD based predictions. 
\end{enumerate}
This is a rather difficult experiment,
because there are no charged tracks.
We must thank our experimentalist colleagues
for their great efforts.

\begin{figure}[htb]
\centerline{\includegraphics*[height=1.5in]{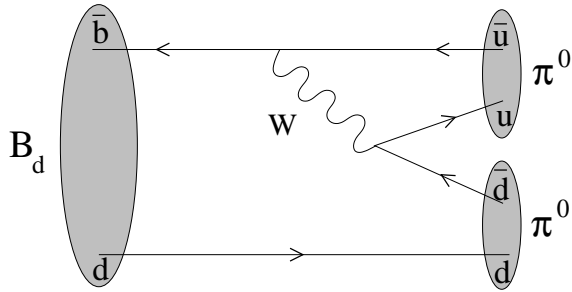}}
\caption{\label{figura colsup-pp}Color suppressed,
tree level diagram for $B_d \rightarrow \pi^0 \pi^0$.
}
\end{figure}

This decay is mediated by the tree level diagram in
FIG.~\ref{figura colsup-pp},
which should be compared with the tree level diagram
in FIG.~\ref{figura treepp}.
There,
the $\bar d u$ quarks coming out of the (color singlet)
$W$ go into the $\pi^+$ quark.
Here,
the $\bar d u$ quarks coming out of the (color singlet)
$W$ go into two distinct mesons:
the quark $\bar d$ must combine with the spectator $d$
to form a (color singlet) meson,
although their colors are initially independent;
and the quark $u$ must combine with the quark $\bar u$
from the other vertex to form a (color singlet) meson,
although their colors are initially independent.
Barring other effects,
this entails a color suppression of the diagram
in FIG.~\ref{figura colsup-pp} 
(named ``color suppressed'')
with respect to that in FIG.~\ref{figura treepp}
(named merely ``tree'').
Because it involves the color suppressed diagram in
FIG.~\ref{figura colsup-pp},
the branching ratio for $B_d \rightarrow \pi^0 \pi^0$ was suspected
to be smaller than it turned out to be.
Indeed,
recent reanalysis including
$\mbox{BR}_{\rm av}^{\rm exp}(\pi^0 \pi^0)$
into the fit,
seem to require a rather large contribution
from the color suppressed amplitude \cite{CKMfitter-04,SU3-04},
and a large strong phase relative to the tree (color allowed)
diagram.

One final note before we proceed.
Sometimes the connection between the two types of global fit mentioned,
SU(3)-based and QCD-based,
is a bit puzzling.
Recently,
in an extremely nice article,
Bauer and Pirjol discussed the relation
between the SU(3) diagrammatic decomposition and the
matrix elements of operators used in the 
soft-collinear effective theory \cite{BP04}.\footnote{I am
admittedly not an expert on this field. But I recommend
this article most vividly.}

\subsubsection{\label{subsec:isospin}The isospin analysis in
$B \rightarrow \pi \pi$ decays}

In this section we discuss the isospin based methods utilized
in the analysis of $B \rightarrow \pi \pi$ decays in more detail,
separating it into small steps.

\textbf{STEP 1:} We start by recalling the definition of $C_f$ and
defining a new quantity, $B^f$,
proportional to the average decay width,
as
\ba
C_f &=&
\frac{|\bar A_f|^2 - |A_f|^2}{|\bar A_f|^2 - |A_f|^2},
\label{C_f_again}\\
B^f &=&
\frac{|\bar A_f|^2 + |A_f|^2}{2}.
\label{B^f}
\ea
The definitions for $C_f$ in Eqs.~(\ref{C_f}) and (\ref{C_f_again})
coincide,
because we are using $|q/p|=1$.
Hence,
the decay amplitudes are determined from the experimentally
quoted values for the average branching ratios and
decay CP violating parameters as\footnote{In going from
the branching ratios to the amplitudes in 
Eq.~(\ref{B^f}) we must take into account the fact that
the lifetimes of $B^+$ and $B_d$ are different.}
\ba
|\bar A_f|^2 &=& B^f (1 + C_f),
\nonumber\\
|A_f|^2 &=& B^f (1 - C_f).
\label{eq:GL-AfromBC}
\ea

\textbf{STEP 2:} Now,
we parametrize the phase of $\lambda_{\pi^+ \pi^-}$ 
by the difference $2 \delta_\alpha$ from the value $2 \alpha$ 
that it would have if there were only tree diagrams:
\be
\lambda_{+-} \equiv 
\lambda_{\pi^+ \pi^-} = |\lambda_{+-}|
e^{2 i (\alpha + \delta_\alpha)}.
\ee
Since Eq.~(\ref{C_f}) leads to
\be
\sqrt{1 - C_{+-}^2} = \frac{2 |\lambda_{+-}|}{1 + |\lambda_{+-}|^2},
\ee
we find
\be
S_{+-}
=
\frac{2 \mbox{Im}( \lambda_{+-} )}{1 + |\lambda_{+-}|^2}
=
\sqrt{1 - C_{+-}^2} \sin{(2 \alpha + 2 \delta_\alpha)}.
\label{eq:GL-GQtrick}
\ee
Surprising it may be,
this is the crucial trick in the analysis by Grossman and Quinn
\cite{GQ}.

\textbf{STEP 3:} In the limit of exact isospin symmetry,
the two pions coming out of $B$ decays must be
in an isospin $I=0$ or $I=2$ combination.
Because gluons are isosinglet,
they can only contribute to the $I=0$ final state.
Therefore,
the amplitude leading into the $I=2$ final state arises
exclusively from tree level diagrams and, thus,
it carries only one weak phase: $\gamma$.
This is the crucial observation behind the Gronau--London
method \cite{GL}.
These issues are discussed in detail in \textbf{(Ex-30)},
from which we take the isospin decomposition
\ba
\frac{1}{\sqrt{2}}
A^{+-}
&\equiv&
\frac{1}{\sqrt{2}}
\langle \pi^+ \pi^- | T | B^0_d \rangle
=
T_2 - A_0,
\nonumber\\
A^{00}
&\equiv&
\langle \pi^0 \pi^0 | T | B^0_d \rangle
=
2\, T_2 + A_0,
\nonumber\\
A^{+0}
&\equiv&
\langle \pi^+ \pi^0 | T | B^+ \rangle
=
3\, T_2,
\label{isospin_decomposition_B}
\ea
where we have used
\ba
A_0 &=& \frac{1}{\sqrt{6}} A_{1/2},
\nonumber\\
T_2 &=& \frac{1}{2\sqrt{3}} A_{3/2},
\ea
and the notation $T_2$ reminds us that this amplitude
carries only the weak phase of tree level diagrams.
Of course,
there is a similar decomposition for the CP conjugated amplitudes:
\ba
\frac{1}{\sqrt{2}}
\bar A^{+-}
&\equiv&
\frac{1}{\sqrt{2}}
\langle \pi^+ \pi^- | T | \overline{B^0_d} \rangle
=
\bar T_2 - \bar A_0,
\nonumber\\
\bar A^{00}
&\equiv&
\langle \pi^0 \pi^0 | T | \overline{B^0_d} \rangle
=
2\, \bar T_2 + \bar A_0,
\nonumber\\
\bar A^{+0}
= A^{-0}
&\equiv&
\langle \pi^- \pi^0 | T | B^- \rangle
=
3\, \bar T_2.
\label{isospin_decomposition_Bbar}
\ea
The first two amplitudes in Eq.~(\ref{isospin_decomposition_B})
add up to the third one,
the same happening with Eq.~(\ref{isospin_decomposition_Bbar}).
This can be visualized as two triangles in the complex plane.

\begin{figure}[htb]
\centerline{\includegraphics*[height=2.45in]{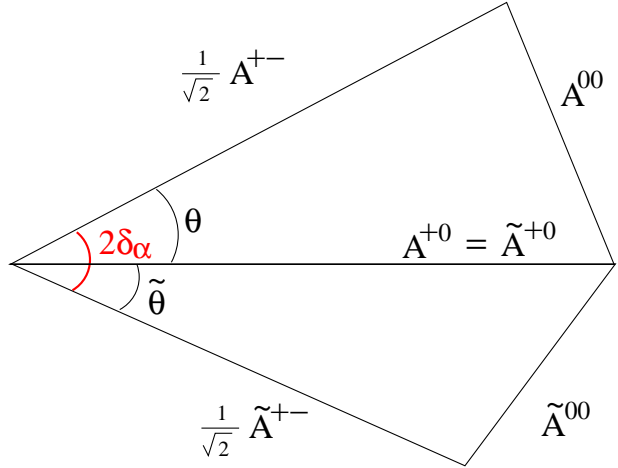}}
\caption{\label{fig:isospin-triangles}Isospin triangles utilized
in the Gronau--London method.
}
\end{figure}

We will now follow the presentation of the Gronau--London
method contained in \cite{GLSS}.
Because $T_2$ only carries the weak phase $\gamma$,
we may write it as
\be
T_2 = |T_2| e^{i \vartheta} e^{i \gamma}, 
\ee
where $\vartheta$ is a strong phase.
As a result
\be
\left( e^{2 i \gamma} \right)
\bar T_2
=
\left(e^{2 i \gamma} \right)
|T_2| e^{i \vartheta} e^{- i \gamma}
=
T_2.
\ee
This means that,
rotating all sides of the CP conjugated triangle by the
phase $2 \gamma$,
\ba
\tilde{A}^{+-} &=&
\left( e^{2 i \gamma} \right)  \bar A^{+-},
\nonumber\\
\tilde{A}^{00} &=&
\left(e^{2 i \gamma} \right) \bar A^{00},
\nonumber\\
\tilde{A}^{+0} &=&
\left(e^{2 i \gamma} \right) \bar A^{+0}
= A^{+0},
\ea
makes the sides $\tilde{A}^{+0} = A^{+0}$ coincide.
This is shown in FIG.~\ref{fig:isospin-triangles},
where we define the angle $\theta$ ($\tilde{\theta}$) between
$A^{+-}$ ($\tilde{A^{+-}}$) and $A^{+0}$.
The isospin prediction that $|A^{+0}| = |\bar A^{+0}|$ can
be probed by looking for CP violation in the decay
$B^+ \rightarrow \pi^+ \pi^0$.\footnote{This measurement 
constrains possible contributions from
electroweak penguins which,
although they break isospin,
are expected to be very small in these channels.}

Clearly,
\ba
\cos{\theta} &=&
\frac{|A^{+0}|^2 + \frac{1}{2} |A^{+-}|^2 - |A^{00}|^2}{
\sqrt{2} |A^{+0}| |A^{+-}|},
\nonumber\\
\cos{\tilde{\theta}} &=&
\frac{|\tilde{A}^{+0}|^2 +  \frac{1}{2} 
|\tilde{A}^{+-}|^2 - |\tilde{A}^{00}|^2}{
\sqrt{2} |\tilde{A}^{+0}| |\tilde{A}^{+-}|},
\label{eq:GL-costhetathetatilde}
\ea
from which we can extract $\sin{\theta}$ and $\sin{\tilde{\theta}}$,
up to their signs.

\textbf{STEP 4:} FIG.~\ref{fig:isospin-triangles} also shows
the phase $2 \delta_\alpha$ as being the
angle between $A^{+-}$ and $\tilde{A}^{+-}$.
Indeed,
\ba
\lambda_{+-} &=&
|\lambda_{+-}| e^{2i(\alpha + \delta_\alpha)}
=
\frac{q_B}{p_B} \frac{\bar A^{+-}}{A^{+-}}
\nonumber\\
&=&
e^{-2i \beta} \left( e^{-2i \gamma} e^{2i \gamma} \right)
\frac{\bar A^{+-}}{A^{+-}}
= 
e^{2i \alpha}
\frac{\tilde{A}^{+-}}{A^{+-}},
\ea
proving that assertion.

Since any of two triangles in FIG.~\ref{fig:isospin-triangles}
could be inverted,
$|\delta_\alpha|$ might equal $|\theta \pm \tilde{\theta}|$.
It will be important below to note that
the deviation of the phase of $\lambda_{+-}$ from
the weak CKM phase $2 \alpha$ is maximized
when the two triangles lie on opposite sides,
as in FIG.~\ref{fig:isospin-triangles}.
So, we consider that case,
for which $\delta_\alpha = \theta + \tilde{\theta}$,
and
\be
1 - \sin^2{\delta_\alpha} =
\cos{2 \delta_\alpha} =
\cos{\theta} \cos{\tilde{\theta}}
-
\sin{\theta} \sin{\tilde{\theta}}
=
\mbox{function} \left( B^{+0}, B^{+-}, C_{+-}, B^{00}, C_{00} \right).
\label{eq:GL-delta-alpha}
\ee

We are now ready to understand the Gronau-London method \cite{GL}.
Eqs.~(\ref{eq:GL-AfromBC}) and (\ref{eq:GL-costhetathetatilde})
imply that the measurements of
$B^{+0}$, $B^{+-}$, $C_{+-}$, $B^{00}$, and $C_{00}$
determine (up to discrete ambiguities)
$\theta$, $\tilde{\theta}$ and, thus,
$\delta_\alpha$, through Eq.~(\ref{eq:GL-delta-alpha}).
Combining this with the additional measurement of
the interference CP violation in the decay 
$B_d \rightarrow \pi^+ \pi^-$,
$S_{+-}$,
into Eq.~(\ref{eq:GL-GQtrick}),
\be
\sin(2 \alpha + 2 \delta_\alpha)
=
\frac{S_{+-}}{\sqrt{1 - C^2_{+-}}},
\ee
yields $\alpha$.
This is the Gronau-London method.


We may now ask what would happen if we were not able to
measure $C_{00}$.
In that case,
we would have to assume the worse case scenario 
(maximum value) for $\delta_\alpha$.
This means that we must take
$\delta_\alpha = \theta + \tilde{\theta}$,
as done above,
and minimize the function in Eq.~(\ref{eq:GL-delta-alpha})
with respect to $C_{00}$.
One obtains \cite{GLSS}
\be
{C_{00}}_{\mbox{\small minimize}}
=
\frac{C_{+-}}{2}
\frac{B^{+-} 
\left( \frac{1}{2} B^{+-} - B^{+0} - B^{00} \right)}{
B^{00} 
\left( \frac{1}{2} B^{+-} + B^{+0} - B^{00} \right)},
\ee
from which
\be
\cos{2 \delta_\alpha} \geq
\frac{\left( \frac{1}{2} B^{+-} + B^{+0} - B^{00} \right)^2
- B^{+-} B^{+0}}{
B^{+-} B^{+0} \sqrt{1 - C^2_{+-}}}.
\label{eq:GL-GLSS-v1}
\ee
This constraint on the deviation of $S_{+-} / \sqrt(1 - C^2_{+-})$
from $\sin{2 \alpha}$ constitutes the Gronau-London-Sinha-Sinha
bound,
and it is the best we can do to constrain $\alpha$
if $C_{00}$ is not known.

This bound may be rewritten as
\be
\cos{2 \delta_\alpha} \geq
\frac{1 - 2 B^{00}/B^{+0}}{\sqrt{1 - C^2_{+-}}}
+
\frac{\left( \frac{1}{2} B^{+-} - B^{+0} + B^{00} \right)^2}{
B^{+-} B^{+0} \sqrt{1 - C^2_{+-}}}.
\label{eq:GL-GLSS-v2}
\ee
Since the second term is positive,
we reach \cite{GQ,Charles}
\be
\cos{2 \delta_\alpha} \geq
\frac{1 - 2 B^{00}/B^{+0}}{\sqrt{1 - C^2_{+-}}},
\label{eq:GL-GQ-Charles}
\ee
which is an improvement due to Charles on the earlier
result appearing in reference \cite{GQ}
\be
\cos{2 \delta_\alpha} \geq
1 - 2 B^{00}/B^{+0}.
\label{eq:GL-GQ}
\ee
This is the famous Grossman-Quinn bound,
although their article also quotes another version of this bound,
more refined \cite{GQ}.

The Grossman-Quinn bound in Eq.~(\ref{eq:GL-GQ}) would be
extremely useful if it turned out that $B^{00}$ were
very small,
as originally expected due to the ``color suppression''
mentioned in connection with FIG.~\ref{figura colsup-pp}.
That some such bound was possible is very easy to see
by looking back at Eqs.~(\ref{isospin_decomposition_B}).
Indeed,
in the exact limit $|A^{00}|=0$, we have
$A_0 = - 2 T_2$,
from which we conclude that $A^{+-} = 3 T_2$ only
carries the weak CKM phase $\gamma$.
Thus,
the deviation of $S_{+-} / \sqrt(1 - C^2_{+-})$
from $\sin{2 \alpha}$ is intimately connected
with how large $B^{00}$ is,
\textit{i.e.},
$B^{00}$ sets an upper bound on the
penguin contribution.
Including their recent measurement of
$C_{\pi^0 \pi^0} = -0.12 \pm 0.56 \pm 0.06$,
BABAR finds $|\delta_\alpha| < 30^\circ $ at $90\%$ 
C.~L.~\cite{BABAR-Cpi0pi0}.

\subsubsection{The decay $B_d \rightarrow \rho^+ \rho^-$}

In principle,
the isospin analysis discussed in subsection~\ref{subsec:isospin},
including the Grossman-Quinn type bounds,
is also applicable to the decays $B_d \rightarrow \rho \rho$.
This could be complicated by the fact that the $\rho$ has
three helicities,
but it turns out that experiments \textit{measure} the final
state to be completely longitudinally polarized
\cite{Giorgi}.
Such a final state is CP even,
and the analysis can proceed as before.

The decays $B_d \rightarrow \rho \rho$ have 
one very important advantage over
their $B_d \rightarrow \pi \pi$ counterparts;
the stringent upper bound on $B_d \rightarrow \rho^0 \rho^0$
means that,
here,
the Grossman-Quinn bound is very effective.
Indeed \cite{Zoltan-ICHEP04},
\be
\frac{\mbox{BR}_{\rm av} (\pi^0 \pi^0)}{
\mbox{BR}_{\rm av} (\pi^+ \pi^-)}
=
0.33 \pm 0.07\ ,
\ee
lead BABAR to $|\delta_\alpha| < 30^\circ$ \cite{BABAR-Cpi0pi0}, 
while
\be
\frac{\mbox{BR}_{\rm av} (\rho^0 \rho^0)}{
\mbox{BR}_{\rm av} (\rho^+ \rho^-)}
< 0.04\ ,
\ee
at the $90\%$ C.L.,
already implies that $|\delta_\alpha| < 11^\circ$
in the $B_d \rightarrow \rho \rho$ decays \cite{Giorgi}.

However,
there are also two additional difficulties.
Firstly,
due to the finite width of the $\rho$ resonance,
the identical particle symmetry invoked for
$\pi \pi$ in order to exclude $I=1$ as a possible
final state configuration needs to be altered
\cite{FLNQ}.
In principle,
the corrections will be of the order of
$(\Gamma_\rho/m_\rho) \sim 4 \%$,
but this effect deserves further attention.
Secondly,
there may be interference with non-resonant contributions
to $B$ meson decays into four pions,
and with other resonances yielding the same final state.
These effects may be modeled and fitted for.

\subsubsection{Dealing with discrete ambiguities}

\begin{figure}[h!]
\centerline{\includegraphics*[height=2.8in]{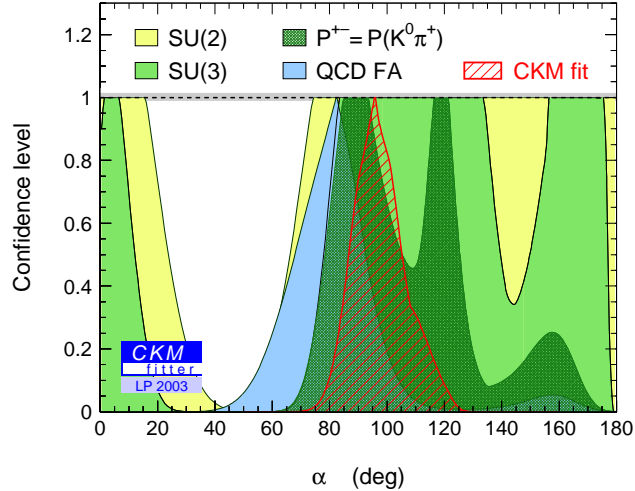}}
\caption{\label{alpha_pipi_03}Extraction of $\alpha$
from $B_d \rightarrow \pi^+ \pi^-$ with a variety of theoretical
assumptions,
compared with the result from the CKM fit (in red).
}
\end{figure}

The CKMfitter's FIG.~\ref{alpha_pipi_03},
shows bounds on $\alpha$ at the time of LP2003,
when no measurement of $C_{\pi^0 \pi^0}$ was available.
\begin{figure}[h!]
\centerline{\includegraphics*[height=2.8in]{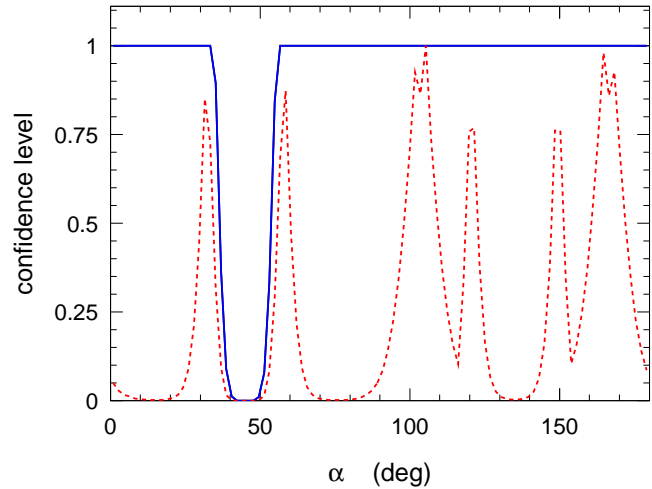}}
\caption{\label{with_Cpi0pi0_Hoecker}Toy exercise illustrating
an SU(2) fit for $\alpha$, using putative, future experimental
results excluding (red) or including (blue)
a precise measurement of $C_{\pi^0 \pi^0}$.
(Extreme courtesy of A.\ H\"{o}cker \cite{Hoecker-toy}.)
}
\end{figure}
The SU(2) approach gave a very loose bound,
corresponding roughly to the Grossman-Quinn bound.
This is partly due to the lack of a precise measurement of
$C_{\pi^0 \pi^0}$.
To illustrate the impact that a good measurement on this
quantity would have,
H\"{o}cker performed a very elucidative toy exercise
just after LP2003 \cite{Hoecker-toy},
shown in FIG.~\ref{with_Cpi0pi0_Hoecker}.
He used the world averaged central values for
$\mbox{BR}_{\rm av}(\pi^+ \pi^-)$,
$\mbox{BR}_{\rm av}(\pi^0 \pi^0)$,
and $\mbox{BR}_{\rm av}(\pi^+ \pi^0)$
and the BABAR central values for
$C_{\pi^+ \pi^-}$,
$S_{\pi^+ \pi^-}$,
taking the errors to improve by a factor of five.
Applying the Gronau-London method yields the curve in blue.
Now,
assume that the measurement of $C_{\pi^0 \pi^0}$ was very precise,
corresponding to $|\lambda_{B_d \rightarrow \pi^0 \pi^0}| = 1.00 \pm 0.08$,
just to see its impact.
(All the numbers used in this exercise were utilized exclusively
for illustrative purposes and need not be realistic.)
Applying the Gronau-London method yields the curve in red.
One sees the dramatic effect that a good measurement of
$C_{\pi^0 \pi^0}$ will have;
it will allow us to see the eight discrete
ambiguous solutions that the method yields for
$\alpha$ in the interval $[0^\circ, 180^\circ]$.
Fortunately,
improved results for this observable are expected.
This is the good news!

The bad news is that there are 1, 2, 3, \dots 8,
discretely ambiguous solutions for $\alpha$
in this interval.
For the sake of argument,
if all these solutions were separated,
and each had an error of $10^\circ$,
then we would have a $80^\circ$ allowed region for
$\alpha$ in the interval $[0^\circ, 180^\circ]$.
This is a major stumbling block in our search for new physics;
new physics effects could be hiding behind any of these
solutions and we wouldn't know it.
So,
there are really two difficulties which must be dealt with
in our search for new physics:
hadronic ``messy'' elements;
and discrete ``smokescreen'' ambiguities.

\begin{figure}[h!]
\centerline{\includegraphics*[height=3in]{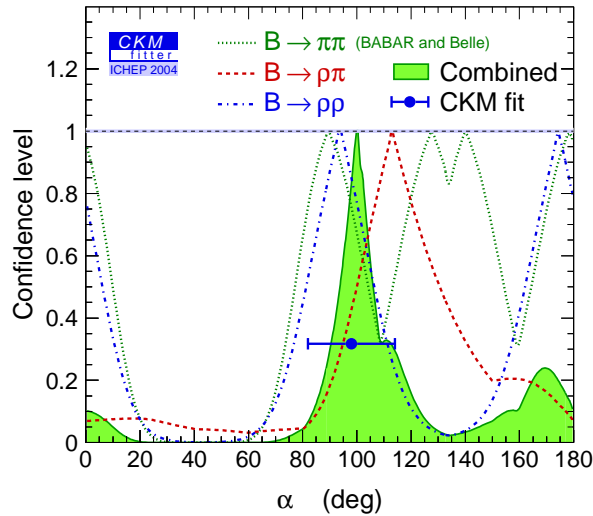}}
\caption{\label{all_alpha_Hoecker}Extraction of 
$\alpha = \pi - \beta - \gamma$
from the decays $B_d \rightarrow \pi \pi$,
$B_d \rightarrow \rho \pi$,
and $B_d \rightarrow \rho \rho$.
(Courtesy of A.\ H\"{o}cker.)
}
\end{figure}
Now comes one of those intellectual twists that makes CP
violation such an exciting field to work on:
the presence of the first difficulty (hadronic effects)
may help us resolve the second difficulty (discrete
ambiguities).
The simplified idea is the following:
if there were no hadronic effects associated with the
presence of the penguin diagram 
(which carries a weak phase that differs from the one in the tree
level diagram),
the CP violating asymmetry in the decay $B_d \rightarrow \pi^+ \pi^-$
would probe only $S_{\pi^+ \pi^-}$,
which would equal $- \sin(2 \beta + 2 \gamma)$.
From this,
one can extract $\beta + \gamma$,
up to a four-fold discrete ambiguity.
Were it not for the presence of hadronic effects,
related decays might also provide $\sin(2 \beta + 2 \gamma)$
and the same ambiguity might remain.
Fortunately,
the presence of hadronic effects shifts (and, in some cases,
reduces the number of)
the discrete ambiguities in the extraction of $\beta + \gamma$
in all decays.
And this occurs differently for different decays,
such as $B_d \rightarrow \rho \pi$ and
$B_d \rightarrow \rho \rho$;
each experiment gives a different set of discretely
ambiguous solutions.
Since the true solution to $\beta + \gamma$ must
be common to all sets,
we are able to exclude a number of ``wrong'' solutions.
This can be seen clearly in FIG.~\ref{all_alpha_Hoecker}
from the CKMfitter group \cite{CKMfitter-site},
which combines the results from the decays
$B_d \rightarrow \pi \pi$,
$B_d \rightarrow \rho \pi$,
and $B_d \rightarrow \rho \rho$,
at the time of ICHEP2004.
Notice the removal of many discrete ambiguities.
Incidentally,
this figure also shows that the extraction of
$\beta + \gamma$ from these $b \rightarrow u$ decays
is already competitive with the determination
of $\beta + \gamma$ performed with the standard CKM fit.

\subsection{\label{sec:B-Kpi}$B \rightarrow K \pi$ decays}

\subsubsection{Diagrammatic decomposition and experimental results}

\begin{figure}[htb]
\centerline{\includegraphics*[height=2.45in]{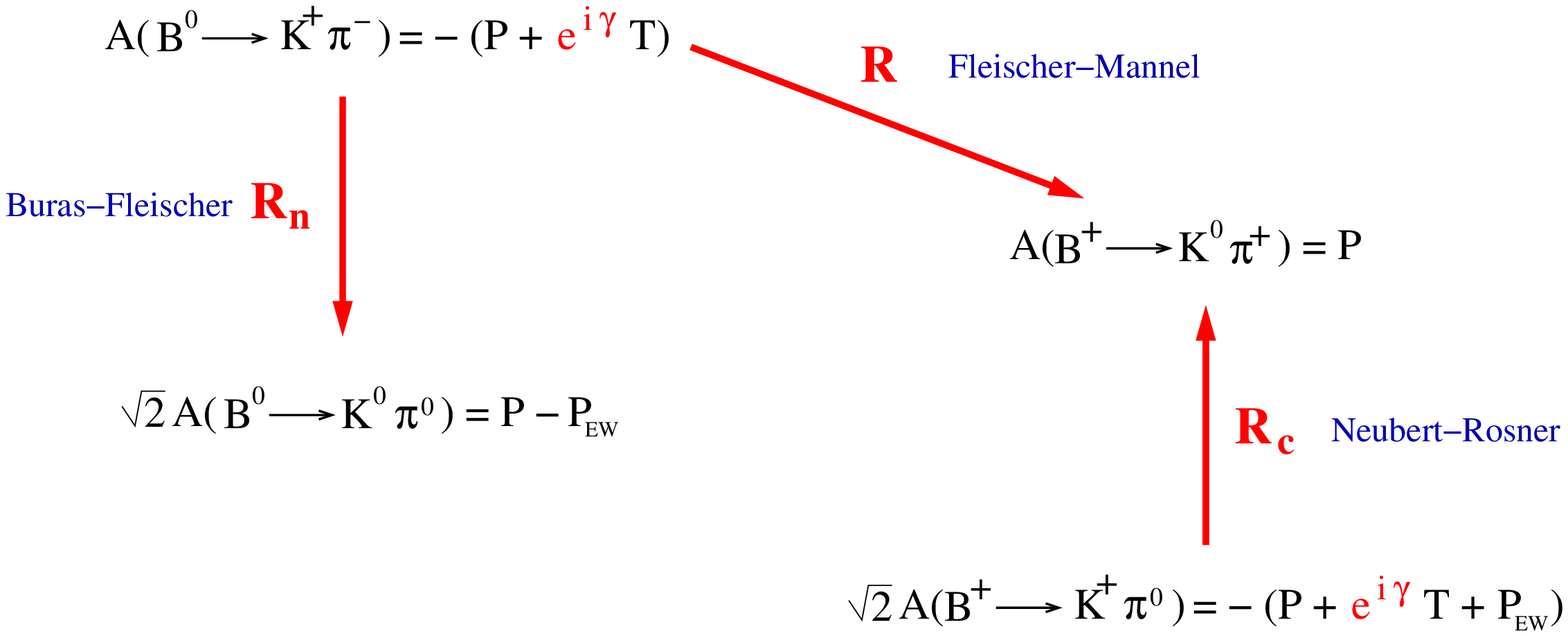}}
\caption{\label{fig:B_Kpi_panel}Simplified diagrammatic decomposition
of $B \rightarrow K \pi$ decays.
See text for details.
}
\end{figure}

In this section,
we will concentrate on the decays in 
FIG.~\ref{fig:B_Kpi_panel},
which shows the diagrammatic decomposition
discussed by Gronau and Rosner
in \cite{SU3-GR-Kpi-1,SU3-GR-Kpi-2}.
Here,
$T$ and $P$ stand for the tree and gluonic penguin diagrams
discussed above,
while $P_{\rm EW}$ is an electroweak penguin;
it is similar to the gluonic penguin,
but with the gluon substituted by the $Z$-boson or the
photon.\footnote{To be precise,
the corresponding (gauge invariant) calculations must include also
a $WW$ box diagram.}
Only the weak phase (which is $\gamma$) has been factored
out explicitly.
In analyzing exclusively $B \rightarrow K \pi$ decays,
references \cite{SU3-GR-Kpi-1,SU3-GR-Kpi-2} include color suppressed
(tree and electroweak penguins)
through the redefinitions
$P_{\rm EW} + C$, $T + P_{\rm EW}^c$, and $P - 1/3\,P_{\rm EW}^c$.
In the SU(3) decomposition,
there exist other diagrams
(with annihilation and exchange topologies)
which have been neglected.
A few features of FIG.~\ref{fig:B_Kpi_panel} are immediately
noticeable:
\begin{itemize}
\item if there were only gluonic penguin diagrams, then
the decays with neutral pions in the final state would have
about half the decay rate of those with charged pions
in the final state;
\item CP violation comes in with the tree diagram through the
weak phase $\gamma$;
\item the decays into neutral pions involve also the
electroweak penguins $P_{\rm EW}$.
\end{itemize}

A lot can be learned from experiment by comparing these
decays with each other.
It is useful to define the ratios
\ba
R &=&
\frac{\Gamma[B^0_d \rightarrow K^+ \pi^-] + 
\Gamma[\overline{B^0_d} \rightarrow K^- \pi^+]}{
\Gamma[B^+ \rightarrow K^0 \pi^+] + 
\Gamma[B^- \rightarrow \overline{K^0} \pi^-]
}
=
\frac{\tau_0}{\tau_+} 
\frac{\mbox{BR}_{\rm av}(K^+ \pi^-)}{
\mbox{BR}_{\rm av}(K^0 \pi^+)},
\label{R}
\\
R_c &=& 2
\frac{\Gamma[B^+ \rightarrow K^+ \pi^0] + 
\Gamma[B^- \rightarrow K^- \pi^0]}{
\Gamma[B^+ \rightarrow K^0 \pi^+] + 
\Gamma[B^- \rightarrow \overline{K^0} \pi^-]
}
=
2 \frac{\mbox{BR}_{\rm av}(K^+ \pi^0)}{
\mbox{BR}_{\rm av}(K^0 \pi^+)},
\label{Rc}
\\
R_n &=& \frac{1}{2}
\frac{\Gamma[B^0_d \rightarrow K^+ \pi^-] + 
\Gamma[\overline{B^0_d} \rightarrow K^- \pi^+]}{
\Gamma[B^0_d \rightarrow K^0 \pi^0] + 
\Gamma[\overline{B^0_d} \rightarrow \overline{K^0} \pi^0]
}
=
\frac{1}{2}
\frac{\mbox{BR}_{\rm av}(K^+ \pi^-)}{
\mbox{BR}_{\rm av}(K^0 \pi^0)},
\label{Rn}
\ea
first introduced by
Fleischer and Mannel \cite{FM},
Neubert and Rosner \cite{NR},
and Buras and Fleischer \cite{BF},
respectively.

The experimental results for the corresponding branching ratios
(averaged over CP conjugated channels)
and CP asymmetries at the time of LP2003,
quoted in reference \cite{SU3-GR-Kpi-2},
are shown in Table~\ref{table:BKpi}.
Also shown, between parenthesis,
are the updated results from ICHEP 2004 quoted in \cite{Zoltan-ICHEP04}.
%
\begin{table}[hbt]
\begin{centering}
\begin{tabular}{|c|c|c|}
\hline
Decay mode                                               &
$10^{6} \times \mbox{BR}_{\rm av}$
$\begin{array}{c}
\mbox{at LP2003}\\
\mbox{(at ICHEP2004)}
\end{array}$                                              &
$A_{\rm CP}$                                             \\
\hline
$B^+ \rightarrow K^0 \pi^+$                              &
$\begin{array}{c}
21.78 \pm 1.40\\
(24.1 \pm 1.3)
\end{array}$                                              &   
$\begin{array}{c}
0.016 \pm 0.057\\
(-0.02 \pm 0.03)
\end{array}$                                             \\
\hline
$B^+ \rightarrow K^+ \pi^0$                              &
$\begin{array}{c}
12.82 \pm 1.07\\
(12.1 \pm 0.8)
\end{array}$                                              &   
$\begin{array}{c}
0.00 \pm 0.12\\
(0.04 \pm 0.04)
\end{array}$                                             \\
\hline
$B^0_d \rightarrow K^+ \pi^-$                            &
$\begin{array}{c}
18.16 \pm 0.79\\
(18.2 \pm 0.8)
\end{array}$                                             &   
$\begin{array}{c}
-0.095 \pm 0.029\\
(-0.11 \pm 0.02)
\end{array}$                                             \\
\hline
$B^0_d \rightarrow K^0 \pi^0$                            &
$\begin{array}{c}
11.92 \pm 1.44\\
(11.5 \pm 1.0)
\end{array}$                                             &   
$\begin{array}{c}
0.03 \pm 0.37\\
(0.01 \pm 0.16)
\end{array}$                                             \\
\hline
\end{tabular}
\caption{\label{table:BKpi}Experimental measurements
of $B \rightarrow K \pi$ branching ratios
(averaged over CP conjugated modes) and
CP violating asymmetries.
Values from LP2003 as quoted in \cite{SU3-GR-Kpi-2}.
The values included between parenthesis
were presented by Ligeti at ICHEP2004 in \cite{Zoltan-ICHEP04}.
A similar averaging of ICHEP2004 results by Giorgi
leads to $A_{\rm CP}(K^+ \pi^-) = -0.114 \pm 0.020$
and $A_{\rm CP}(K^+ \pi^0) = +0.049 \pm 0.040$
\cite{Giorgi}.
}
\end{centering}
\end{table}
%
A simple glance at the table is enough to convince oneself
that penguin diagrams have indeed been observed and that they
are dominant features in these decays.
Those results imply that
\ba
R &=&
0.898 \pm 0.071
\hspace{1cm}
(0.82 \pm 0.06)\ ,
\label{Bound_R}
\\
R_c &=&
1.18 \pm 0.12
\hspace{1.4cm}
(1.00 \pm 0.08)\ ,
\label{Bound_Rc}
\\
R_n &=&
0.76 \pm 0.10
\hspace{1.4cm}
(0.79 \pm 0.08)\ ,
\label{Bound_Rn}
\ea
where the numbers without (within) parenthesis refer to
the LP2003 values quoted in reference \cite{SU3-GR-Kpi-2}
(ICHEP2004 values quoted in reference \cite{Zoltan-ICHEP04}).
Next we comment on the usefulness of these results for the extraction
of the CKM phase $\gamma$.

\subsubsection{\label{subsec:FM}Using $R$ to
learn about the CKM phase $\gamma$}

For the moment,
let us concentrate on the decay $B_d \rightarrow K^+ \pi^-$,
normalized to the decay $B^+ \rightarrow K^0 \pi^+$.
In the decays into two pions,
the tree diagram was believed to be dominant,
and we defined the ratio ``penguin over tree''.
Here,
the penguin dominates and we define instead
the ``tree over penguin'' ratio as
\be
r e^{i \delta} = \frac{T}{P}\ ,
\ee
where $\delta$ is the relative strong phase.
Trivially \textbf{(Ex-31)},
\be
R = 1 - 2 r \cos{\gamma} \cos{\delta} + r^2.
\label{eq:FM-1}
\ee
Now comes the beautiful argument by Fleischer and Mannel:
imagine that $R < 1$;
then,
it is clear that $\gamma$ cannot possibly be $\pi/2$,
\textit{regardless of the exact values} of $r$.
Recall that,
since $r$ and $\delta$ are defined as the ratio
of two hadronic matrix elements,
\textit{c.f.\/} Eq.~(\ref{eq:FM-1}),
they suffer from hadronic uncertainties.
Still,
the simple trigonometric argument put forth by Fleischer and
Mannel \cite{FM} means that,
despite this problem,
we can get some information on $\gamma$.\footnote{There
were later some discussions on the impact of rescattering
on this method \cite{Neu98,GerWey99},
but I still find this one of the nicest
arguments in $B \rightarrow K \pi$ decays;
one seems to get something clean out of a mess.}
Their bound is $\sin^2{\gamma} \leq R$ which,
of course,
has no impact if $R \geq 1$.

\begin{figure}[htb]
\centerline{\includegraphics*[height=3in]{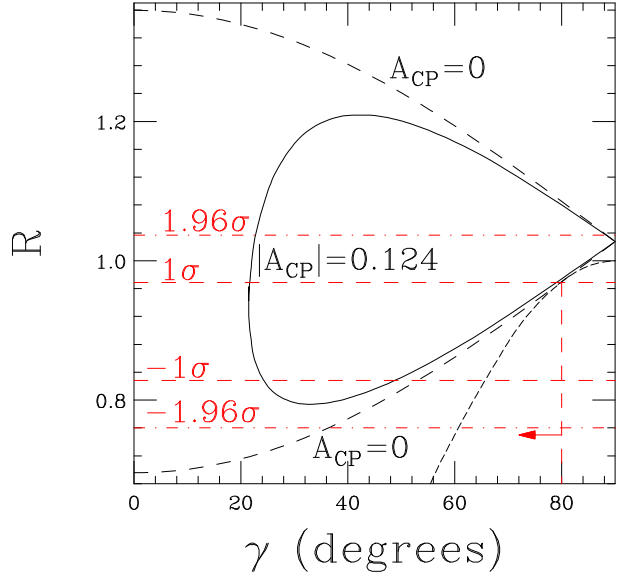}}
\caption{\label{fig:GR_Racp0311}
Behavior of R for $r=0.166$ and
$|A_{\rm CP}| = 0$ (dashed curve) or 
$|A_{\rm CP}| = 0.124$ (solid curve),
as a function of $\gamma$.
The short-dashed curve shows the Fleischer-Mannel bound.
Taken from reference \cite{SU3-GR-Kpi-2}.
}
\end{figure}

As always,
if we knew $r$ and $\delta$,
extracting $\gamma$ would be straightforward.
Gronau and Rosner improved on this method
by noting that the CP asymmetry
\be
A_{\rm CP} =
\frac{\Gamma[\overline{B^0_d} \rightarrow K^- \pi^+]
- \Gamma[B^0_d \rightarrow K^+ \pi^-]}{
\Gamma[\overline{B^0_d} \rightarrow K^- \pi^+]
+ \Gamma[B^0_d \rightarrow K^+ \pi^-]}
=
- \frac{2 r}{R} \sin{\gamma} \sin{\delta},
\label{eq:GR_FM}
\ee
may be used to extract $\delta$.
Therefore,
using $r$ from some related decay,
we can extract a value for $\gamma$.
This is shown schematically in 
FIG.~\ref{fig:GR_Racp0311}.
This figure was drawn for a very specific value,
$r = 0.166$.
For that value,
and using the LP2003 $1 \sigma$ bounds on $R$ and $|A_{\rm CP}|$,
Gronau and Rosner find $49^\circ < \gamma < 80^\circ$
\cite{SU3-GR-Kpi-2}.

The lower bound does not remain if one
allows for lower values of $r$.
Also,
both bounds disappear if we take the $2 \sigma$
ranges for $R$.
But the most striking feature of 
FIG.~\ref{fig:GR_Racp0311}
is yet a third one.
The Gronau-Rosner method restricts the solutions to
lie in the region within the solid curve $|A_{\rm CP}|=0.124$ and
the dashed curve $|A_{\rm CP}|=0$.
For the sake of argument,
let us now take $r=0.166$ and the $1 \sigma$ horizontal bounds on
$R$ (dashed, red horizontal lines).
Then,
there is a very restricted allowed region in the
$R - \gamma$ plane and we get the bound mentioned above:
$49^\circ < \gamma < 80^\circ$.
But this still leaves a $31^\circ$ uncertainty on $\gamma$.
How could we improve on this?
The answer is surprising.
Because the curves for $|A_{\rm CP}|=0.124$ and
$|A_{\rm CP}|=0$ lie so close to each other in the
region of interest,
improving the precision on this (CP violating) measurement does not
improve much the precision on (the CP violating phase) $\gamma$.
In contrast,
improving the precision on the (CP conserving) observable
$R$ can improve considerably the precision on 
(the CP violating phase) $\gamma$,
as one can see by imagining that the lower $1 \sigma$ horizontal 
line moves up slightly.

Thus
\begin{itemize}
\item a more precise constraint on $r$ is required in order
to assess the effectiveness of this method;
\item a better measurement of $R$ is needed;
\item and, surprisingly, an improvement on the
precision of the measurement of the CP conserving
observable $R$ will be much more effective in constraining
the CP violating phase $\gamma$ than an improvement
on the CP violating observable $A_{\rm CP}$.
\end{itemize}

The details of the $B \rightarrow K \pi$ analysis in general,
and of FIG.~\ref{fig:GR_Racp0311} in particular,
depend crucially on the exact values for the
observed branching ratios and CP asymmetries\footnote{To see this,
compare FIG.~\ref{fig:GR_Racp0311},
which was taken from the article \cite{SU3-GR-Kpi-2} by Gronau and Rosner,
with a similar figure drawn a few months before
by the same authors in \cite{SU3-GR-Kpi-1},
based on the earlier experimental results
$R = 0.948 \pm 0.074$ and 
$A_{\rm CP} = -0.088 \pm 0.040$.},
which are still in a state of flux.
This is even more important for the assessment of
electroweak penguins to be performed in the next
subsection.
Nevertheless,
our interest here is on general methods and not on the precise
numerics.
The ideas presented here will remain on our collective
toolbox,
even if the specific examples themselves turn out to be
numerically uninteresting.

\subsubsection{Searching for enhanced $\Delta I = 1$
contributions}

The amplitudes discussed in the previous section may also
be decomposed in terms of isospin amplitudes,
according to \cite{isospin_Kpi} \textbf{(Ex-32)}
\ba
A (B^0_d \rightarrow K^+ \pi^-)
& = &
-B_{1/2} + A_{1/2} + A_{3/2},
\nonumber\\
\sqrt{2} 
A (B^+ \rightarrow K^+ \pi^0)
& = &
-B_{1/2} - A_{1/2} + 2 A_{3/2},
\nonumber\\
A (B^+ \rightarrow K^0 \pi^+)
& = &
B_{1/2} + A_{1/2} + A_{3/2},
\nonumber\\
\sqrt{2} 
A (B^0 \rightarrow K^0 \pi^0)
& = &
B_{1/2} - A_{1/2} + 2 A_{3/2},
\label{eq:isospin_NirQuinn}
\ea
where $A$ and $B$ are $\Delta I = 1$
and $\Delta I = 0$ amplitudes,
respectively,
and the subscripts indicate the isospin of
$K \pi$.
Comparing Eq.~(\ref{eq:isospin_NirQuinn}) with
FIG.~\ref{fig:B_Kpi_panel} \cite{GR-LSR},
we conclude that \textbf{(Ex-33)}
\ba
B_{1/2}
& = &
P + \frac{1}{2} T,
\nonumber\\
A_{1/2}
& = &
\frac{1}{3} P_{\rm EW} - \frac{1}{6} T,
\nonumber\\
A_{3/2}
& = &
- \frac{1}{3} P_{\rm EW} - \frac{1}{3} T.
\label{eq:relation_isospin_diagrammatics}
\ea

Using both decompositions,
one can show that \textbf{(Ex-34)}
\be
R_c - R_n = 
\bigcirc 
\left( 
\frac{P_{\rm EW}, T}{P}
\right)^2
\ \ \ \mbox{in a } \Delta I = 1 \mbox{ combination.}
\label{Rc-Rn}
\ee
Therefore,
this observable highlights $\Delta I = 1$ combinations
which do not involve the gluonic penguin.
In the SM,
the right hand side of Eq.~(\ref{Rc-Rn}) is expected to be small.
Hence,
the fact that $R_c - R_n = 0.42 \pm 0.22$ in 2003 could be
seen as an indication of new physics in electroweak penguins.

This question was first pointed out by 
Gronau and Rosner \cite{GR-LSR} and by Lipkin \cite{Lipkin-LSR},
utilizing a slightly different quantity \textbf{(Ex-35)}
\be
R_L - 1= 2
\frac{\Gamma_{\rm av}[B^+ \rightarrow K^+ \pi^0] + 
\Gamma_{\rm av}[B^0 \rightarrow K^0 \pi^0]}{
\Gamma_{\rm av}[B^+ \rightarrow K^0 \pi^+] + 
\Gamma_{\rm av}[B^0 \rightarrow K^+ \pi^-]} - 1
=
\frac{R_c + R/R_n}{1+R} - 1,
\label{R_L}
\ee
which shares the features on the right hand side of
Eq.~(\ref{Rc-Rn}) \cite{Yuval}.
Averaging over CP conjugated decays is implicit in
the notation $\Gamma_{\rm av}$ used on the
first equality.
It has been shown that it is possible to fit
the LP2003 values for $R_c$ and $R_n$ as long as 
$P_{\rm EW} \sim i P/2$ \cite{large_PEW}.
Curiously,
the search for new physics through
such enhancements of $\Delta I = 1$
electroweak penguins had been proposed sometime before,
and those effects had been named 
``Trojan penguins'' \cite{trojan_penguins}.

More recently,
global analysis to all $B \rightarrow K \pi$ data were
performed in \cite{CKMfitter-04} and \cite{SU3-04},
finding consistency with the SM and,
in particular,
uncovering no unequivocal sign of enhanced electroweak penguins.
One could fear that,
because these analysis involve a fit to many observables,
most of which have nothing to do with possible
enhanced $\Delta I = 1$ pieces,
some dilution might occur.
However,
the latest experimental results,
presented at ICHEP2004,
also seem to be moving in a way which removes this signal:
$R_c - R_n = 0.21 \pm 0.16$
\cite{Zoltan-ICHEP04}.
Whether this signal will remain is unclear,
but even if it does not,
we have learned of a new way to constrain
theories with enhanced electroweak penguins.

\subsubsection{CP asymmetries in $B \rightarrow K \pi$
decays}

Perhaps the most important result in 2004 has been the
improvement by BABAR \cite{BABAR-K+pi-} and Belle \cite{Belle-K+pi-}
of their measurements of the direct CP
asymmetry in $B_d \rightarrow K^\pm \pi^\mp$.
The average became
$A_{\rm CP}(K^+ \pi^-) = -0.114 \pm 0.020$ \cite{Giorgi}.
This constitutes the first observation of 
direct CP violation in the $B$ system agreed upon by both groups,
in analogy to $\epsilon^\prime_K$ in the kaon system.

At ICHEP2004 a new puzzle emerged in $B \rightarrow K \pi$
decays,
coming from
\be
A_{\rm CP}(K^+ \pi^-) - 
A_{\rm CP}(K^+ \pi^0)
=
-0.163 \pm 0.060\ ,
\ee
which constitutes a $3.6\sigma$ signal \cite{Giorgi}.
Looking back at FIG.~\ref{fig:B_Kpi_panel},
we recognize that this can only be due to electroweak penguins.

So,
in 2004,
the rave in $B \rightarrow K \pi$ decays moved from
$R_c - R_n$ to the direct CP asymmetries.
This is
testimony to the fact that, after decades of 
experimental stagnation,
CP violation has moved from theoretical exercises
into a full fledged experimental endeavor.
Since errors are still large,
quite a number of additional interesting hints are to be expected.

\subsection{Other decays of interest}

In the coming years
the programs developed at the $B$-factories and at hadronic
facilities will greatly improve our knowledge of the
CKM mechanism,
and they hold the possibility to uncover new physics effects.
It is impossible to mention all the decays of interest in a
pedagogical review of this size.
Some further possibilities for the $B_d$ system include:
\begin{itemize}
\item Snyder-Quinn method: determining 
$\beta + \gamma$ from the decays $B_d \rightarrow \rho \pi$
\cite{SQ};
\item Determining $\gamma$ from $B \rightarrow D$ decays.
These ideas started with the Gronau-London-Wyler method,
where one determines $\gamma$ through a triangle relation
among the decays
$B^+ \rightarrow K^+ D^0$,
$B^+ \rightarrow K^+ \overline{D^0}$,
$B^+ \rightarrow K^+ (f_{\rm cp})_D$
\cite{GLW}.
A very long list of improvements and related
suggestions was spurred by the Atwood-Dunietz-Soni method
\cite{ADS}.
\item Determining $2 \beta + \gamma$ from $B_d \rightarrow D$ decays.
This interesting class of methods started with
a proposal by Dunietz and Sachs and beats the phase
of $B^0_d - \overline{B^0_d}$ mixing
($2 \beta$ in the usual phase convention)
against the phase in $b \rightarrow u$ transitions
($\gamma$ in the usual phase convention) \cite{2b+g_DS}.  
\end{itemize}

Many interesting pieces of information will also come
from experiments on the $B_s$ system performed at hadronic facilities.
Indeed:
\begin{itemize}
\item A measurement of $\Delta m_s$ will
reduce the hadronic uncertainties involved in
extracting $|V_{td}|$
(\textit{i.e.\/}, $R_t = \sqrt{(1-\rho)^2 + \eta^2}$)
from $\Delta m_d$,
thus improving our determination of $\rho$ and $\eta$;
\item Since the angle $\chi$ is involved
in $B^0_s - \overline{B^0_s}$ mixing
(in the usual phase convention),
it would be interesting to determine it,
for example, from $B_s \rightarrow D_s^+ D_s^-$ decays.
Precisely because in the SM this asymmetry is expected to be small,
this is a perfect channel to look for new physics;
\item There are a variety of interesting CP asymmetries
in $B_s$ decays.
In the Aleksan-Dunietz-Kayser method one determines
$\gamma$ through the decays $B_s \rightarrow D_s^+ K^-$,
$B_s \rightarrow D_s^- K^+$
\cite{ADK};
\item One may invoke SU(3) symmetry to compare
$B_d$ and $B_s$ decays.
For example,
the Silva-Wolfenstein method utilizing $U$-spin
to determine the penguin pollution in $B_d \rightarrow \pi^+ \pi^-$
by its relation with $B_d \rightarrow K^+ \pi^-$ \cite{SW1}
may be adapted
to relate the penguin pollution in $B_d \rightarrow \pi^+ \pi^-$
by its relation with $B_s \rightarrow K^+ K^-$ instead
\cite{Flei-Dun}.
\end{itemize}

This ``Brave New World'' will provide us with many
new tests of the SM and, if we are lucky enough,
the uncovering of new physics.

\section*{Acknowledgments}

I would like to thank the organizers of the
Central European School in Particle Physics for inviting
me to give these lectures,
and the students for the great atmosphere.
I am extremely grateful to Prof.\ Ji\v{r}\'{\i}
Ho\v{r}ej\v{s}\'{\i} for the invitation,
for making my stay so enjoyable,
and for making me feel so welcomed in Prague.
My lectures benefited considerably from close
coordination with V.\ Sharma,
who taught the course on
``Recent experimental results on CP violation'',
and from whom I learned a great deal.
Part of the outline for these lectures was first
prepared for a review talk given at the 19th International Workshop
on Weak Interactions and Neutrinos, which took place in
Wisconsin during October 6-11, 2003.
I am grateful to U.\ Nierste and to
J.\ Morfin for that kind invitation.
I am indebted to 
M.\ Gronau,
J.\ Rosner, 
A.\ H\"{o}cker,
HFAG,
and the CKMfitter group for allowing me to
reproduce their figures.
The spirit of these lectures owes much to 
G.\ C.\ Branco, I.\ Dunietz,
B.\ Kayser, L.\ Lavoura, Y.\ Nir, H.\ R.\ Quinn,
and to L.\ Wolfenstein.
This work is supported in part by project POCTI/37449/FNU/2001,
approved by the Portuguese FCT and POCTI,
and co-funded by FEDER.


\newpage

\appendix


\section{\label{appendix-CPT}Neutral
meson mixing including CPT violation}

\subsection{The mixing matrix}

In this appendix we discuss the mixing in the neutral meson
systems in the presence of CPT violation and we will
continue to assume the Lee-Oehme-Yang approximation \cite{controversy}.
The eigenvector equation~(\ref{PaPb}) becomes generalized into
\begin{eqnarray}
\left(
\begin{array}{c}
| P_H \rangle \\ | P_L \rangle
\end{array}
\right)
=
\left(
\begin{array}{cc}
p_H & - q_H\\ p_L & q_L
\end{array}
\right)
\
\left(
\begin{array}{c}
| P^0 \rangle \\ | \overline{P^0} \rangle
\end{array}
\right)
= 
\mbox{\boldmath $X$}^T\ 
\left(
\begin{array}{c}
| P^0 \rangle \\ | \overline{P^0} \rangle
\end{array}
\right).
\label{PaPbCPT}
\end{eqnarray}
We should be careful with the explicit choice
of $-q_H$ and $+q_L$ in Eq.~(\ref{PaPbCPT});
the opposite choice has been made in references \cite{BLS,reciprocal}.

The relation between these mixing parameters 
($p_H$, $q_H$, $p_L$, and $q_L$),
the eigenvalues of $\mbox{\boldmath $H$}$ in
Eq.~(\ref{mass_eigenvalues}),
and the matrix elements
of $\mbox{\boldmath $H$}$ written in the flavor basis
is still obtained through the diagonalization
\begin{equation}
\mbox{\boldmath $X$}^{-1} \mbox{\boldmath $H$} \mbox{\boldmath $X$} = 
\left(
\begin{array}{cc}
\mu_H & 0\\ 0 & \mu_L
\end{array}
\right),
\label{diagonalizationCPT}
\end{equation}
but now
\begin{equation}
\mbox{\boldmath $X$}^{-1} = \frac{1}{p_H q_L+p_L q_H}
\left(
\begin{array}{cc}
q_L & - p_L\\ q_H & p_H
\end{array}
\right)
\label{X-1CPT}
\end{equation}
substitutes Eq.~(\ref{X-1}).

We may write the mixing matrix 
$\mbox{\boldmath $X$}$ 
in terms of new parameters \cite{Lav91}
\begin{equation}
\theta =
\frac{\frac{q_H}{p_H} - \frac{q_L}{p_L}}{
\frac{q_H}{p_H} + \frac{q_L}{p_L}},
\label{theta}
\end{equation}
and
\begin{equation}
\frac{q}{p} = - \sqrt{\frac{q_H q_L}{p_H p_L}}
\label{ratio}
\end{equation}
We may define $\delta$ in this more general setting
through the first equality in Eq.~(\ref{delta_new_expressions}),
leading to $|q/p| = \sqrt{\frac{1-\delta}{1+\delta}}$.
With this notation the mixing matrix may be re-written as
\begin{equation}
\mbox{\boldmath $X$} =
\left(
\begin{array}{cc}
1 & 1\\
- \frac{q}{p} \sqrt{\frac{1+\theta}{1-\theta}} &
\frac{q}{p} \sqrt{\frac{1-\theta}{1+\theta}}
\end{array}
\right)
\left(
\begin{array}{cc}
p_H & 0\\
0 & p_L
\end{array}
\right),
\label{X-parametCPT}
\end{equation}
\begin{equation}
\mbox{\boldmath $X$}^{-1} =
\left(
\begin{array}{cc}
p_H^{-1} & 0\\
0 & p_L^{-1}
\end{array}
\right)
\left(
\begin{array}{cc}
\frac{1-\theta}{2} & - \frac{p}{q} \frac{\sqrt{1-\theta^2}}{2}\\
\frac{1+\theta}{2} & \frac{p}{q} \frac{\sqrt{1-\theta^2}}{2}
\end{array}
\right).
\label{X-1-parametCPT}
\end{equation}

The fact that the trace and determinant are invariant under the general
similarity transformation in Eq.~(\ref{diagonalizationCPT}) implies that
\begin{eqnarray}
\mu &=& (H_{11}+H_{22})/2,
\nonumber\\
\Delta \mu &=& \sqrt{4 H_{12} H_{21} + (H_{22} - H_{11})^2}.
\label{eigenvaluesCPT}
\end{eqnarray}
Moreover,
from
\begin{eqnarray}
\left( \begin{array}{cc} H_{11} & H_{12} \\ H_{21} & H_{22} \end{array} \right)
\left( \begin{array}{c} p_H \\ -q_H \end{array} \right)
&=& \mu_H
\left( \begin{array}{c} p_H \\ -q_H \end{array} \right),
\nonumber\\
\left( \begin{array}{cc} H_{11} & H_{12} \\ H_{21} & H_{22} \end{array} \right)
\left( \begin{array}{c} p_L \\ q_L \end{array} \right)
&=& \mu_L
\left( \begin{array}{c} p_L \\ q_L \end{array} \right).
\label{eigenvector-equationsCPT}
\end{eqnarray}
we find that
\begin{eqnarray}
\frac{q_H}{p_H}
&=& \frac{H_{11} - \mu_H}{H_{12}}
= \frac{H_{21}}{H_{22} - \mu_H},
\nonumber\\
\frac{q_L}{p_L}
&=& \frac{\mu_b - H_{11}}{H_{12}}
= \frac{H_{21}}{\mu_b - H_{22}},
\label{eigenvectorsCPT}
\end{eqnarray}
leading to
\begin{eqnarray}
\theta &=&
\frac{H_{22} - H_{11}}{\mu_H - \mu_L},
\nonumber\\
\delta &=&
\frac{|H_{12}|-|H_{21}|}{|H_{12}|+|H_{21}|},
\label{mixing-observablesCPT}
\end{eqnarray}
and $q/p = - \sqrt{H_{21}/H_{12}}$.
We see from Eqs.~(\ref{effect_of_discrete_symmetries})
that $\mbox{Re}\, \theta$ and $\mbox{Im}\, \theta$ are
CP and CPT violating,
while $\delta$ is CP and T violating.

Although $\mbox{\boldmath $H$}$ contains eight real numbers,
only seven are physically meaningful.
Indeed,
one is free to change the phase
of the kets $| P^0 \rangle$,
$| \overline{P^0} \rangle$,
$| P_H \rangle$,
and $| P_L \rangle$,
as\footnote{See also appendix~\ref{appendix-rephasing}.}
\begin{eqnarray}
| P^0 \rangle & \rightarrow & e^{i \gamma} | P^0 \rangle,
\nonumber\\
| \overline{P^0} \rangle & \rightarrow & e^{i \overline{\gamma}}
| \overline{P^0} \rangle,
\nonumber\\
| P_H \rangle & \rightarrow & e^{i \gamma_H} | P_H \rangle,
\nonumber\\
| P_L \rangle & \rightarrow & e^{i \gamma_L} | P_L \rangle.
\label{ket rephasingCPT}
\end{eqnarray}
Under these transformations
\begin{eqnarray}
H_{12} & \rightarrow & e^{i \left( \overline{\gamma} - \gamma \right)} H_{12},
\nonumber\\
H_{21} & \rightarrow & e^{i \left( \gamma - \overline{\gamma} \right)} H_{21},
\nonumber\\
q/p & \rightarrow & e^{i \left( \gamma - \overline{\gamma} \right)} q/p,
\label{rephasingCPT}
\end{eqnarray}
while $H_{11}$, $H_{22}$, $\mu$, $\Delta \mu$,
$\theta$, and $\delta$ do not change.
Therefore,
the relative phase between $H_{12}$ and $H_{21}$ is physically
meaningless and $\mbox{\boldmath $H$}$ contains only seven observables.
Similarly,
\textit{the phase of $q/p$ is also unphysical}.
As a result,
we have four observables in the eigenvalues,
$\mu$ and $\Delta \mu$,
and three in the mixing matrix,
$\theta$ and $\delta$ (or, alternatively, $|q/p|$).

Eqs.~(\ref{eigenvaluesCPT}) and (\ref{mixing-observablesCPT})
give the measurable mixing and eigenvalue parameters in
terms of the $H_{ij}$ matrix elements which one can calculate
in a given model.
Given the current and upcoming experimental probes of the
various neutral meson systems,
it seems much more appropriate to do precisely the opposite;
that is, to give the $H_{ij}$ matrix elements in terms of the
experimentally accessible quantities.
Such expressions would give $M_{ij}$ and $\Gamma_{ij}$ in
a completely model independent way,
with absolutely no assumptions.
One could then calculate these quantities in any given model;
if they fit in the allowed ranges the model would be viable.

Surprisingly,
this is not is done in most expositions of the
$P^0 - \overline{P^0}$ mixing.
The reason is simple.
Eqs.~(\ref{eigenvaluesCPT}) and (\ref{mixing-observablesCPT})
are non-linear in the $H_{ij}$ matrix elements.
Thus,
inverting them by brute force would entail a tedious calculation.
With the matrix manipulation discussed here this inversion
is straightforward.
Indeed,
Eq.~(\ref{diagonalizationCPT}) can be trivially transformed into
\cite{Alv99,reciprocal}
\begin{eqnarray}
\mbox{\boldmath $H$} &=&
\mbox{\boldmath $X$}
\left(
\begin{array}{cc}
\mu_H & 0\\
0 & \mu_L
\end{array}
\right)
\mbox{\boldmath $X$}^{-1}
\nonumber\\
&=&
\left(
\begin{array}{cc}
\mu - \frac{\Delta \mu}{2} \theta \ &
- \frac{p}{q} \frac{\sqrt{1-\theta^2}}{2} \Delta \mu
\\*[2mm]
- \frac{q}{p} \frac{\sqrt{1-\theta^2}}{2} \Delta \mu \ &
\mu + \frac{\Delta \mu}{2} \theta
\end{array}
\right),
\label{masterCPT}
\end{eqnarray}
where we have used Eqs.~(\ref{X-parametCPT}) and (\ref{X-1-parametCPT})
\textbf{(Ex-36)}.
Although rarely seen,
this equation is very interesting because it expresses in a very compact form 
the relation between the quantities which are experimentally accessible
and those which are easily calculated in a given theory.
The full power of Eq.~(\ref{masterCPT}) can be seen when considering
the propagation of a neutral meson system in matter
\textbf{(Ex-37)}.

\subsection{Time evolution}
\label{sec:time-evolution-CPT}

To find the time evolution of the neutral meson system 
we start from Eq.~(\ref{ev-operator}) proved as an exercise.
Then,
using Eqs.~(\ref{PtildeaPtildeb}), (\ref{PaPbCPT}),
(\ref{X-parametCPT}), and (\ref{X-1-parametCPT}),
we find
\begin{eqnarray}
\exp{(- i {\cal H} t)}
&=&
\left(
\begin{array}{cc}
| P^0 \rangle, & | \overline{P^0} \rangle
\end{array}
\right)
\mbox{\boldmath $X$}
\left(
\begin{array}{cc}
e^{-i \mu_H t} & 0\\
0 & e^{-i \mu_L t}
\end{array}
\right)
\mbox{\boldmath $X$}^{-1}
\left(
\begin{array}{c}
\langle P^0 | \\
\langle \overline{P^0} |
\end{array}
\right)
\nonumber\\*[3mm]
&=&
\left(
\begin{array}{cc}
| P^0 \rangle, & | \overline{P^0} \rangle
\end{array}
\right)
\left(
\begin{array}{cc}
g_+(t) + \theta\, g_-(t)\  & \frac{p}{q} \sqrt{1  - \theta^2} g_-(t)\\*[2mm]
\frac{q}{p} \sqrt{1  - \theta^2} g_-(t)\  & g_+(t) - \theta\, g_-(t)
\end{array}
\right)
\left(
\begin{array}{c}
\langle P^0 | \\
\langle \overline{P^0} |
\end{array}
\right),
\label{evolution-in-generalCPT}
\end{eqnarray}
where the functions $g_\pm(t)$ are those already defined in 
Eq.~(\ref{g+-}).
This corresponds to the usual expressions for the time evolution 
of a state which starts out as $P^0$ or $\overline{P^0}$,
\begin{eqnarray}
| P^0(t) \rangle
= \exp{(- i {\cal H} t)} | P^0 \rangle
&=&
\left[ g_+(t) + \theta\, g_-(t) \right] | P^0 \rangle
+
\frac{q}{p} \sqrt{1  - \theta^2} g_-(t)\, | \overline{P^0} \rangle,
\nonumber\\*[3mm]
| \overline{P^0}(t) \rangle
= \exp{(- i {\cal H} t)} | \overline{P^0} \rangle
&=&
\frac{p}{q} \sqrt{1  - \theta^2} g_-(t) | P^0 \rangle
+
\left[ g_+(t) - \theta\, g_-(t) \right] | \overline{P^0} \rangle,
\label{usual-time-evolutionCPT}
\end{eqnarray}
respectively.
At this point it is important to emphasize the fact that,
in deriving this result,
no assumptions were made about the form of the original matrix 
$\mbox{\boldmath $H$}$.
This observation will become important once we consider
the evolution in matter \textbf{(Ex-37)}.

\section{\label{appendix-rephasing}Phase transformations and
CP conservation}

This appendix contains a detailed description of the
phase transformations and of the conditions implied by
CP conservation which we have used in chapter~\ref{ch:producao}
in order to identify the relevant CP violating parameters.

\subsection{Phase transformations}

As mentioned,
any ``ket'' may be redefined by an arbitrary phase
transformation \cite{Dirac},
\begin{eqnarray}
\label{eq:rephazing-kets}
| i \rangle \rightarrow e^{i\gamma_i} |i \rangle\ ,
&\hspace{5mm}&
|\bar i\rangle \rightarrow
e^{i\overline{\gamma}_i} |\bar i \rangle\ ,
\nonumber\\
| P^0 \rangle \rightarrow e^{i\gamma_P} | P^0 \rangle\ ,
&\hspace{5mm}&
| \overline{P^0} \rangle \rightarrow
e^{i\overline{\gamma}_P} | \overline{P^0} \rangle\ ,
\nonumber\\
| f \rangle \rightarrow e^{i\gamma_f} | f \rangle\ ,
&\hspace{5mm}&
| \bar f \rangle \rightarrow
e^{i\overline{\gamma}_f} | \bar f \rangle\ .
\end{eqnarray}
These phase transformations modify the mixing parameters
and the transition amplitudes, according to
\begin{eqnarray}
\label{eq:rephazing-quantities}
\frac{q}{p} \rightarrow 
e^{i(\gamma_P - \overline{\gamma}_P)} \frac{q}{p}\ ,
&\hspace{5mm}&
\nonumber\\*[2mm]
A_{i \rightarrow P^0} \rightarrow
e^{i(\gamma_i-\gamma_P)} A_{i \rightarrow P^0}\ ,
&\hspace{5mm}&
A_{\bar i \rightarrow P^0} \rightarrow
e^{i(\overline{\gamma}_i-\gamma_P)} A_{\bar i \rightarrow P^0}\ ,
\nonumber\\*[2mm]
A_{i \rightarrow \overline{P^0}} \rightarrow
e^{i(\gamma_i-\overline{\gamma}_P)} A_{i \rightarrow \overline{P^0}}\ ,
&\hspace{5mm}&
A_{\bar i \rightarrow \overline{P^0}}\rightarrow
e^{i(\overline{\gamma}_i-\overline{\gamma}_P)}
A_{\bar i \rightarrow \overline{P^0}}\ ,
\nonumber\\*[2mm]
A_f \rightarrow
e^{i(\gamma_P - \gamma_f)} A_f \ ,
&\hspace{5mm}&
\bar A_f \rightarrow
e^{i(\overline{\gamma}_P - \gamma_f)} \bar A_f\ ,
\nonumber\\*[2mm]
A_{\bar f} \rightarrow
e^{i(\gamma_P - \overline{\gamma}_f)} A_{\bar f}\ ,
&\hspace{5mm}&
\bar A_{\bar f} \rightarrow
e^{i(\overline{\gamma}_P - \overline{\gamma}_f)}
\bar A_{\bar f}\ .
\end{eqnarray}
Only those quantities which remain invariant
under these redefinitions may have physical meaning.
Clearly,
the magnitudes of all the quantities in
Eq.~(\ref{eq:rephazing-quantities})
satisfy this criterion.

Besides these,
there are quantities which remain invariant under phase
redefinitions and which arise from the ``interference''
between the parameters describing the mixing and those
describing the transitions:
\begin{eqnarray}
\lambda_f \equiv \frac{q}{p}
\frac{\bar A_f}{A_f}\ ,
& \hspace{5mm} &
\lambda_{\bar f} \equiv \frac{q}{p}
\frac{\bar A_{\bar f}}{A_{\bar f}}\ ,
\label{def:nova:int-2}
\\ 
\xi_{i \rightarrow P} \equiv
\frac{A_{i \rightarrow \overline{P^0}}}{A_{i \rightarrow P^0}}
\frac{p}{q}\ ,
& \hspace{5mm} &
\xi_{\bar i \rightarrow P} \equiv
\frac{A_{\bar i \rightarrow \overline{P^0}}}{
A_{\bar i \rightarrow P^0}}
\frac{p}{q}\ .
\label{def:nova:int-3}
\end{eqnarray}
The parameters in Eq.~(\ref{def:nova:int-2})
describe the interference between
between mixing in the neutral meson system and
\textit{its subsequent decay} into the final states $f$ and $\bar f$.
In contrast,
the parameters in Eq.~(\ref{def:nova:int-3})
describe the
interference between the \textit{production of the neutral meson
system} and the mixing in that system.

\subsection{Conditions for CP conservation}

If CP were conserved,
then there would exist phases $\xi_i$, $\xi$ and $\xi_f$,
as well as a CP eigenvalue, $\eta_P=\pm1$,
such that
\begin{eqnarray}
{\cal CP} | i \rangle & = & e^{i\xi_i} | \bar i \rangle\ ,
\nonumber\\
{\cal CP} | P^0 \rangle & = & e^{i\xi} | \overline{P^0} \rangle\ ,
\nonumber\\
{\cal CP} | f \rangle & = & e^{i\xi_f} | \bar f \rangle\ ,
\label{eq:CP1}
\end{eqnarray}
and
\begin{eqnarray}
{\cal CP} | P_H \rangle &=& \eta_P | P_H \rangle\ ,
\nonumber\\
{\cal CP} | P_L \rangle &=& - \eta_P | P_L \rangle\ ,
\label{eq:CP2}
\end{eqnarray}
where, as usual $H$ ($L$) refers to the ``heavy''
(``light'') eigenvalue.
Here, we use the convention $\Delta m >0$.
With this convention,
it is the sign of $\eta_P$ which must be determined
by experiment.
For example,
we know from experiment that the heavier kaon also has
the longest lifetime.
Moreover,
if there were no CP violation in the mixture of neutral kaons,
this state would be CP odd.
As a result,
we must use $\eta_K = -1$ whenever we neglect the
mixing in the neutral kaon system.

On the other hand,
the CP transformation of the multi-particle intermediate
state $X P$ is given by
\begin{equation}
{\cal CP} | X P^0 \rangle  =  \eta_X e^{i\xi} 
| \overline{X}\,  \overline{P^0} \rangle\ .
\label{eq:CP3}
\end{equation}
Here,
$\eta_X$ contains the CP transformation properties of the state $X$,
as well as the parity properties corresponding to the
relative orbital angular momentum between $X$ and $P$.

From Eqs.~(\ref{eq:CP1}), (\ref{eq:CP2}), and (\ref{eq:CP3}),
we derive the conditions required for CP invariance,
\begin{equation}
\frac{q}{p} = - \eta_P e^{i \xi}
\label{CP_invariance_q/p}
\end{equation}
and
\begin{eqnarray}
A_{i \rightarrow P^0} =
\eta_X
e^{i(\xi_i - \xi)} A_{\bar i \rightarrow \overline{P^0}}\ ,
&\hspace{5mm}&
A_{i \rightarrow \overline{P^0}} =
\eta_X
e^{i(\xi_i + \xi)} A_{\bar i \rightarrow P^0}\ ,
\nonumber\\*[2mm]
A_{f} =
e^{i(\xi - \xi_f)} \bar A_{\bar f}\ ,
&\hspace{5mm}&
A_{\bar f} =
e^{i(\xi + \xi_f)} \bar A_{f}\ .
\label{CP_invariance_amplitudes}
\end{eqnarray}
Therefore,
if CP were a good symmetry
we would have
\begin{eqnarray}
\left| \frac{q}{p} \right| = 1\ ,
&\hspace{5mm}&
\nonumber\\*[2mm]
\left| A_{i \rightarrow P^0} \right| =
\left| A_{\bar i \rightarrow \overline{P^0}} \right|\ ,
&\hspace{5mm}&
\left| A_{i \rightarrow \overline{P^0}} \right| =
\left| A_{\bar i \rightarrow P^0} \right|\ ,
\nonumber\\*[2mm]
\left| A_{f}  \right| = 
\left| \bar A_{\bar f} \right|\ ,
&\hspace{5mm}&
\left| A_{\bar f} \right| =
\left| \bar A_{f} \right|\ .
\end{eqnarray}
Also,
the parameters describing interference CP violation
would become related by
\begin{eqnarray}
\lambda_{f}\, \lambda_{\bar f}
& = & 1\ ,
\nonumber\\
\xi_{i \rightarrow P}\, \xi_{\bar i \rightarrow P}
& = & 1\ .
\end{eqnarray}
This means that, if CP were conserved, then
$\arg \lambda_f + \arg \lambda_{\bar f}$
and
$\arg \xi_{i \rightarrow P} + \arg \xi_{\bar i \rightarrow P}$
would vanish.
We may use Eqs.~(\ref{CP_invariance_q/p}) and
(\ref{CP_invariance_amplitudes})
in order to find more complicated
conditions for CP invariance,
such as
\begin{equation}
A_{\bar i \rightarrow \overline{P^0}}
\bar A_{f}
(A_{\bar i \rightarrow P^0})^\ast
(A_{f})^\ast
=
A_{i \rightarrow P^0}
A_{\bar f}
(A_{i \rightarrow \overline{P^0}})^\ast
(\bar A_{\bar f})^\ast.
\end{equation}

A very important particular case occurs when the final
state $f$ is an eigenstate of CP.
In that case,
$\eta_f \equiv e^{i \xi_f} = \pm 1$,
and the conditions for CP invariance become
\begin{equation}
\label{CP inv with CP eigen}
\left| A_{f} \right| = 
\left| \bar A_{f} \right|
\ \ \mbox{e }\ \ 
\lambda_{f} = \eta_P \eta_f.
\end{equation}

As mentioned in the main text,
we have just found the usual three types of
CP violation
\begin{enumerate}
\item $|q/p| - 1$ describes CP violation in the
mixing of the neutral meson system;
\item $|A_{i \rightarrow P^0}| - |A_{\bar i \rightarrow \overline{P^0}}|$
and $|A_{i \rightarrow \overline{P^0}}| - |A_{\bar i \rightarrow P^0}|$,
on the one hand,
and $|A_{f}| - |\bar A_{\bar f}|$ and $|A_{\bar f}| - |\bar A_{f}|$,
on the other hand,
describe the CP violation present directly in the
production of the neutral meson system and in its decay,
respectively;
\item $\arg \lambda_{f} + \arg \lambda_{\bar f}$
measures the CP violation arising from the interference
between mixing in the neutral meson system and
\textit{its subsequent decay} into the final states $f$ and $\bar f$.
We call this the ``interference CP violation: first mix, then decay''. 
When $f=f_{cp}$ is an CP eigenstate,
this CP violating observable $\arg \lambda_{f} + \arg \lambda_{\bar f}$,
becomes proportional to $\mbox{Im} \lambda_{f}$.
\end{enumerate}

However,
we can also identify a new type of CP violating observable
\be
\arg \xi_{i \rightarrow P} + \arg \xi_{\bar i \rightarrow P}.
\ee
This observable measures the CP violation arising from the
interference between the \textit{production of the neutral meson
system} and the mixing in that system.
We call this the ``interference CP violation: first produce, then mix''.

\vspace{1cm}

\section{\label{appendixEx}Exercises}

\vspace{3mm}

\begin{itemize}
\item[\textbf{Ex-1:}]
Prove Eqs.~(\ref{CP-of-pipi}).
For one single pion,
${\cal CP} \pi^\pm = - \pi^\mp$,
${\cal CP} \pi^0 = - \pi^0$.
You should check in an introductory book that
these properties may be inferred, for example,
from the processes
$\pi^- + d \rightarrow n + n$
and
$\pi^0 \rightarrow \gamma \gamma$,
and require an implicit phase convention
(see, for example, reference \cite{BLS}).
For multiple pions we must worry about the
relative angular momentum,
since the orbital wave functions change with
P as $(-1)^L$.
Clearly,
the system of two pions originating from a
kaon decay must be in a relative S wave.
\vspace{0.5cm}
\item[\textbf{Ex-2:}]
Using Eqs.~(\ref{PaPb}) and (\ref{X-1}),
check that $\mbox{\boldmath $X$} \mbox{\boldmath $X$}^{-1} =
\mbox{\boldmath $X$}^{-1} \mbox{\boldmath $X$} =
\mbox{\boldmath $1$}$.
\vspace{0.5cm}
\item[\textbf{Ex-3:}]
Prove the last equality of Eq.~(\ref{diagonalization_inverted}).
\vspace{0.5cm}
\item[\textbf{Ex-4:}]
Using Eqs.~(\ref{real_dmu_2})--(\ref{delta_new_expressions}),
show that
\ba
|M_{12}|^2 & = &
\frac{4 (\Delta m)^2 + \delta^2 (\Delta \Gamma)^2}{
16 (1 - \delta^2)},
\nonumber\\
|\Gamma_{12}|^2 & = &
\frac{(\Delta \Gamma)^2 + 4 \delta^2 (\Delta m)^2}{
4 (1 - \delta^2)},
\ea
and
\ba
(\Delta m)^2 & = &
\frac{4 |M_{12}|^2 - \delta^2 |\Gamma_{12}|^2}{
1 + \delta^2},
\nonumber\\
(\Delta \Gamma)^2 & = &
\frac{4 |\Gamma_{12}|^2 - 16 \delta^2 |M_{12}|^2}{
1 + \delta^2}.
\ea
Show also that
\ba
- \frac{q}{p} \Gamma_{12} & = &
\frac{y + i \delta x}{1 + \delta} \Gamma,
\nonumber\\
- \frac{q}{p} M_{12} & = &
\frac{x - i \delta y}{2(1 + \delta)} \Gamma,
\ea
where $x$ and $y$ are defined in Eq.~(\ref{eq:xandy}).
\vspace{0.5cm}
\item[\textbf{Ex-5:}]
To reach this conclusion in a different way,
use Eq.~(\ref{diagonalization_inverted})
to prove that
\be
\left[ \mbox{\boldmath $H$}, \mbox{\boldmath $H$}^\dagger \right]
=
\left| \Delta \mu \right|^2
\frac{\delta}{ 1 - \delta^2}
\left(
\begin{array}{cc}
1 & 0\\
0 & -1
\end{array}
\right),
\ee
meaning that this commutator is directly proportional to the
CP violating observable $\delta$.
If you need to use
\be
\left| \frac{q}{p} \right|^2 = \frac{1 - \delta}{1 + \delta},
\ee
prove it first.
\vspace{0.5cm}
\item[\textbf{Ex-6:}]
Here you prove 
Eqs.~(\ref{g+-})--(\ref{standard-time-evolution-1})
fully in matrix form,
starting from Eq.~(\ref{decomposition-mass}).
Use
Eqs.~(\ref{definition-reciprocal}) and (\ref{partition-of-unity}),
together with the fact that
$|P_H \rangle \langle \tilde P_H|$ and
$|P_L \rangle \langle \tilde P_L|$ are projection operators,
in order to show that
\begin{eqnarray}
\exp{(- i {\cal H} t)}
&=&
e^{-i \mu_H t} |P_H \rangle \langle \tilde P_H|
+
e^{-i \mu_L t} |P_L \rangle \langle \tilde P_L|.
\nonumber\\*[3mm]
&=&
\left(
\begin{array}{cc}
| P_H \rangle, & | P_L \rangle
\end{array}
\right)
\left(
\begin{array}{cc}
e^{-i \mu_H t} & 0\\
0 & e^{-i \mu_L t}
\end{array}
\right)
\left(
\begin{array}{c}
\langle \tilde P_H | \\
\langle \tilde P_L |
\end{array}
\right).
\label{ev-operator}
\end{eqnarray}
In fact, 
so far you have just taken a rather long path to
prove the trivial statements in Eqs.~(\ref{mass-basis-time-evolution}).
Now,
use Eqs.~(\ref{PaPb}),
(\ref{X-1}),
and (\ref{PtildeaPtildeb}),
to show that
\begin{eqnarray}
\exp{(- i {\cal H} t)}
&=&
\left(
\begin{array}{cc}
| P^0 \rangle, & | \overline{P^0} \rangle
\end{array}
\right)
\mbox{\boldmath $X$}
\left(
\begin{array}{cc}
e^{-i \mu_H t} & 0\\
0 & e^{-i \mu_L t}
\end{array}
\right)
\mbox{\boldmath $X$}^{-1}
\left(
\begin{array}{c}
\langle P^0 | \\
\langle \overline{P^0} |
\end{array}
\right)
\nonumber\\*[3mm]
&=&
\left(
\begin{array}{cc}
| P^0 \rangle, & | \overline{P^0} \rangle
\end{array}
\right)
\left(
\begin{array}{cc}
g_+(t)\  & \frac{p}{q} g_-(t)\\*[2mm]
\frac{q}{p} g_-(t)\  & g_+(t)
\end{array}
\right)
\left(
\begin{array}{c}
\langle P^0 | \\
\langle \overline{P^0} |
\end{array}
\right).
\label{evolution-in-time}
\end{eqnarray}
Get Eq.~(\ref{standard-time-evolution-1}) from this.
\vspace{0.5cm}
\item[\textbf{Ex-7:}]
Prove Eqs.~(\ref{g+-quad}).
\vspace{0.5cm}
\item[\textbf{Ex-8:}]
Check Eqs.~(\ref{GpmG+-})--(\ref{eq:xandy}).
\vspace{0.5cm}
\item[\textbf{Ex-9:}]
Prove all the equalities in Eq.~(\ref{eq:A_M}).
\vspace{0.5cm}
\item[\textbf{Ex-10:}]
Prove Eqs.~(\ref{master_Bs})--(\ref{S_f}).
\vspace{0.5cm}
\item[\textbf{Ex-11:}]
Prove Eqs.~(\ref{Lf_from_DCS}) and (\ref{D_C_S}).
\vspace{0.5cm}
\item[\textbf{Ex-12:}]
To practice with this compact notation,
write the $h.c.$ (hermitian conjugated) terms  of
Eq.~(\ref{L_Yukawa}) explicitly.
Clearly,
the final expression is a number,
and one may decide to rewrite the expression by taking its
transpose.
If one does this,
one must include an explicit minus sign,
which arises from the fact that,
when taking the transpose,
one is interchanging the position of two fermion fields
which, as such, anti-commute.
This detail will be important for 
\textbf{(Ex-16, Ex-17, Ex-20)}.
\vspace{0.5cm}
\item[\textbf{Ex-13:}]
Defining
\ba
W_\mu^\pm &=&
\frac{W_\mu^1 \mp i W_\mu^2}{\sqrt{2}},
\nonumber\\
\tau_\pm &=&
\frac{\tau_1 \pm i \tau_2}{\sqrt{2}},
\nonumber\\
e &=&
g \sin{\theta_W} = g^\prime \cos{\theta_W},
\ea
and\footnote{Notice that Eq.~(\ref{eq:Q}) makes sense for
doublet fields,
if you think of $Q$ and $Y$ multiplying the unit $2 \times 2$ matrix.
For singlet fields, take $\tau_3 \rightarrow 0$.}
\be
Q = \frac{1}{2} \tau_3 + Y, 
\label{eq:Q}
\ee
show that
\ba
 & &
- \frac{g}{2}\,
\vec{\tau}.\vec{W}_\mu
- g^\prime\, Y B_ \mu 
=
\nonumber\\
 & &
\hspace{7mm}
- \frac{g}{2}
\left( \tau_+ W_\mu^+ + \tau_- W_\mu^- \right)
- e Q A_\mu
- \frac{g}{\cos{\theta_W}}
\left( \frac{1}{2} \tau_3 - Q \sin^2{\theta_W} \right) Z_\mu.
\ea
Use this to prove that
Eqs.~(\ref{LW_weak}) and (\ref{LZ_weak})
follow from Eqs.~(\ref{L_kinetic}) and (\ref{cov_der}).
\vspace{0.5cm}
\item[\textbf{Ex-14:}]
Show that Eq.~(\ref{LH_mass}) follows from Eqs.~(\ref{L_Yukawa}) and
(\ref{eq:Higgs_basis}),
with the basis transformations in Eqs.~(\ref{mass_basis}).
\vspace{0.5cm}
\item[\textbf{Ex-15:}]
Show that Eqs.~(\ref{LW_mass})--(\ref{LZ_mass})
follow by applying the basis transformations in Eqs.~(\ref{mass_basis})
to Eqs.~(\ref{LW_weak})--(\ref{LZ_weak}).
\vspace{0.5cm}
\item[\textbf{Ex-16:}]
Recalling that $\gamma^0 \gamma^\mu \gamma^0 = \gamma_\mu$,
$C^{-1} \gamma_\mu C = - \gamma_{\mu}^T$,
and the other trivialities about Dirac $\gamma$-matrices,
use the CP transformations in Eq.~(\ref{CP_with_rephasing}),
to prove that:
\ba
\left( \cp \right) \bar u d \left( \cp \right)^\dagger
&=&
e^{i (\xi_d - \xi_u)}\,
\bar d u,
\nonumber\\
\left( \cp \right) \bar u \gamma_5 d \left( \cp \right)^\dagger
&=&
- e^{i (\xi_d - \xi_u)}\,
\bar d \gamma_5 u,
\nonumber\\
\left( \cp \right) \bar u \gamma^\mu d \left( \cp \right)^\dagger
&=&
- e^{i (\xi_d - \xi_u)}\,
\bar d \gamma_\mu u,
\nonumber\\
\left( \cp \right) \bar u \gamma^\mu \gamma_ 5 d \left( \cp \right)^\dagger
&=&
- e^{i (\xi_d - \xi_u)}\,
\bar d \gamma_\mu \gamma_ 5 u,
\ea
where an extra minus sign appears when taking the transpose,
because the two fermion fields anti-commute.
Also,
\be
\left( \cp \right) 
\left[ \bar u \gamma^\mu (1 - \gamma_ 5) d 
\right]
\left( \cp \right)^\dagger
=
- e^{i (\xi_d - \xi_u)}
\left[
\bar d \gamma_\mu (1- \gamma_ 5) u
\right].
\ee
\vspace{0.5cm}
\item[\textbf{Ex-17:}]
Verify that the Lagrangian in Eq.~(\ref{LW_expanded})
is invariant under the CP transformations in Eq.~(\ref{CP_with_rephasing})
if and only if Eq.~(\ref{afinal}) holds.
\vspace{0.5cm}
\item[\textbf{Ex-18:}]
Define
\be
Q_{\alpha i \beta j} = 
V_{\alpha i} V_{\beta j} V_{\alpha j}^\ast V_{\beta i}^\ast,
\ee
and show that $Q_{\alpha i \beta j} = Q_{\beta j \alpha i}= 
Q_{\alpha j \beta i}^\ast = Q_{\beta i \alpha j}^\ast$.
Thus,
$\mbox{Im} Q_{\alpha i \beta j}$ may change sign 
under a reshuffling of the indexes,
and the magnitude is useful in Eq.~(\ref{J_CKM}).
\vspace{0.5cm}
\item[\textbf{Ex-19:}]
Prove that \cite{Nir01}
\be
\mbox{Im}
\left(
V_{\alpha i} V_{\beta j} V_{\alpha j}^\ast V_{\beta i}^\ast
\right)
=
J_{\rm CKM}
\sum_{\gamma = 1}^3
\sum_{k = 1}^3
\epsilon_{\alpha \beta \gamma}
\epsilon_{i j k}.
\ee
\vspace{0.5cm}
\item[\textbf{Ex-20:}]
Verify that the Yukawa Lagrangian in Eq.~(\ref{L_Yukawa})
is invariant under the CP transformations in
Eq.~(\ref{CP_with_WBT})
if and only if Eq.~(\ref{afinal3}) holds.
\vspace{0.5cm}
\item[\textbf{Ex-21:}]
Show that
\be
\mbox{Im}
\left\{
\mbox{Tr} \left( H_u H_d \right)
\right\}
= 0 =
\mbox{Im}
\left\{
\mbox{Tr} \left( H_u^2 H_d^2 \right)
\right\}.
\ee
\vspace{0.5cm}
\item[\textbf{Ex-22:}]
Show that
\ba
&&
\mbox{Im}
\left\{
\mbox{Tr} \left( H_u H_d H_u^2 H_d^2 \right)
\right\}
=
\sum_{\alpha,\beta = u,c,t}
\,
\sum_{i,j = d,s,b}
m_{u_\alpha}^2
m_{d_i}^2
m_{u_\beta}^4
m_{d_j}^4
\mbox{Im} \left( Q_{\alpha i \beta j} \right)
\nonumber\\
&&
= (m_t^2 - m_c^2)(m_t^2 - m_u^2)(m_c^2 - m_t^2)
(m_b^2 - m_s^2)(m_b^2 - m_d^2)(m_s^2 - m_d^2)
J_{\rm CKM}.
\ea
You may need the result in \textbf{(Ex-19)}.
\vspace{0.5cm}
\item[\textbf{Ex-23:}]
Show that the Chau--Keung parametrization in
Eq.~(\ref{Chau-Keung}) results from
\be
V =
\left(
\begin{array}{ccc}
  1 & 0 & 0\\
  0 & c_{23} & s_{23}\\
  0 & - s_{23} & c_{23}
\end{array}
\right)
\,
\left(
\begin{array}{ccc}
  c_{13} & 0 & s_{13} e^{-i \delta_{13}}\\
  0 & 1 & 0\\
  - s_{13} e^{i \delta_{13}} & 0 & c_{13}
\end{array}
\right)
\,
\left(
\begin{array}{ccc}
  c_{12} & s_{12} & 0\\
  - s_{12} & c_{12} & 0\\
  0 & 0 & 1
\end{array}
\right).
\ee
\vspace{0.5cm}
\item[\textbf{Ex-24:}]
Prove that the definitions of $\alpha$,
$\beta$, and $\gamma$ in Eqs.~(\ref{alpha})--(\ref{gamma})
imply that $\alpha + \beta + \gamma = \mbox{arg}(-1)$,
leading to Eq.~(\ref{a+b+g}).\footnote{I wouldn't put this 
trivial exercise here,
were it not for the fact that some misleading statements are
sometimes made about this.}
\vspace{0.5cm}
\item[\textbf{Ex-25:}]
Use the unitarity of the CKM matrix in order to 
prove Eqs.~(\ref{Rb_Rt_and_sines}).
\vspace{0.5cm}
\item[\textbf{Ex-26:}]
Prove that all the triangles in
Eqs.~(\ref{triangle_ds})--(\ref{triangle_ut})
have the same area $J_{\rm CKM}/2$.
\vspace{0.5cm}
\item[\textbf{Ex-27:}]
Obtain Eq.~(\ref{rescalled_triangle}) from
Eq.~(\ref{triangle_db}) and the 
definitions in Eqs.~(\ref{R_b})--(\ref{gamma}).
Use the Wolfenstein parametrization,
through Eqs.~(\ref{R_t_Wolf}) and (\ref{R_b_Wolf})
to show that this represents a triangle which has an
apex at coordinates $(\rho, \eta)$ and area $|\eta|/2$.
Check also how much simpler this gets if one uses instead
the redundant parametrization in Eq.~(\ref{crazy_V}).
\vspace{0.5cm}
\item[\textbf{Ex-28:}]
Expand Eq.~(\ref{Lf_r_and_delta}) to first order in
$r$.
Substituting into Eqs.~(\ref{C_f}) and (\ref{S_f}),
verify Eqs.~(\ref{def:adir}) and (\ref{def:aint}).
\vspace{0.5cm}
\item[\textbf{Ex-29:}]
The diagram in 
FIG.~\ref{figura treeJK} is proportional to
$V_{cb}^\ast V_{cs} \sim (A \lambda^2)(1)$ and,
thus, it carries no weak phase in the
standard phase convention for the CKM matrix.
Use this,
together with Eqs.~(\ref{simple_q/p_B}) and (\ref{simple_q/p_K}),
to show that
\be
\lambda_{B_d \rightarrow J/\psi K_S}
=
- e^{- 2 i(\tilde{\beta} - \chi^\prime)}.
\ee
In the absence of new physics in $B^0_d - \overline{B^0_d}$
mixing,
$\tilde{\beta} = \beta$,
and we recover the result on the last line of
Eq.~(\ref{lambda_psiKs_SM}).
Now we know why most people ignore the spurious phases $\xi$.
\vspace{0.5cm}
\item[\textbf{Ex-30:}]
Here we study the isospin decomposition of the decay amplitudes
for $B \rightarrow \pi \pi$ in some detail.
\begin{itemize}
\item[\textbf{a)}] Since the pions are spinless,
they must arise from $B$ decays in an $s$ wave and they must be in
an overall symmetric state.
This implies a symmetric isospin configuration.
Use addition of angular momenta to show that the resulting
final states are:
\ba
\langle \pi^0 \pi^0 | &=&
\sqrt{\frac{2}{3}} \langle 2, 0 | 
- \sqrt{\frac{1}{3}} \langle 0, 0 |,
\nonumber\\
\langle \pi^+ \pi^- |
&\equiv&
\frac{1}{\sqrt{2}} \left(
\langle \pi_1^+ \pi_2^- | + \langle \pi_1^- \pi_2^+ |
\right)
=
\sqrt{\frac{1}{3}} \langle 2, 0 | 
+ \sqrt{\frac{2}{3}} \langle 0, 0 |,
\nonumber\\
\langle \pi^+ \pi^0 |
&\equiv&
\frac{1}{\sqrt{2}} \left(
\langle \pi_1^+ \pi_2^0 | + \langle \pi_1^0 \pi_2^+ |
\right)
=
\langle 2, 1 | .
\ea
\item[\textbf{b)}] The first two channels are reached by
$| B^0_d \rangle = |1/2, -1/2 \rangle $,
the third by $| B^+ \rangle = |1/2, 1/2 \rangle $.
In general,
the transition matrix has $\Delta I = 1/2$,
$\Delta I = 3/2$,
and $\Delta I = 5/2$ pieces.
Use the Wigner-Eckart theorem to show that
\ba
\langle \pi^0 \pi^0 | T | B^0_d \rangle
&=&
- \sqrt{\frac{1}{3}} A_{1/2}
+ \sqrt{\frac{1}{6}} A_{3/2}
- \sqrt{\frac{1}{6}} A_{5/2},
\nonumber\\
\langle \pi^+ \pi^- | T | B^0_d \rangle
&=&
\sqrt{\frac{1}{6}} A_{1/2}
+ \sqrt{\frac{1}{3}} A_{3/2}
- \sqrt{\frac{1}{3}} A_{5/2},
\nonumber\\
\langle \pi^+ \pi^0 | T | B^+ \rangle
&=&
\frac{\sqrt{3}}{2} A_{3/2}
+ \sqrt{\frac{1}{3}} A_{5/2},
\label{isospin_decomposition}
\ea
where $A_k$ are the relevant reduces matrix elements.
\item[\textbf{c)}] In the SM,
\begin{itemize}
\item tree level diagrams --- contribute to $A_{1/2}$ and  $A_{3/2}$;
\item gluonic penguin diagrams --- contribute only to $A_{1/2}$;
\item electroweak penguin diagrams --- are expected to be small;
\item $A_{5/2} \sim \alpha A_{1/2}$ --- arise from $A_{1/2}$
together with the $\Delta I = 2$ electromagnetic rescattering
of the two pions in the final state \cite{Don}.
Because of the $\Delta I = 1/2$ rule in place for
the decay $K \rightarrow \pi \pi$,
the contribution from $A_{5/2}$ is detectable and has been measured
in the kaon system \cite{5/2_BLS},
but it is expected to be negligible in
$B \rightarrow \pi \pi$ decays.
\end{itemize}
Neglecting $A_{5/2}$,
use Eqs.~(\ref{isospin_decomposition}) to prove
Eq.~(\ref{GL_isospin_triangle}).
\end{itemize}
\vspace{0.5cm}
\item[\textbf{Ex-31:}]
Do the trivial exercise to get 
Eq.~(\ref{eq:FM-1}).
\vspace{0.5cm}
\item[\textbf{Ex-32:}]
Derive the isospin decomposition in
Eq.~(\ref{eq:isospin_NirQuinn}).
\vspace{0.5cm}
\item[\textbf{Ex-33:}]
Check that Eq.~(\ref{eq:relation_isospin_diagrammatics})
follows trivially by comparing Eq.~(\ref{eq:isospin_NirQuinn}) with
FIG.~\ref{fig:B_Kpi_panel}.
See \cite{GR-LSR} for a generalization,
including other diagrams neglected here.
\vspace{0.5cm}
\item[\textbf{Ex-34:}]
Use the diagrammatic decomposition in FIG.~\ref{fig:B_Kpi_panel}
to show the first equality of Eq.~(\ref{Rc-Rn}).
Use the isospin decomposition of
Eq.~(\ref{eq:isospin_NirQuinn})
to show that the leading terms in
$R_c - R_n$ are proportional to a quadratic
combination of $A_{1/2}$ and $A_{3/2}$
over the square of $B_{1/2}$,
thus explaining the written comment in
Eq.~(\ref{eq:isospin_NirQuinn}).
\vspace{0.5cm}
\item[\textbf{Ex-35:}]
Prove the last equality in
Eq.~(\ref{R_L}).
\vspace{0.5cm}
\item[\textbf{Ex-36:}]
Prove the last equality in Eq.~(\ref{masterCPT})
of appendix~\ref{appendix-CPT}.
\vspace{0.5cm}
\item[\textbf{Ex-37:}]
When a neutral meson system propagates through matter,
it is subject to additional strangeness-preserving interactions
which may be parametrized by
\begin{equation}
\mbox{\boldmath $H$}_{\rm nuc} =
\left(
\begin{array}{cc}
\chi & 0\\
0 & \bar \chi
\end{array}
\right),
\label{nuc}
\end{equation}
which are written in the $P^0 - \overline{P^0}$ rest frame
and must be added to the Hamiltonian in vacuum.
The full Hamiltonian in matter becomes
\begin{equation}
\mbox{\boldmath $H$}^\prime = 
\mbox{\boldmath $H$} + \mbox{\boldmath $H$}_{\rm nuc},
\label{hh-in-matter}
\end{equation}
where we denote matrices, matrix elements and eigenvalues
in vacuum by unprimed quantities and their analogues
in matter by primed quantities.

Now,
we have already studied the most general effective Hamiltonian,
and Eq.~(\ref{masterCPT}) relates such an Hamiltonian written in the
flavor basis with the corresponding eigenvalues and mixing parameters.
Use Eqs.~(\ref{masterCPT}), (\ref{nuc}) and (\ref{hh-in-matter})
to show that \cite{reciprocal}
\begin{equation}
\left(
\begin{array}{cc}
\mu^\prime - \frac{\Delta \mu^\prime}{2} \theta^\prime \ &
- \frac{p^\prime}{q^\prime} \frac{\sqrt{1-\theta^{\prime 2}}}{2}
\Delta \mu^\prime
\\*[2mm]
- \frac{q^\prime}{p^\prime} \frac{\sqrt{1-\theta^{\prime 2}}}{2}
\Delta \mu^\prime \ &
\mu^\prime + \frac{\Delta \mu^\prime}{2} \theta^\prime
\end{array}
\right)
=
\left(
\begin{array}{cc}
\mu - \frac{\Delta \mu}{2} \theta \ &
- \frac{p}{q} \frac{\sqrt{1-\theta^2}}{2} \Delta \mu
\\*[2mm]
- \frac{q}{p} \frac{\sqrt{1-\theta^2}}{2} \Delta \mu \ &
\mu + \frac{\Delta \mu}{2} \theta
\end{array}
\right)
+
\left(
\begin{array}{cc}
\chi & 0\\
0 & \bar \chi
\end{array}
\right).
\label{master-matter}
\end{equation}

Now prove the following results:
\begin{itemize}
\item[\textbf{a)}] Clearly,
$H^\prime_{12}= H_{12}$, $H^\prime_{21}= H_{21}$,
and $q^\prime/p^\prime=q/p$.
This means that
the CP and T violating parameter $\delta$,
which depends on $|q^\prime/p^\prime|=|q/p|$,
is the same in vacuum and in the presence of matter.
\item[\textbf{b)}] Prove that
the parameters in vacuum and in matter are related through,
\begin{eqnarray}
\mu^\prime &=& \mu + \frac{\chi + \bar \chi}{2},
\nonumber\\
\Delta \mu^\prime &=& \sqrt{(\Delta \mu)^2 + 
2 \theta\, \Delta \mu\, \Delta \chi + (\Delta \chi)^2}
= \Delta \mu\, \sqrt{1 + 4 r\, \theta + 4 r^2},
\nonumber\\
\theta^\prime &=&
\frac{\Delta \mu\, \theta + \Delta \chi}{
\sqrt{(\Delta \mu)^2 + 2 \theta\, \Delta \mu\, \Delta \chi + 
(\Delta \chi)^2}}
= \frac{\theta + 2 r}{\sqrt{1 + 4 r\, \theta + 4 r^2}},
\label{matter-vacuum-general}
\end{eqnarray}
where $\Delta \chi = \bar \chi - \chi$,
and we have introduced the `regeneration parameter'
$r=\Delta \chi/(2 \Delta \mu)$.
\item[\textbf{c)}] Infer from Eqs.~(\ref{master-matter})
and (\ref{matter-vacuum-general})
that the flavor-diagonal matter effects considered here
act just like violations of CPT.
\item[\textbf{d)}] Since we expect the matter effects to
be much larger than any (necessarily small) CPT-violation
that there might be already present in vacuum,
set $\theta=0$ to get
\begin{eqnarray}
\mu^\prime &=& \mu + \frac{\chi + \bar \chi}{2},
\nonumber\\
\Delta \mu^\prime &=& \sqrt{(\Delta \mu)^2 + (\bar \chi - \chi)^2}
= \Delta \mu\, \sqrt{1 + 4 r^2},
\nonumber\\
\theta^\prime &=&
\frac{\bar \chi - \chi}{\sqrt{(\Delta \mu)^2 + (\bar \chi - \chi)^2}}
= \frac{2 r}{\sqrt{1 + 4 r^2}}.
\label{matter-vacuum-particular}
\end{eqnarray}
\item[\textbf{e)}]
Because Eq.~(\ref{masterCPT}) is completely general,
so is the time evolution in Eq.~(\ref{usual-time-evolutionCPT}).
Therefore,
obtain the time-evolution in matter simply by
substituting the unprimed quantities in
Eq.~(\ref{usual-time-evolutionCPT}) by primed
quantities.\footnote{This solution had been
found for the kaon system by Good \cite{Goo57},
building on earlier work by Case \cite{Cas56},
but the authors write a new evolution equation obtained by combining
the diagonalized form of $\mbox{\boldmath $H$}$ with the new term 
$\mbox{\boldmath $H$}_{\rm nuc}$
written in the $\{K_L, K_S\}$ basis.
Thus,
they would seem to be solving a new complicated set of equations:
the so-called `Good equations'.
In the method presented here,
we have made no reference to `new' differential equations.
We had already solved the most general evolution equation once and for all,
Eqs.~(\ref{usual-time-evolutionCPT});
and we had seen how $\mbox{\boldmath $H$}$ 
could be written in terms of observables,
Eq.~(\ref{masterCPT}).
All we had to do was to refer back to those results.}
We stress that the primed quantities which refer to the propagation
in matter are obtained from the properties in
vacuum, from $\chi$, and from $\bar \chi$ through
Eqs.~(\ref{matter-vacuum-particular}).
\end{itemize}
\end{itemize}



\begin{thebibliography}{99}
\bibitem{complicacao}
For historical reasons,
one usually introduces two new ``experimental''
quantities when studying the neutral meson system,
defined by
%
\ba
\epsilon_K &\equiv&
\frac{2 \eta_{+-} + \eta_{00}}{3},
\nonumber\\
\epsilon^\prime_K &\equiv&
\frac{\eta_{+-} - \eta_{00}}{3},
\ea
%
where
%
\ba
\eta_{+-} &\equiv& \frac{\langle \pi^+ \pi^- | T | K_L \rangle}{
\langle \pi^+ \pi^- | T | K_S \rangle},
\nonumber\\
\eta_{00} &\equiv& \frac{\langle \pi^0 \pi^0 | T | K_L \rangle}{
\langle \pi^0 \pi^0 | T | K_S \rangle}.
\ea
%
Unfortunately,
the literature contains several distinct definitions
for the ``theoretical'' parameters $\epsilon_K$ and $\epsilon^\prime_K$.
Although the fact that $\epsilon_K$ and $\epsilon^\prime_K$
are very small implies that all those definitions
are operationally equivalent,
a careful analysis of this issue lies well outside the
scope of these lectures.
More details may be found in \cite{BLS,Nir01}.
Fortunately, these definitions are not necessary for our
purposes.
On the one hand,
the calculation of $\epsilon^\prime_K$ involves always
large hadronic uncertainties.
On the other hand,
for all the definitions of $\epsilon_K$
%
\be
\delta_K \approx \sqrt{2} |\epsilon_K|.
\ee
%
As a result we may refer only to $\delta_K$,
whose definition is clear and universal.
For a good explanation of the relation of $\epsilon_K$
and $\epsilon^\prime_K$ with the parameters
$|q/p|-1$, $|\bar A_f| - |A_f|$, and $\mbox{Im} \lambda_f$ used
in these lectures, see reference \cite{Nir01}.
\bibitem{appendixA-BLS}
The reconciliation of Eq.~(\ref{bad_q/p}) with Eq.~(\ref{good_q/p_1})
was explained for the first time and in great detail in 
appendix A of \cite{BLS}.
\bibitem{garbage}
Some authors state that what matters is
the size of the phase difference $\phi_1 - \phi_2$
between the two diagrams.
Others concentrate on the ratio $r_{\rm ckm}$ of the
magnitudes of the matrix elements.
Strictly speaking,
this is poor wording.
In fact,
for any particular decay,
we can use the unitarity of the CKM matrix in order
to change the two phases we concentrate on,
thus changing the sizes of $r_{\rm ckm}$ and $\phi_1 - \phi_2$.
It is best to talk about
$r_{\rm ckm} \sin{(\phi_1 - \phi_2)}$,
although the issue is complicated because the relevant
operators (and their matrix elements) also change.
For a specific example see section 33.4.1 of \cite{BLS}.
\bibitem{controversy}
There is considerable discussion on how to correctly
parametrize CPT once one wishes to go beyond the
Lee-Oehme-Yang approximation.
This issue is way beyond the level of these lectures
and highly controversial.
One would do best to enlist the help of an expert
on these delicate questions before attempting to
read the literature,
a reason why none is mentioned here.
\end{thebibliography}
\end{document}